\pdfoutput=1
\documentclass[showpacs,superscriptaddress, preprintnumbers,amsmath,amssymb]{revtex4}

\usepackage{graphicx}
\usepackage{dcolumn}
\usepackage{bm}
\include{epsf}

\begin{document}


\title{On World Religion Adherence Distribution Evolution}

\author{M. Ausloos}\email{marcel.ausloos@ulg.ac.be}
\affiliation{GRAPES, B5 Sart-Tilman, B-4000 Li\`ege, Euroland
}

\author{F. Petroni}\email{filippo.petroni@ulg.ac.be}
\affiliation{GRAPES, Universit\'e de Li\`ege, B5 Sart-Tilman, B-4000 Li\`ege, Belgium
}
\affiliation{Universit\'a dell'Aquila, I-67010, L'Aquila, Italy}%


\begin{abstract}
Religious adherence can be considered as a {\it degree of freedom}, in a statistical physics sense, for a human agent belonging to a population. The distribution, performance  and life time of religions can thus be studied having in mind heterogeneous interacting agent modeling in mind. We present a comprehensive analysis of 58 so called religion (to be better defined in the main text) evolutions, as measured through their number of adherents between 1900 and 2000, - data taken from the World Christian Encyclopedia:  40 are considered to be ''presently growing'' cases, including 11 turn overs in the XX century;  18 are ''presently  decaying'', among which 12 are found to have had a recent maximum, in the XIX or the XX century. 
The Avrami-Kolmogorov differential equation which usually describes solid state transformations, like crystal growth, is used in each case in order to obtain the preferential attachment parameter introduced previously \cite{religion1}. It is often found close to unity, indicating a smooth evolution. However large values suggest the occurrence of extreme cases which we conjecture are controlled by so called external fields. A few cases indicate the likeliness of a detachment process.  We discuss different growing and decaying religions, and illustrate various fits. Some cases seem to indicate the lack of reliability of the data. Others, departure from Avrami law. We point out two difficulties in the analysis : (i) the ''precise'' original time of apparition of a religion, (ii) the time of its maximum, both informations being necessary for integrating reliably any evolution equation. Moreover the Avrami evolution equation might be surely improved, in particular, and somewhat obviously, for the decaying religion cases.  
\noindent

\end{abstract}



\maketitle

\section{Introduction}

Religion like sex, age, wealth, political affiliation, language, ... can be considered to characterize a group or an individual status. Whence religion or sex, age, wealth, political affiliation, language distributions can be studied as a function of time, space, auto-correlated, or correlated with any other variable or ''parameter'' characterizing a population,  going toward socio-economic studies pertaining to attitudes, behaviors, opinion formations \cite{holyst}, ... 
Several interesting considerations well known in statistical physics can be found in most sociological systems : the role of  nucleation, growth, aging, death, criticality, self-organization, epidemic spreading, and subsequent avalanches. If some geometric-like transition or some thermodynamical-like transition exists then fluctuations should be seen. 

Recently the dynamics of world's languages, especially on their disappearing due to competition with other languages \cite{Abr+03} has been of interest \cite{Viviane} in such a respect.
One of our aims has been recently to approach a set of similar questions  on religions, through a statistical physics point of view, attempting to quantify religion dynamics as seen from individual adherence distribution functions \cite{religion1}. 
We emphasize that we are not interested here in any religion's
origin,  history or in finding any hierarchy, but rather in the statistical physics-like
aspects of a complex non-equilibrium biological agent based system \cite{complexity1,complexity2}. 

 History is full of examples of individuals or entire groups of people
changing their religion, - for various reasons: following the ''leader'', e.g. Constantinus,  Clovis, ...  or ''external pressure'' , leading to martyrdom, or ''conversely''  like at inquisition time, or following a fatwah, ... or ''internal pressure'' (Khazars, ...) or so called adaptation under proselytism action, e.g. sun worshipping Incas in presence of catholic missionaries,  zoroastrian Persians in presence of muslim arabs,  ... ''Competition'' through interactions or under ''external field conditions'' exist in many cases. In so doing the number of adherents can much evolve due to such various conditions \cite{roach}. However notice that external field conditions can be rather more  drastic in the religious domain than in language history \cite{footnote1}. 
See also Appendix A for some discussion outlining a few aspects, i.e. ''differences'' between languages and religions, from a physics point of view, perspective or input into modeling such sociological features.

We consider as the fundamentally relevant variable the number of adherents of each religion  \cite{footnote2} Only this number is treated as the physics object.
Thus a religion is hereby considered as a (socially based)
variable, like a language or wealth, to be so studied like any other
organizational parameter for defining a thermodynamic-like state. We recognize that a religious state \cite{footnote3}  is more individualistic than a linguistic state.  Thus, in some sense one can better define the religious adherence of an agent than the linguistic one.   Indeed one can hardly be multi-religious but one can be a polyglot. Of course one can switch more easily, i.e. through ''conversion''  from one religious denomination to another than in language cases. Thus the observation time of a religious state needs careful attention in surveys.

From another point of view, time and time scales, one can notice that a religion can seem to appear rather instantaneously, often as a so called sect,  at the beginning, and its number of adherent can grow steadily (see the recent Mormon or Rastafarianism case) or not;  a religion can also ''rather quickly''' disappear (see the Antoinists in some coal mine regions of Western Europe), - in both cases for quite ``interesting'' reasons or causes, actually outside the realm of this paper. Thus the time life, aging, of a religion can be studied through the number of adherents, surely for modern times, - with some caution.

In so doing several pertinent questions can be raised, e.g.  
from a ``macroscopic'' point of view :  (i) how many religions exist at a
given time? (ii) how are they spatially distributed ? ... From a ``microscopic'' view point:  (iii) How many 
adherents belong to one religion?  (iv) Does the number of adherents
increase or not, and how?  - and maybe why? (!), 
(v) Last but not least is there some modelization, ... some agent based model possible?

We recognize that the definition of a religion  or an adherent (or adept) might not be accepted univocally, but the same can be said about languages; we
recognize that there are various denominations which can impair data
gathering and subsequent analysis; like many, we admit to put on the same
footing religions, philosophies, sects and rituals. $Idem$, we do not distinguish between adherents or
adepts;  there are also agnostics, atheists or ``not concerned''. In fact,
a similar set of considerations exists when discussing languages and
dialects, slangs, etc. There are  e.g. three definitions of a  language \cite{klinkenberg}. Similarly one could ''weight'' the level of adherence to a religion, one could try as for languages to define a religion through its rituals, and quantity of practitioners. Many other indicators are possible   (see Appendix B). To consider such variants would lead us too far away from the main stream of the present research and is left for further investigations when possible.

Thus to address some of these issues, we have followed classical scientific steps as in
physics investigations \cite{religion1}. We have $accepted$ as such and subsequently analyzed ''empirical'' data on the number
of adherents of religions. We have discussed in \cite{religion1} two different  freely available data
sets.  The exactness of both data sets from an experimental (laboratory or naturally based) physics point of view is debatable. Some discussion will be rejuvenated in Sec. II.  Yet, it
 has been found in \cite{religion1}  that empirical laws can be deduced for the number of adherent, I.e. the {\it probability distribution function} (pdf).  Two quite different statistical models were proposed, both reproducing well the data, with the same precision, one being a preferential attachment model  \cite{prefatt},  like for  heterogeneous interacting agents on evolving networks, e.g. it is more likely
that one has the religion of one's mother or neighbor..... (leading to a log-normal distribution), another based on a ``time of failure'' argument (leading to a Weibull distribution function). 

\begin{figure}
\centering
\includegraphics[height=5cm,width=5cm]{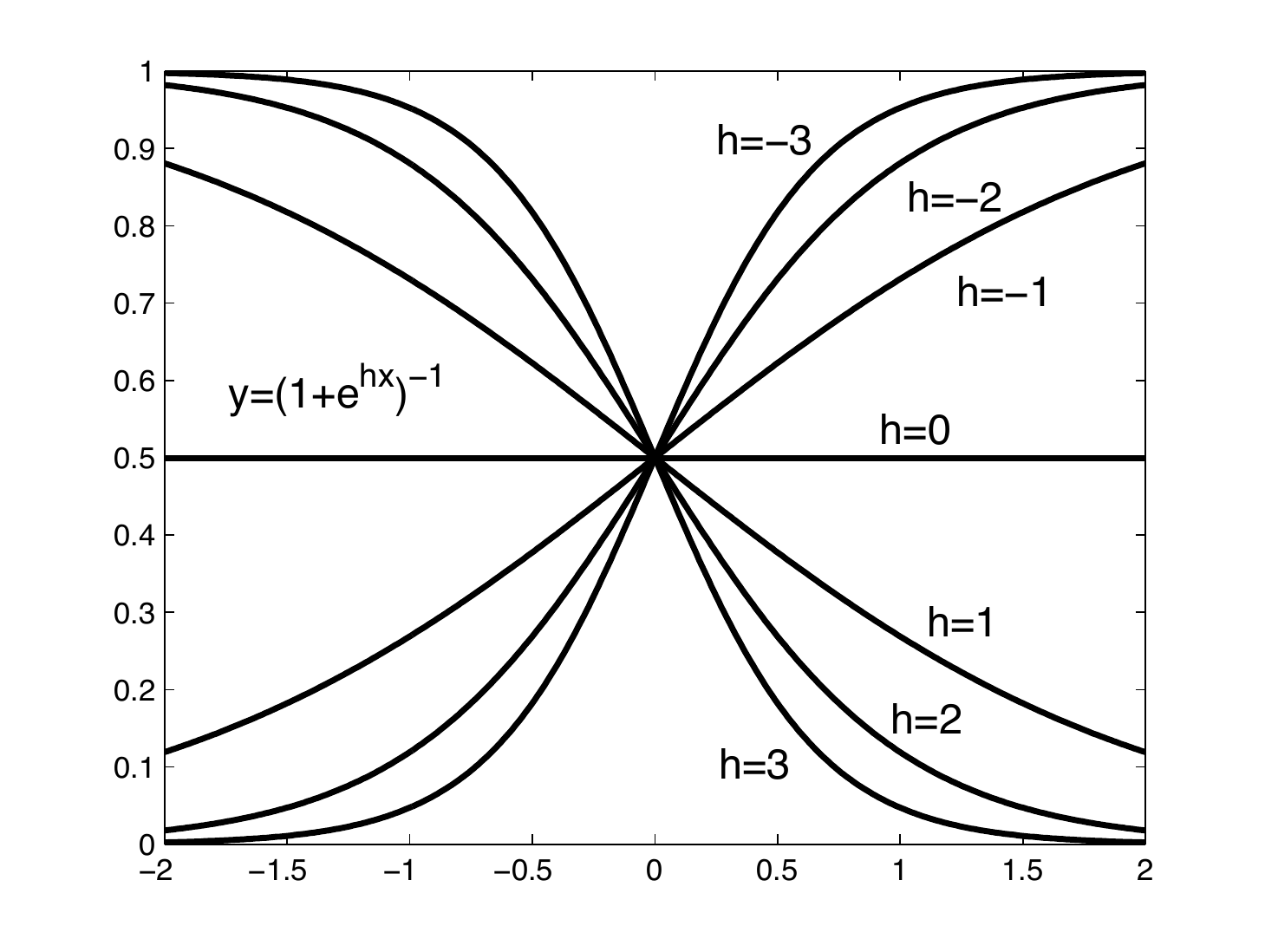}
\includegraphics[height=5cm,width=5cm]{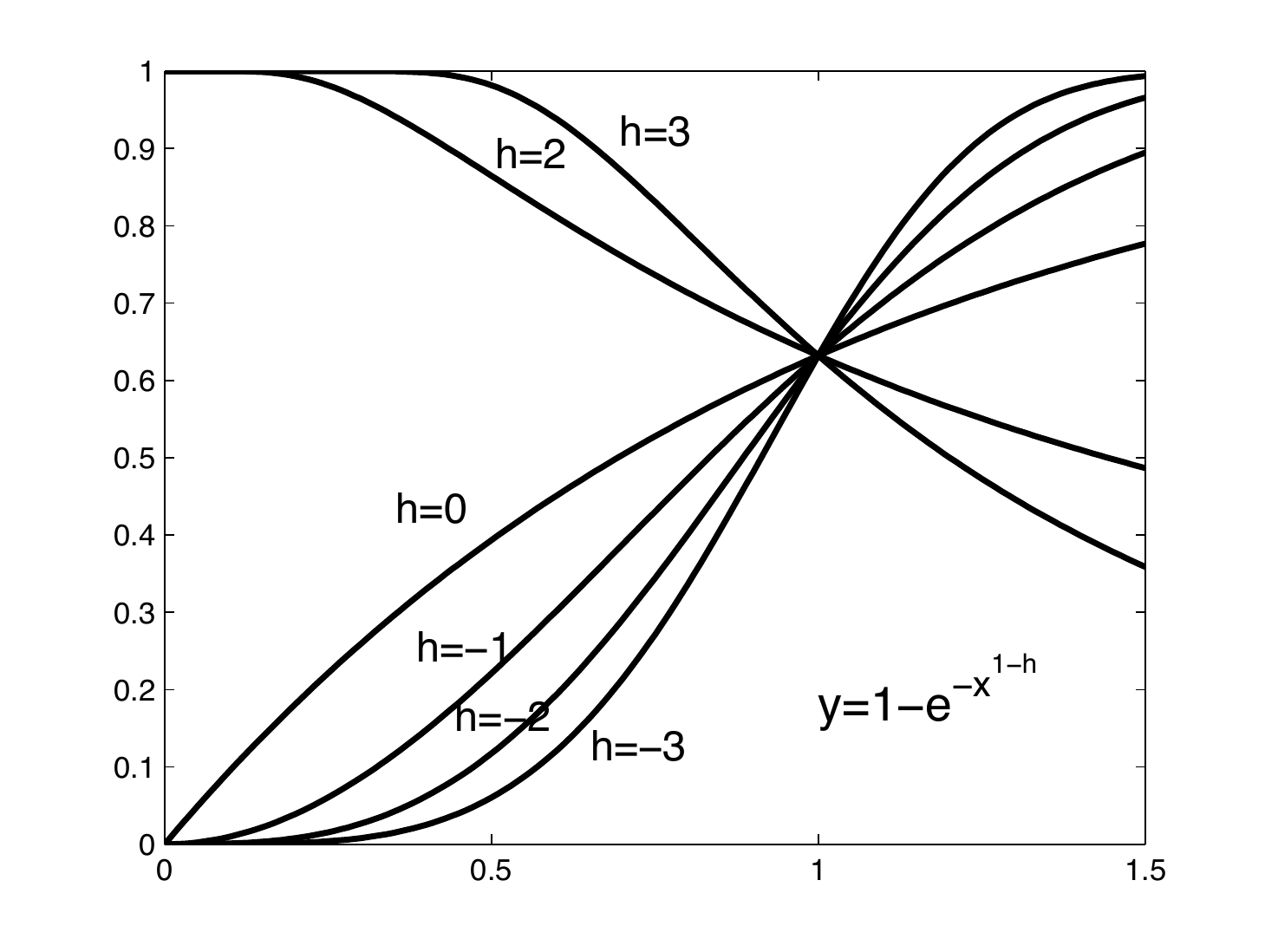}
\caption{\label{avrami_th} (left) Logistic map or Verhulst  law (Eq.(2)) and (right) theoretical behavior of the solution of an Avrami Equation as (Eq.(5) in reduced units for various typical $h$ values}
\end{figure}

Moreover, a
population growth-death equation has been conjectured  to be a plausible modeling of the
evolution dynamics in a continuous time framework, i.e. the time evolution of several ``main'' religions, from a microscopic interpretation is plausible along the lines of
the growth Avrami-Kolmogorov equation describing solid state formation in a continuous time framework,  which solution is usually written as

\begin{equation}\label{Avrami}
F(t)=1- exp[- K t^{n} ]
\end{equation}
where $F(t)$ is the volume fraction being transformed from one phase to another; $K$  and $n$ are adjustable parameters (Fig. \ref{avrami_th}).
For $n=1$, this equation reproduces the loading of a capacitance in series with a resistance $R$, for which the differential equation for the voltage $V(t)$ across the capacitance $C$ reads

\begin{equation}\label{condensdiff}
\frac{d}{dt}V(t)=    \frac{E-V}{RC}
\end{equation}
in terms of the emf $E$, and for which one remembers that one interprets $RC$ as a relaxation time $\tau$. It is also the behavior of the Verhulst logistic map above the inflection point; indicating that this Avrami equation is of interest for so called late stage growth, i.e.
\begin{equation}\label{logisticmap}
\hat F(t)=\frac{1}{1+exp[- K t ]}
\end{equation}

For $n \neq 1$,  Eq.(1) can correspond to complex non linear electronic circuits containing an infinity of elements, or also to an infinite combination of springs and damping elements \cite{kutner} in mechanics.

 {\it A priori} in analogy with crystal growth studies \cite{auslooscrystalgrowth,Gadom}, we have considered that a microscopic-like, continuous time differential equation can be written for the evolution of the number of adherents, in terms of the percentage with respect to the world population,  of the world main religions, as for competing phase entities in Avrami sense
\begin{equation}\label{Avramidiff}
{d \over dt} g(t)=\gamma t^{-h} [1-g(t)].
\end{equation}
It can be solved easily giving the evolution equation for the fraction $g(t)$ of religion adherents  
\begin{equation}\label{Avrami_sol}
g(t)=1-\eta e^{-\frac{\gamma}{1-h} t^{1-h}}
\end{equation}
where, adapting to our case this 
Eq. (4), $\eta$ is related to the initial condition, $\gamma$ is a (positive for growth process) rate (or scaling) parameter to be determined, see discussion in Sect. III,  and $h$ is a parameter  to be
deduced in each case, measuring the attachment-growth (or death) process
in this continuous time approximation.  The {\it relaxation time} $\tau_n$,  since $n\equiv 1-h$, of this stretched exponential growth is 

\begin{equation}\label{tau_n}
\tau_n = (\frac{\gamma}{1-h})^{\frac{-1}{1-h}}
\end{equation}
which is markedly rate ($K$) dependent.
For further consideration, let us explicitly write the ''very (infinitely) slow'' growth case $h$=1, i.e., 
\begin{equation}\label{Avramidiffh1}
{d \over dt} g(t)=\gamma t^{-1} [1-g(t)],
\end{equation}
whence 

\begin{equation}\label{Avrami_sol1}
g(t)=1- \beta t^{-\gamma},
\end{equation}
where $\beta$, being positive (negative) for a growth (decay) case, is set by initial conditions; for $h=1$,  there is no ''relaxation time'', but a scaling time $\tau_1$ = $\beta^{\frac{1}{\gamma}}$, or $\beta = \tau_1^\gamma$.

The $h$-cases which can be illustrated through an Avrami equation are shown in arbitrary time units in Fig.1 for various $h$ values, for $\eta$ =1 and $\gamma= 1-h$. They are compared to the (generalized, i.e. $n \neq 1$) logistic map.

What should be emphasized is the fact that religions have appeared at some time $t_0$ which is somewhat unknown, or to say the least, loosely defined, due to a lack of historical facts but also due to the inherent process of the creation of a religion. 
Yet this initial time is a parameter much more important than in crystal growth, but less well defined. Therefore we rewrite the Avrami equation as
 
\begin{equation}\label{Avrami_t_0}
g(t)=1-e^{-\left({t-t_0 \over t_1}\right)^{1-h}},
\end{equation}
thereby allowing also for a time scaling through $t_1$ related to some growth (or death) rate process. Notice or so that the maximum in such theoretical laws occurs at zero or +/- infinity, - a time information on which there is not much data in the case of religions.

If $(t-t_0)/t_1$ is much smaller than 1,  Eq. (\ref{Avrami_t_0}) can be expanded in Taylor series, taking only the first order, and gives
\begin{equation}\label{Avrami_sol2}
g(t)=\alpha+\left({t\over t_1}\right)^{1-h}
\end{equation}
where we have chosen $t$ starting from $0$ (instead of 1900, as in our data, being this completely arbitrary and purely conventional) and $\alpha$ represent the initial condition, i.e. the value of the number of adherents for $t=0$.  
 
A few examples of religions for which the number of adherents is
increasing (e.g., Islam), decaying (e.g., Ethnoreligions) or
rather stable (e.g.,  Christianity and Buddhism) is already shown in Fig. 4 of  \cite{religion1}.  In such cases we have found that $h$ $\simeq$ -1.8,   6.9, 1.5 and 1.4, respectively in the time range of interest (1900-2050). 
  However in \cite{religion1}  the main denominations were ''loosely grouped''. To be more specific: Christians in \cite{religion1} were the results of grouping together 12 denominations; similarly for Muslims we grouped 15 denominations.  
  
Here we present a more complete and somewhat more detailed analysis of the values of $h$ and its meaning for 58 ''time series'', where 58 means ''55''+''1''+''2'' religions. More precisely there are 56 data sets for specific religions, in the WCE and WCT references \cite{WCE,WCT}, most of them being in the {\it main denomination } bracket, i.e. in the upper part of the pdf as obtained from the surveys  taken between 1900 and 2000.   The "1" refers to some data containing 3000 religions which are put together, as ''$other$ religions'' in the WCT tables. The ''2'' refers to the set of data on  $atheists$ and  $nonreligious$ persons, as mentioned in Table 1-2 of ref. \cite{WCT}.  Thereafter for conciseness, we will also identify/call those three sets as ''religions''. 

Emphasis will be on distinguishing between growing and decaying cases, discussing our ''theoretical'' fit, comparing to the forecasting in ref. \cite{WCT},  for 2025 and later, and observing diverse anomalies, thus raising questions to be further investigated.

\par The remainder of the paper is organized as follows: in Section II the data bank is briefly discussed, - and criticized, though accepted for further research and subsequent analysis along the theoretical and methodological tools used here  which we adapt to the considered time series set. The results are largely presented and discussed in Section III under the form of Tables and graphs for various  groups of religions, grouping according to the apparent behavior. Some concluding remarks are done in Section IV.

\section{Data bank. Theoretical and methodological framework}
The data \cite{footnote4} analyzed here were taken from the World Christian Trends \cite{WCT}. It is fair to say that this is a remarkable compilation work.
Their tables give information on the number of adherents of the world's main religions and their main denominations : 55 specific  (''large'') religious  groups + atheists + nonreligious, plus a set called other religionists made of 3000 religions which however contains, Yezidis and Mandeans which we consider also, so that we examine 53+2= 55 (truly recognized) religions.  From this data set we have also
information on changes during one century of the number of adherents of
each religion from 1900 till 2000 (information in the data set are given for the following years 1900, 1970, 1990, 1995 and 2000)  -  with a
forecast for 2025 and 2050. Let us point out that it is not understood (or barely understandable) how such a forecast is made in the data bank.

A critical view of this data has to follow:  we have already \cite{religion1} noticed a break at $10^7$, in the pdf,  indicating in our view
an overestimation of adepts/adherents  in the most prominent religions, or a lack of
distinctions between denominations, for these, - as can be easily understood either in terms of propaganda or politics, or because of the difficulty of surveying such cases precisely. Yet one paradoxical surprise stems in the apparent  precision of the data. E.g., in several cases in which the religion adherent numbers are reported,  the data in \cite{WCT} seems to be precise up to the last digit  i.e.,  in mid-2000 , there are  1057328093 and 38977 roman catholics and mandeans respectively.  In strong contrast there are 7000000 and  1650000 wahhabites and black muslims respectively, numbers which are quite well rounded. Thus a mere reading of the numbers  warns about the difficulty of fully trusting the data. Nevertheless the analysis is pursued bearing this $caveat$ here below.

\vskip 1cm

\begin{table}\caption{Values of the parameters $h$, $\alpha$, and $ t_1$,  used for fitting the data for ''increasing religions'' with a power law formula; see Eq. (\ref{Avrami_sol2}); religions are hereby  ranked based on the size of the attachment parameter $h$ which can be negative or positive but $\le 1$}
\begin{center} \begin{tabular}{|l|r|r|r|} \hline Religion & $h$ &  $\alpha$ & $t_1$ \\ \hline Shaivites & -5.32 & 0.032 & 239 \\ 
Hanbalites & -4.66 & 0.000305 & 527 \\ 
Hanafites & -3.84 & 0.0629 & 211 \\ 
Zoroastrians & -3.64 & 3.29e-005 & 530 \\ 
Kharijites & -2.88 & 0.000196 & 1.15e+003 \\ 
Afro-Caribbean religionists & -2.75 & -5.06e-007 & 1.8e+003 \\ 
Black Muslims & -2.36 & -8.06e-006 & 1.11e+003 \\ 
Pentecostals/Charismatics & -2.19 & -0.00186 & 208 \\ 
Independents & -1.61 & 0.00427 & 288 \\ 
Shafiites & -1.49 & 0.024 & 528 \\ 
Afro-American spiritists & -1.32 & 6.87e-005 & 4.99e+003 \\ 
Ithna-Asharis & -1.26 & 0.0137 & 812 \\ 
Afro-Brazilian cultists & -1.21 & 5.52e-005 & 2.61e+003 \\ 
Zaydis & -1.10 & 0.000741 & 3.45e+003 \\ 
Alawites & -1.09 & 0.000154 & 7.56e+003 \\ 
Ismailis & -1.04 & 0.00142 & 1.87e+003 \\ 
Yezidis & -1.01 & 1.84e-005 & 2.23e+004 \\ 
High Spiritists & -0.83 & 2.33e-005 & 5.69e+003 \\ 
Sikhs & -0.792 & 0.00182 & 3.13e+003 \\ 
Ahmadis & -0.789 & 4.32e-005 & 4.17e+003 \\ 
Baha$'$is & -0.368 & 4.87e-006 & 1.38e+004 \\ 
Druzes & -0.366 & 4.38e-005 & 8.85e+004 \\ 
Neo-Hindus & -0.212 & 6.19e-005 & 1.28e+004 \\ 
Marginal Christians & -0.206 & 0.000569 & 1e+004 \\ 
Mandeans & -0.0667 & 5e-006 & 3.17e+007 \\ 
\hline
Malikites & 0.0566 & 0.0167 & 6.38e+003 \\ 
Other sectarian Muslims & 0.0929 & 0.000311 & 2.62e+006 \\ 
crypto-Christians & 0.230 & 0.0022 & 1.8e+004 \\ 
Reform Hindus & 0.384 & 0.000154 & 1.78e+007 \\
\hline \end{tabular} \end{center} \label{table1} \end{table}

\newpage
\begin{table}\caption{Values of the parameters $h$,  $t_0$, and $t_1$ used for fitting the data on ''ecreasing religions'' with Eq. (\ref{Avrami_t_0}); $h$ is in this case $\ge 1$}
\begin{center} \begin{tabular}{|l|r|r|r|} \hline Religion & $h$ &  $t_0$ & $t_1$ \\ \hline Chinese folk-religionists & 1.07 & -3.36e-007 & 3.49e-015 \\ 
Orthodox & 1.14 & -0.821 & 1.06e-008 \\ 
Theravada & 1.81 & -242 & 3.27 \\ 
Mahayana & 2.04 & -321 & 16.2 \\ 
Karaites & 2.09 & -99.3 & 0.00226 \\ 
Lamaists & 2.77 & -614 & 29.7 \\ 
\hline \end{tabular} \end{center} \label{table2} \end{table}

\newpage
\begin{table}\caption{Values of the parameter used for fitting data on 12 ''decreasing'' and 11 ''increasing''  religions with the polynomial equation  $Cx^2+Bx+A$; for a warning on the six ''central'' (in the table) religions, see text}
\begin{center} \begin{tabular}{|l|r|r|r|} \hline Religion & $C$ &  $B$ & $A$ \\ \hline Nonreligious & -2.61e-005 & 0.103 & -102 \\ 
Atheists & -1.31e-005 & 0.0514 & -50.3 \\ 
unaffiliated Christians & -4.38e-006 & 0.017 & -16.5 \\ 
Roman Catholics & -4.2e-006 & 0.0165 & -16 \\ 
New-Religionists (Neoreligionists) & -3.88e-006 & 0.0153 & -15 \\ 
Shamanists & -4.87e-007 & 0.00185 & -1.74 \\ 
Confucianists & -2.11e-007 & 0.000831 & -0.815 \\ 
Wahhabites & -5.53e-008 & 0.000215 & -0.208 \\ 
Taoists & -4.4e-008 & 0.000174 & -0.171 \\ 
Other religionists (in 3000 religions) & -3.81e-008 & 0.00015 & -0.148 \\ 
&&&\\
Ashkenazis & -2.46e-008 & 4.26e-005 & 0.0149 \\ 
Oriental Jews & -3.23e-009 & 1.22e-005 & -0.0112 \\ 
\hline
Samaritans & 7.14e-012 & -3.01e-008 & 3.18e-005 \\ 
Sefardis & 6.52e-010 & -2.8e-006 & 0.00315 \\ 
Jains & 8.97e-009 & -3.61e-005 & 0.0371 \\ 
Shintoists & 2.05e-007 & -0.000836 & 0.853 \\ 
&&&\\
Saktists & 2.12e-007 & -0.000824 & 0.806 \\ 
Protestants & 7.66e-007 & -0.00306 & 3.11 \\ 
Anglicans & 9.75e-007 & -0.00386 & 3.83 \\ 
Vaishnavites & 1.25e-006 & -0.00487 & 4.82 \\ 
Sufis & 2.34e-006 & -0.00921 & 9.11 \\ 
Animists & 2.77e-006 & -0.0111 & 11.2 \\ 
Evangelicals & 5.95e-006 & -0.0233 & 22.8 \\ 
\hline \end{tabular} \end{center} \label{table3} \end{table} 

\section{Results}

Results of the $h$-fit to Avrami equation of the WCT surveys \cite{WCT} are summarized in Tables \ref{table1}, \ref{table2} and \ref{table3}: the 58 ''denominations''  of interest are given. The parameters are obtained by a least-square best fit of the data (not considering the WCT forecast) to the equations mentioned in each table caption for the various cases. The ranking in the tables is according to the  fit parameter $h$ or $A$.

The parameter $h$ values and their meaning deserve some short explanation and discussion here. According to the standard growth (Avrami) process $h$ should be positive and less than 1,  since $n\equiv 1-h$;  if it is
greater than 1, this is indicating the possibility for $detachment$.  We consider that if $|h|$ is outside  the $(0,1)$ interval, we have to imagine that the nucleation growth process is heterogeneous and/or conjecture that it is due to {\it external field} influences. Moreover notice that when $h$ is greater than 1, the Avrami equation solution decays, ... from a maximum at the time $t_0$. However it is hardly difficult to know when a religion has attained its maximum number of adherents.  Thus the time scale or the initial appearance time of a religion are questionable points. Another point is obvious from Fig.1. The theoretical expressions do not allow a fit in the vicinity of a maximum of minimum. We should expect deviations, if such a case occurs, whence other empirical functions to be of interest.

 \begin{figure}
\centering
\includegraphics[height=5cm,width=5cm]{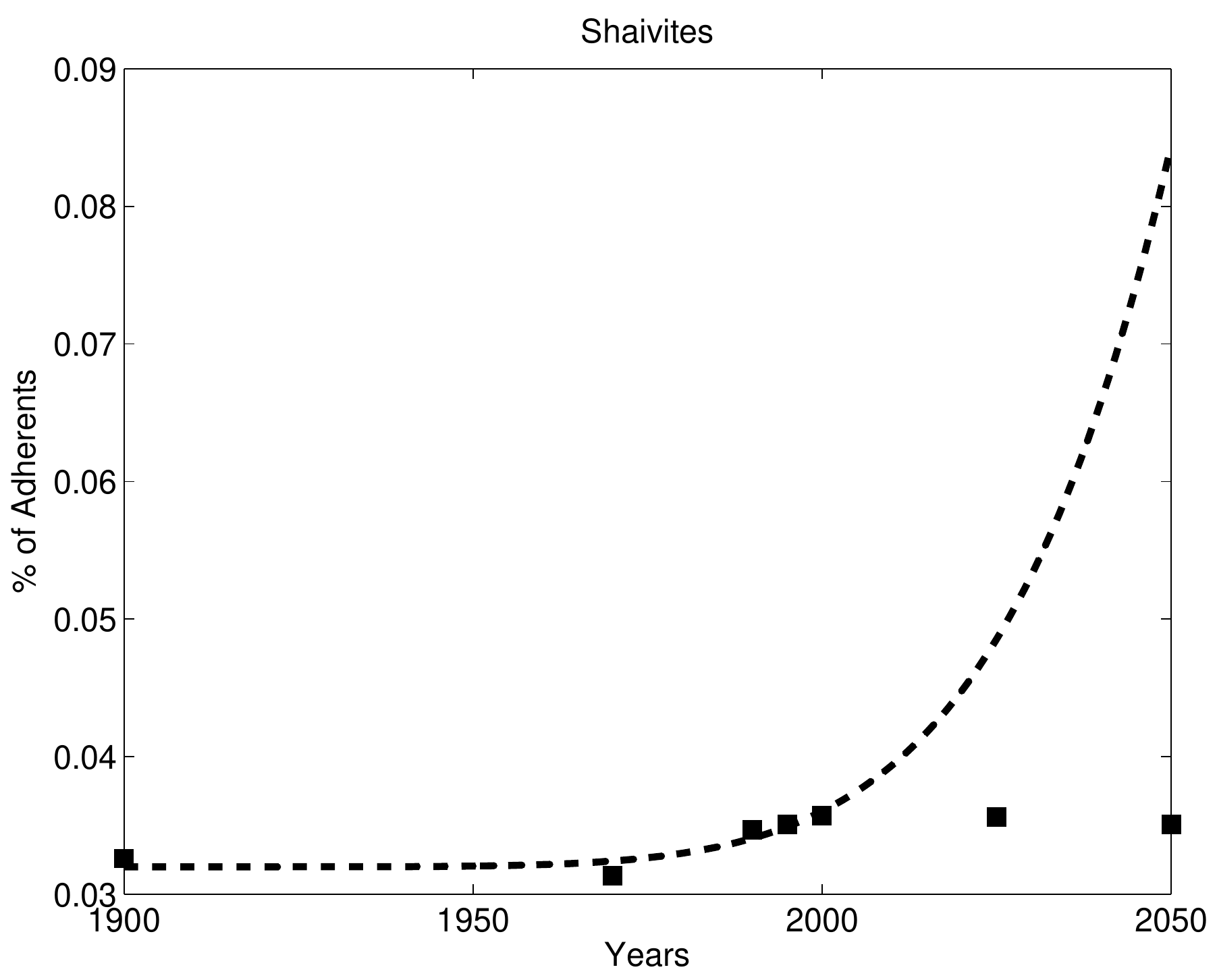}
\includegraphics[height=5cm,width=5cm]{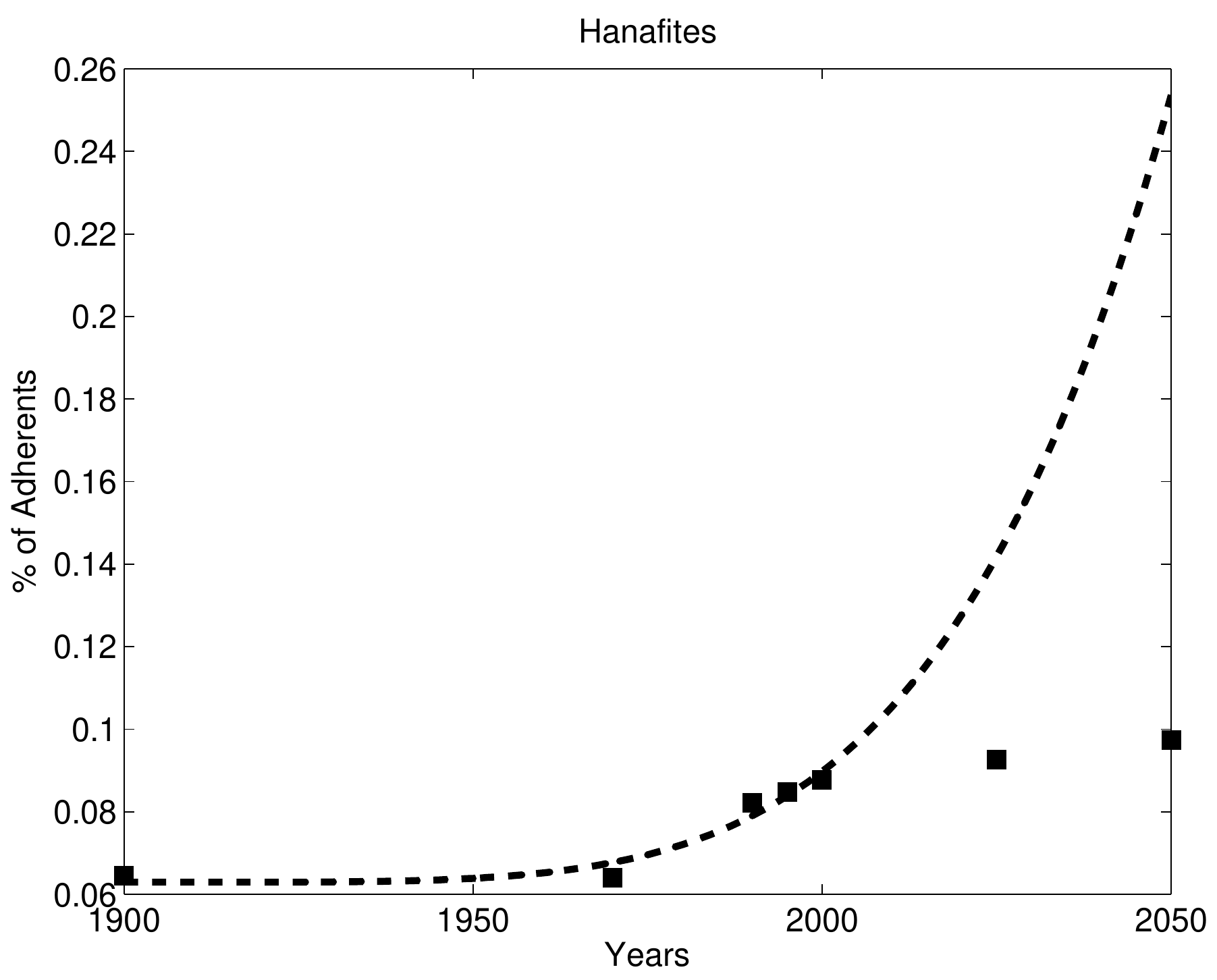}
\includegraphics[height=5cm,width=5cm]{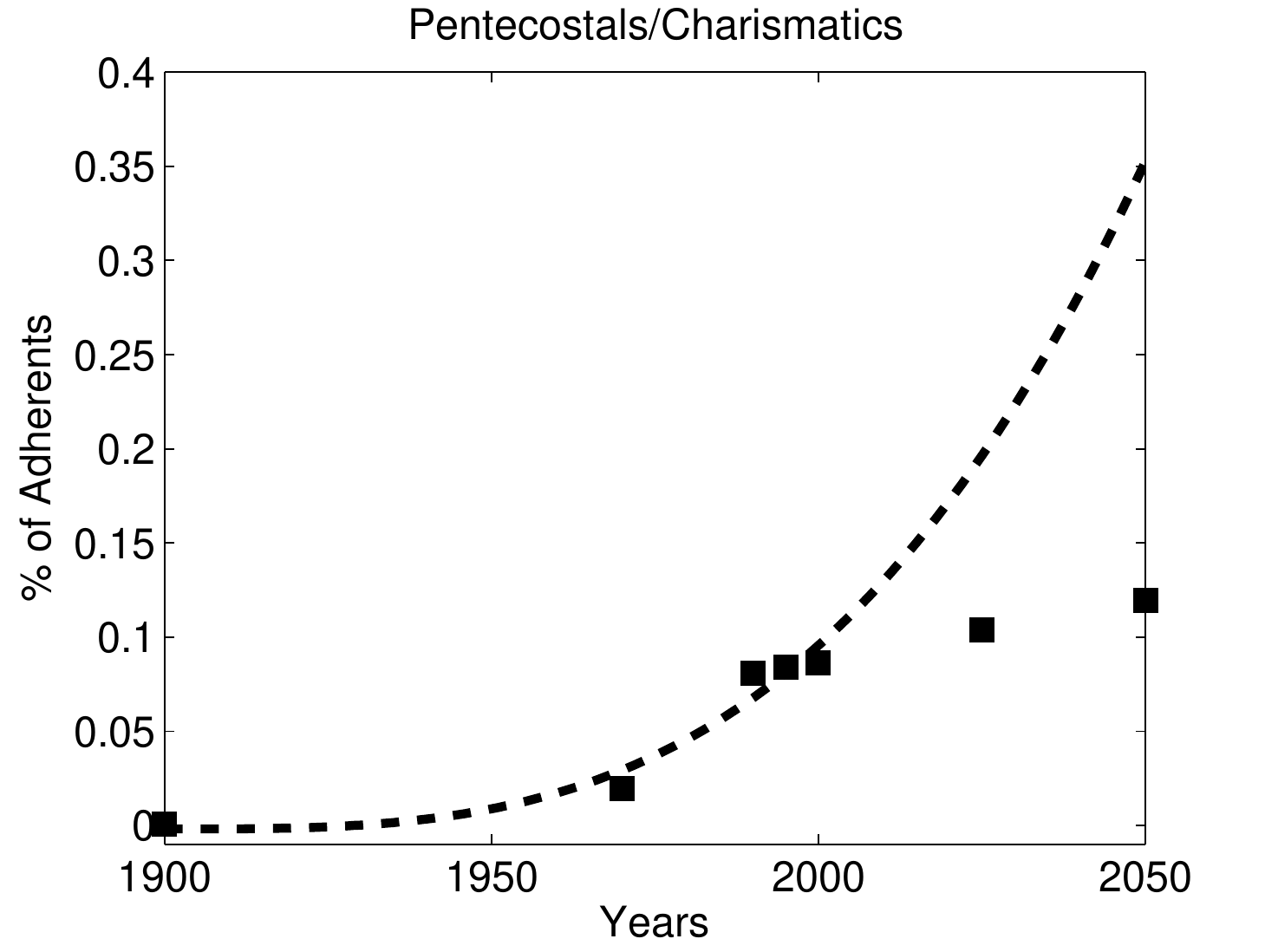}
\includegraphics[height=5cm,width=5cm]{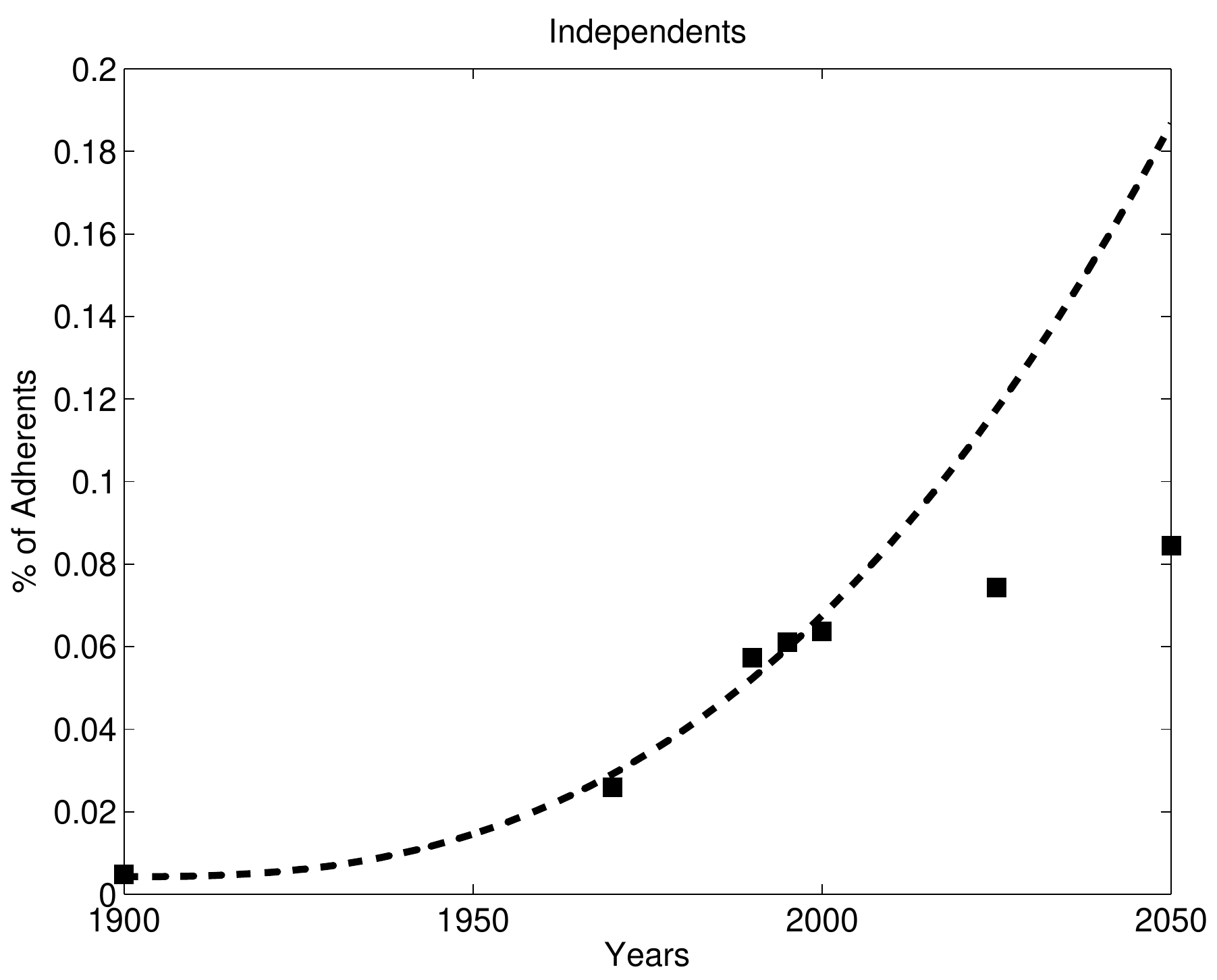}
\includegraphics[height=5cm,width=5cm]{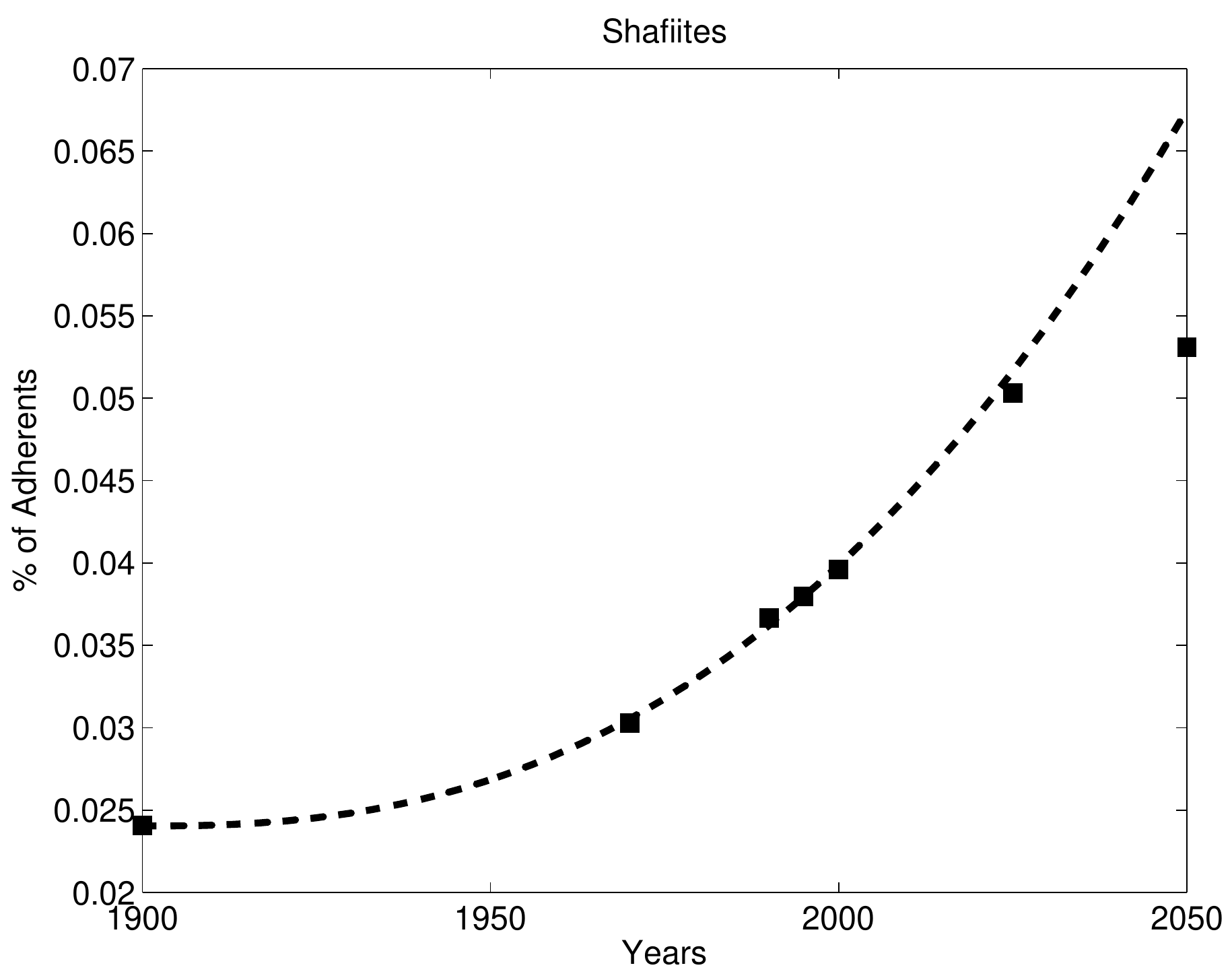}
\includegraphics[height=5cm,width=5cm]{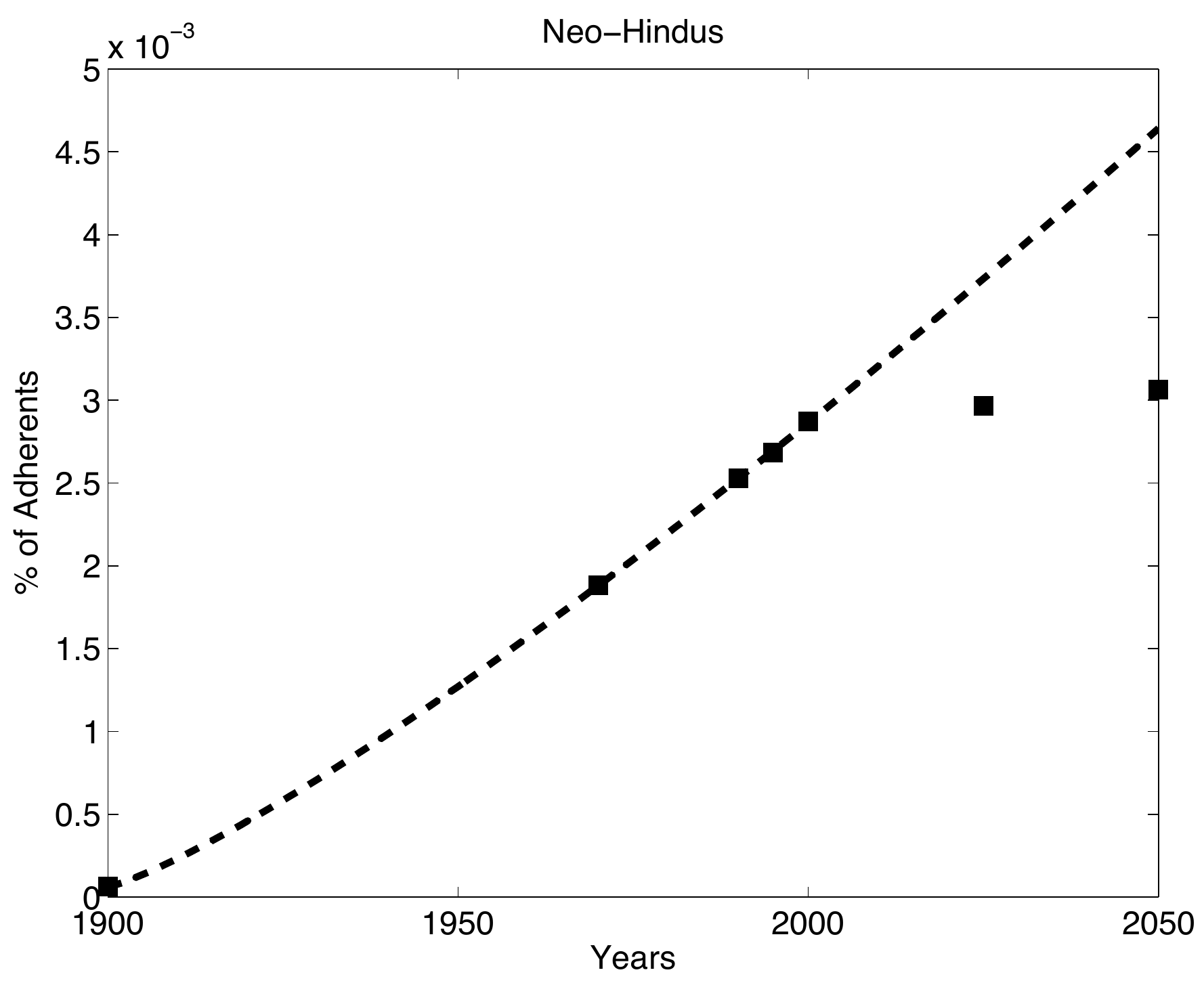}
\caption{\label{Fig5}    Six illustrative cases of actually increasing  religions, ...  with $h \le 0$. Observe the overshooting in the forecast with respect to WCT in all cases}
\end{figure}

 \begin{figure}
\centering
\includegraphics[height=5cm,width=5cm]{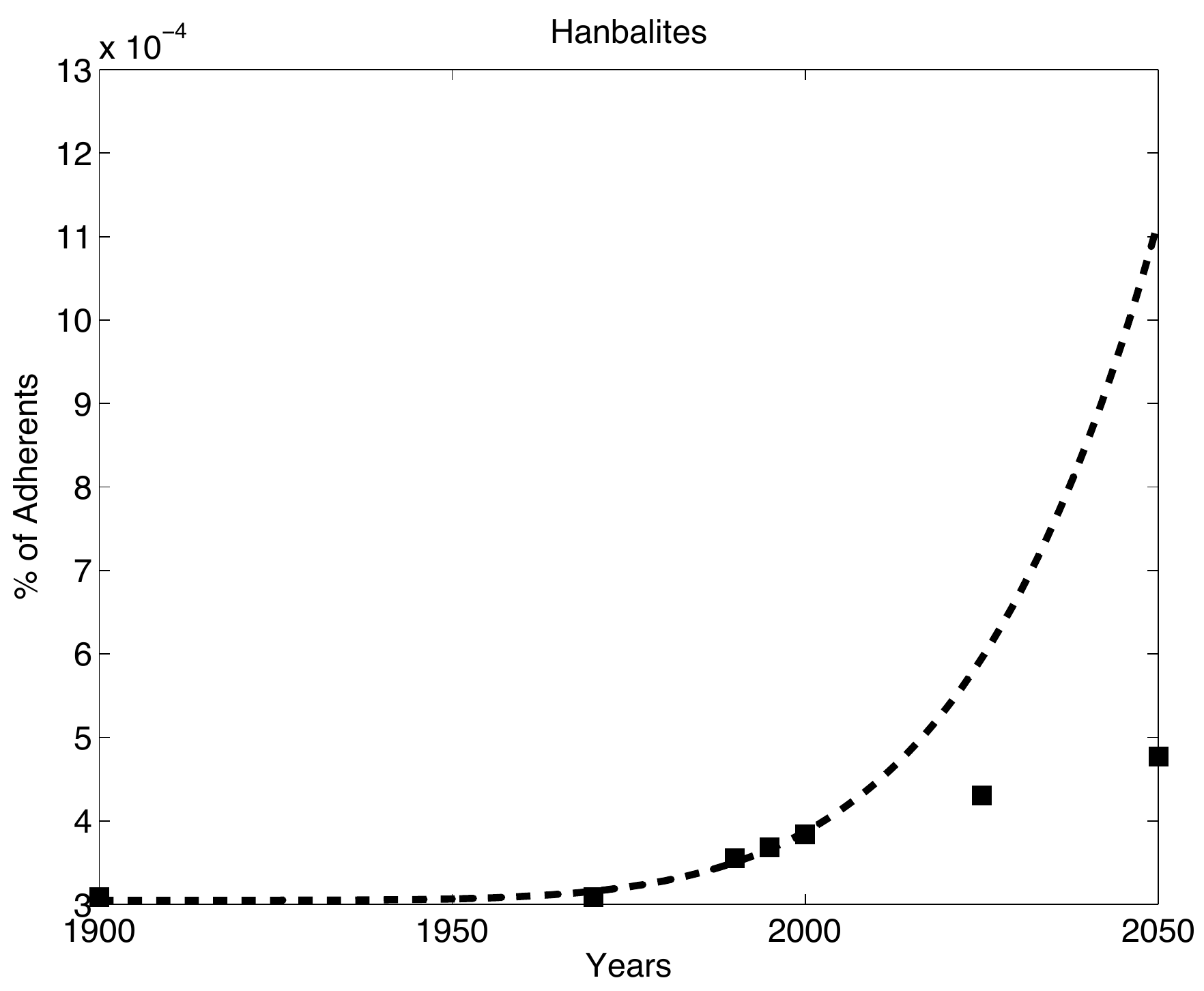}
\includegraphics[height=5cm,width=5cm]{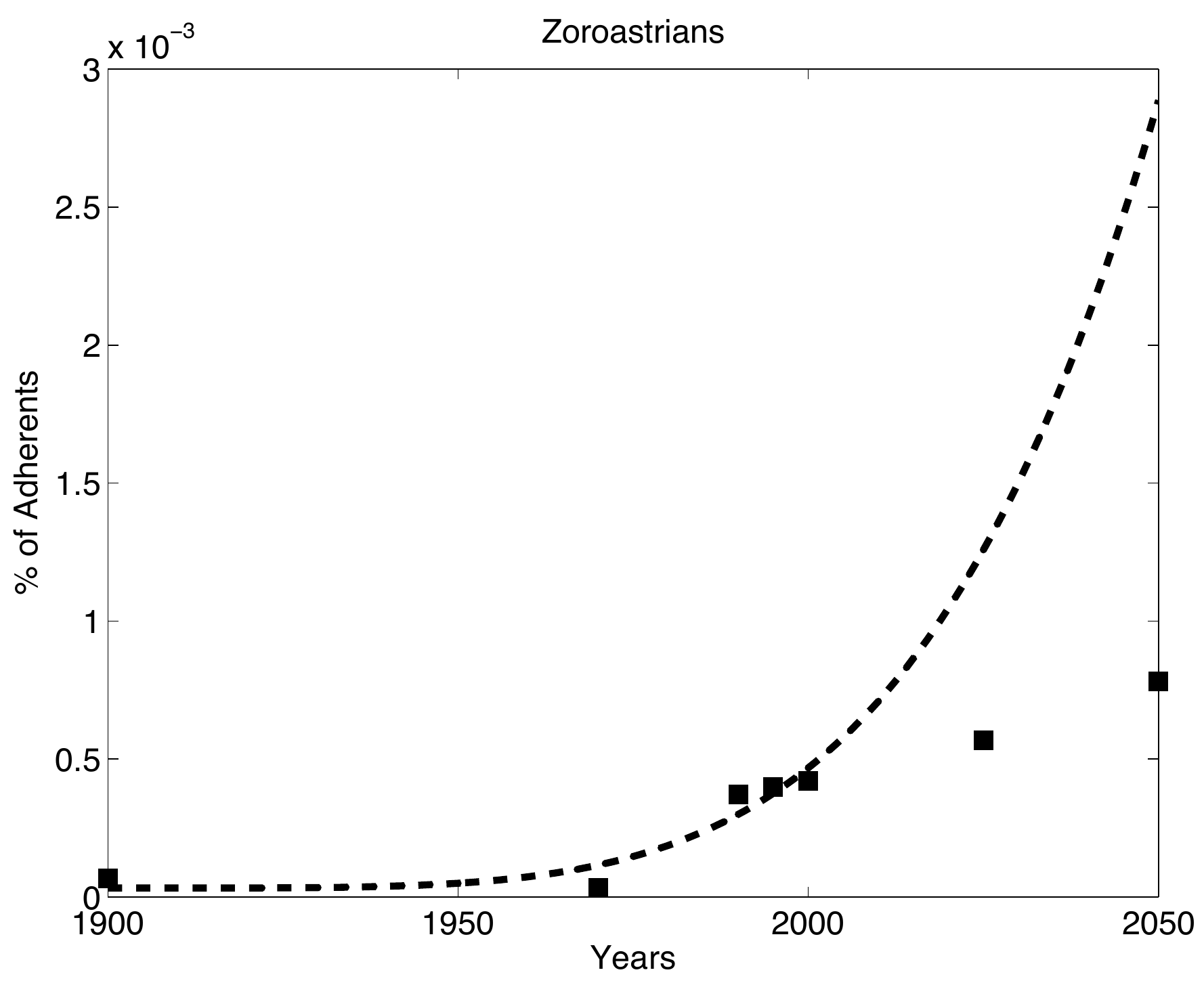}
\includegraphics[height=5cm,width=5cm]{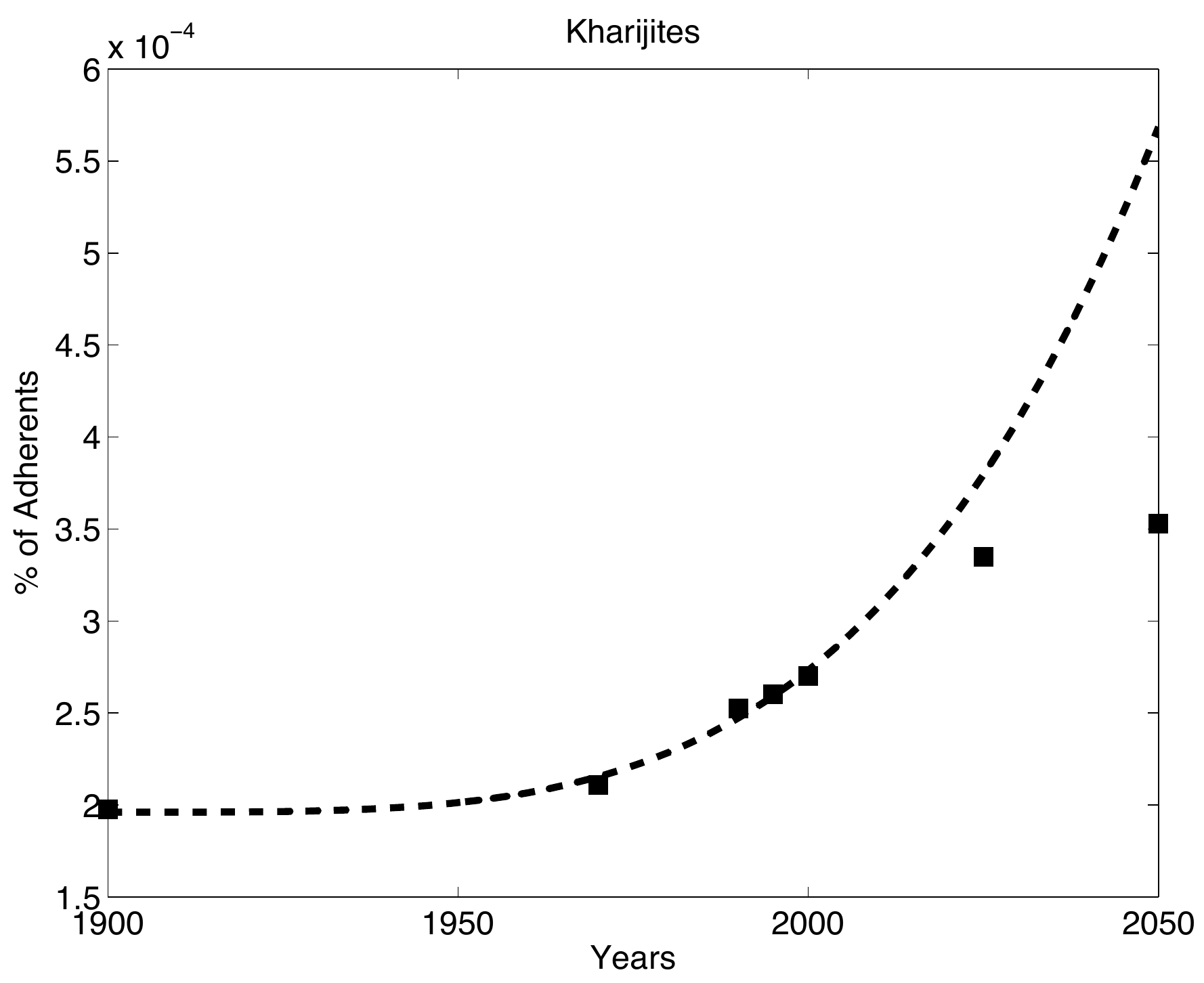}
\includegraphics[height=5cm,width=5cm]{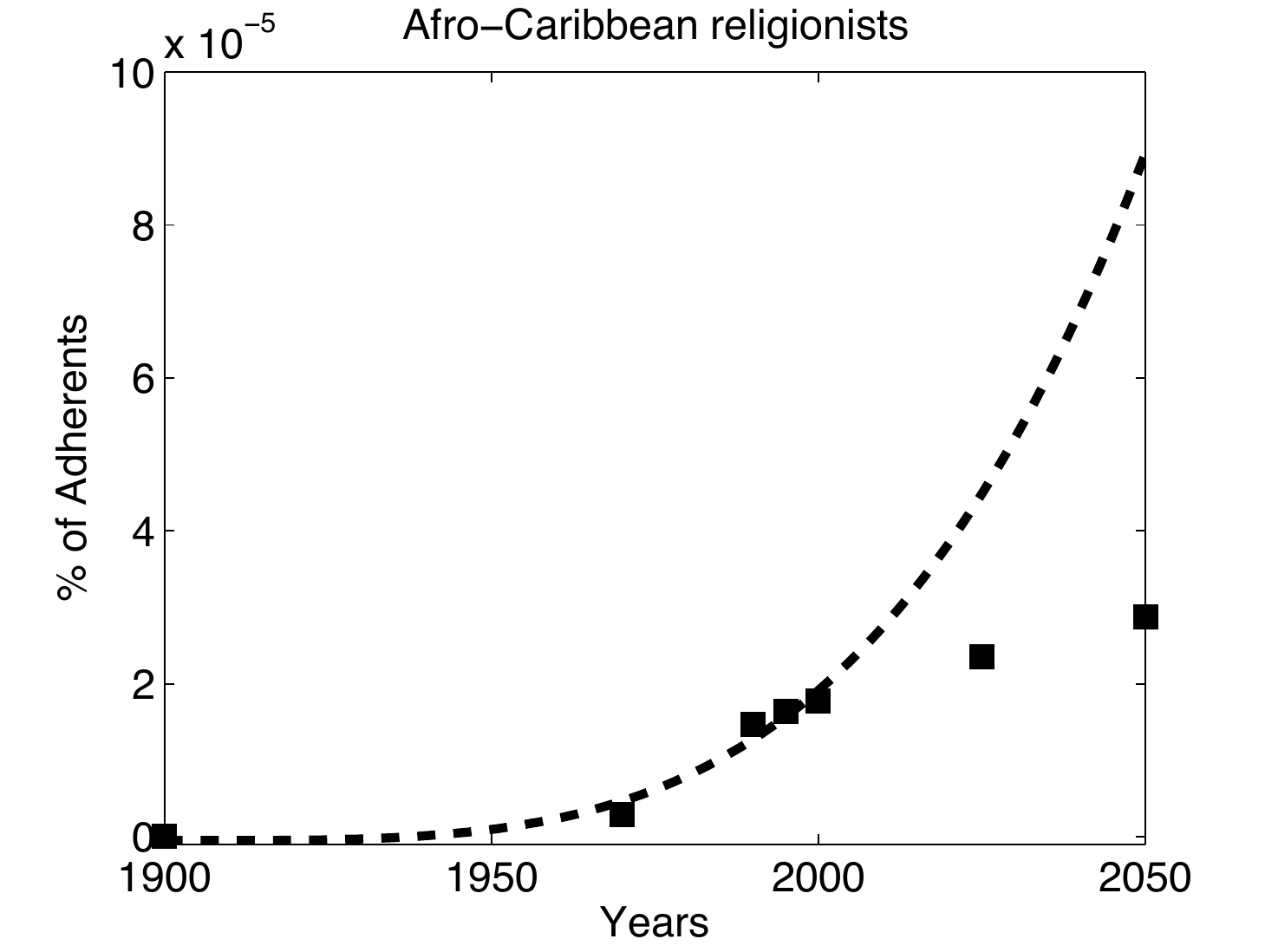}
\includegraphics[height=5cm,width=5cm]{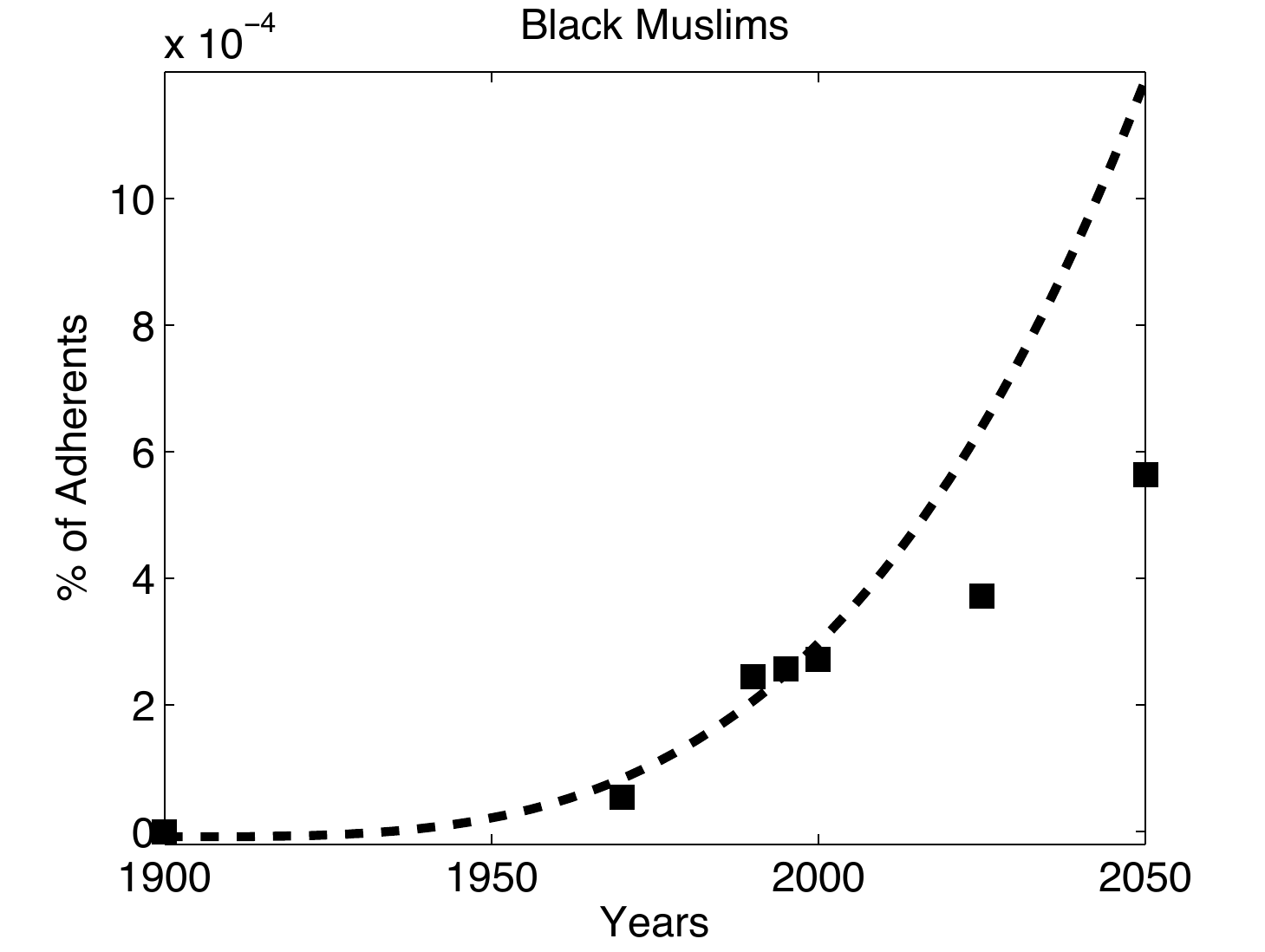}
\includegraphics[height=5cm,width=5cm]{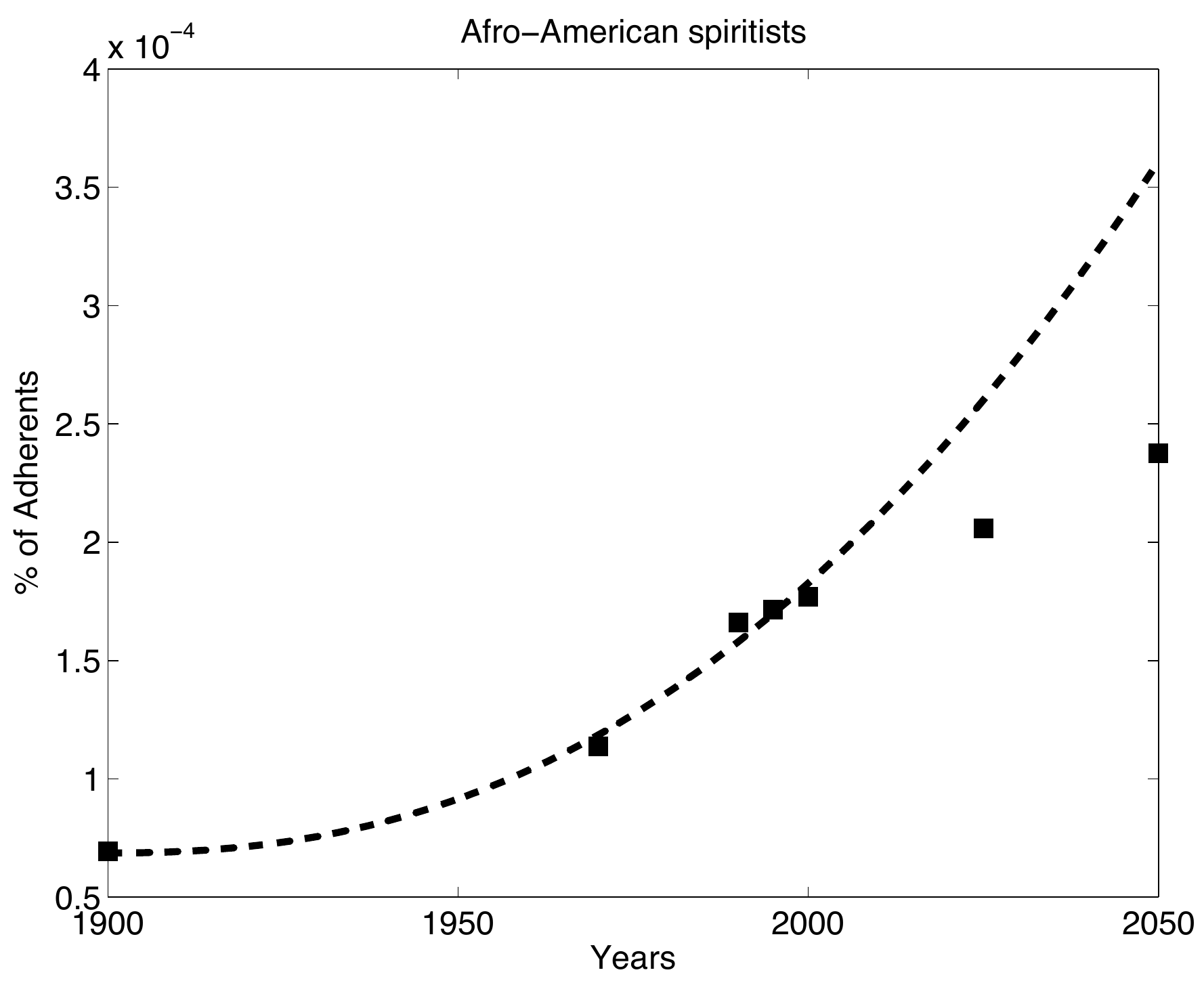}
\includegraphics[height=5cm,width=5cm]{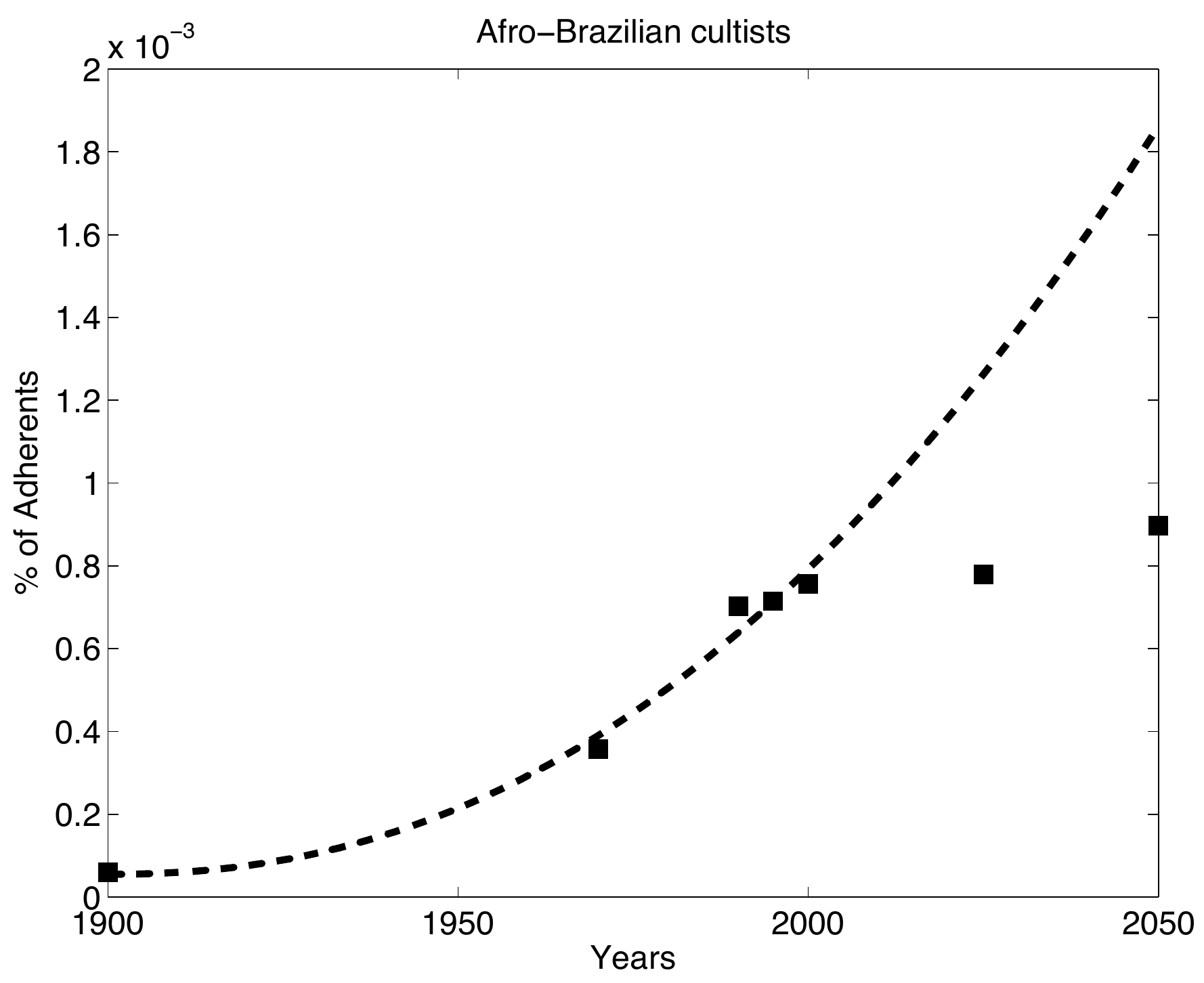}
\includegraphics[height=5cm,width=5cm]{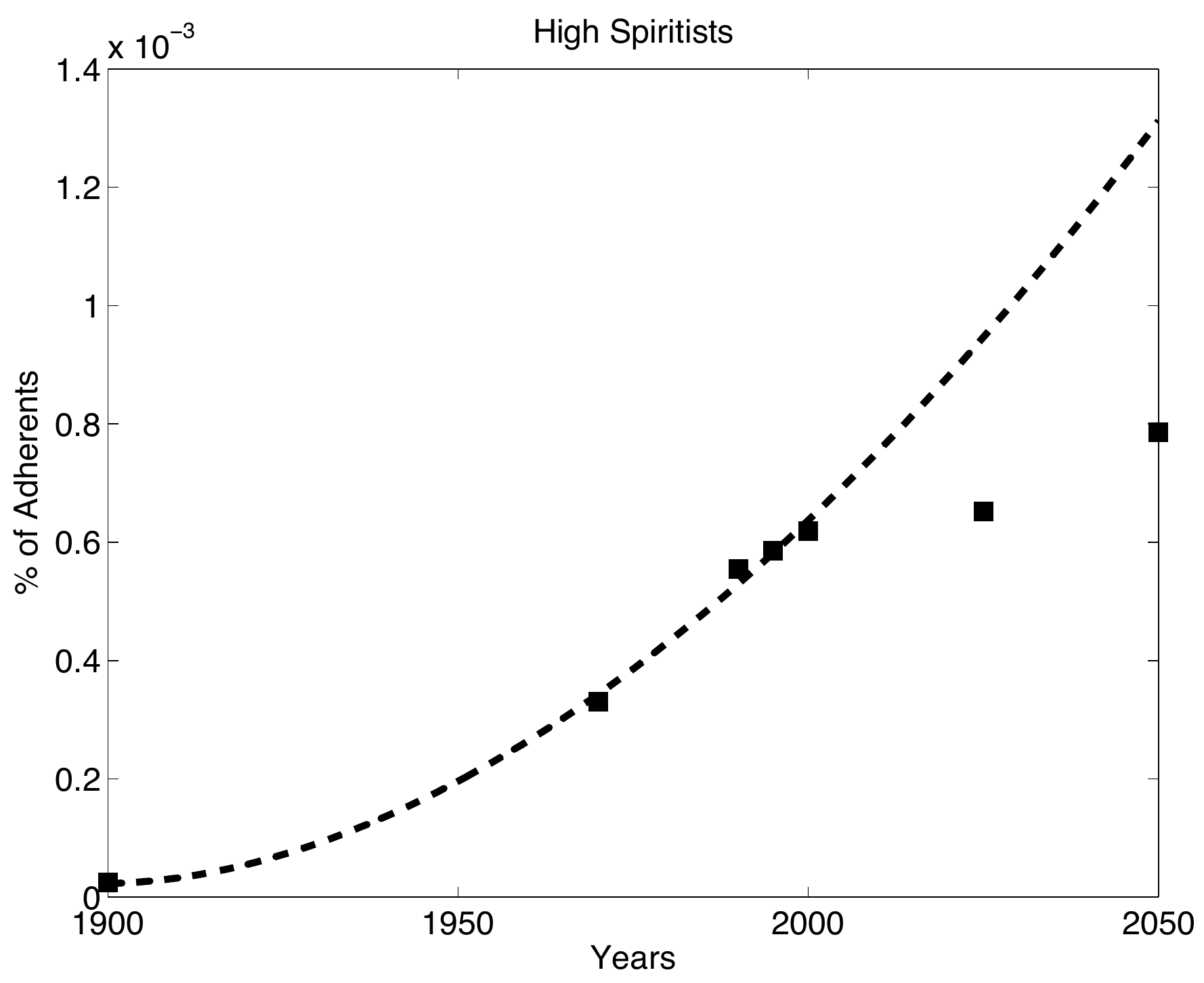}
\includegraphics[height=5cm,width=5cm]{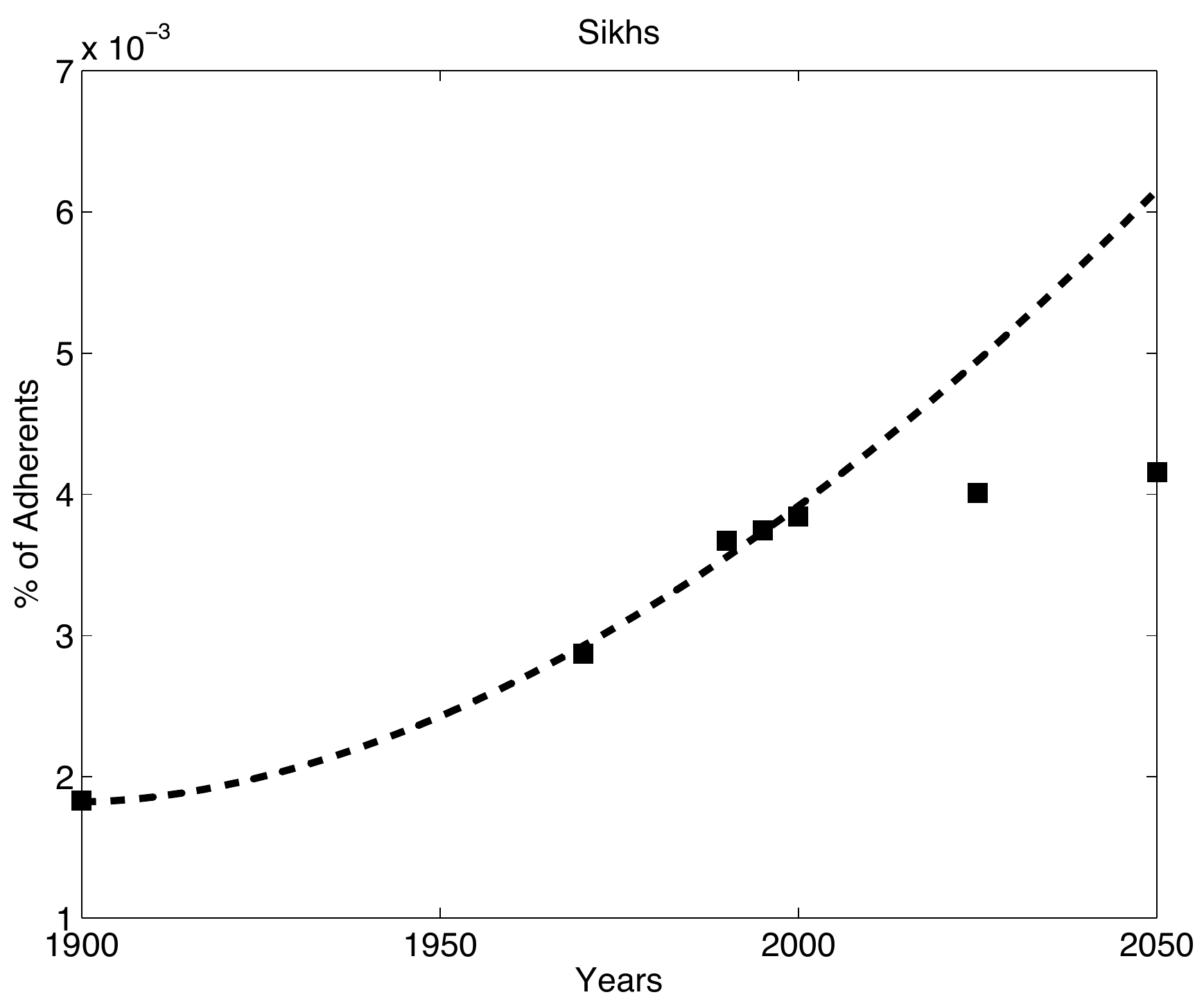}
\caption{\label{Fig4}    Six  illustrative cases of actually increasing (small size)  religions ... with $h \le 0$; our empirical law does not  confirm the WTE forecast in the next years, but overshoots the WTE value}
\end{figure}

\begin{figure}
\centering
\includegraphics[height=5cm,width=5cm]{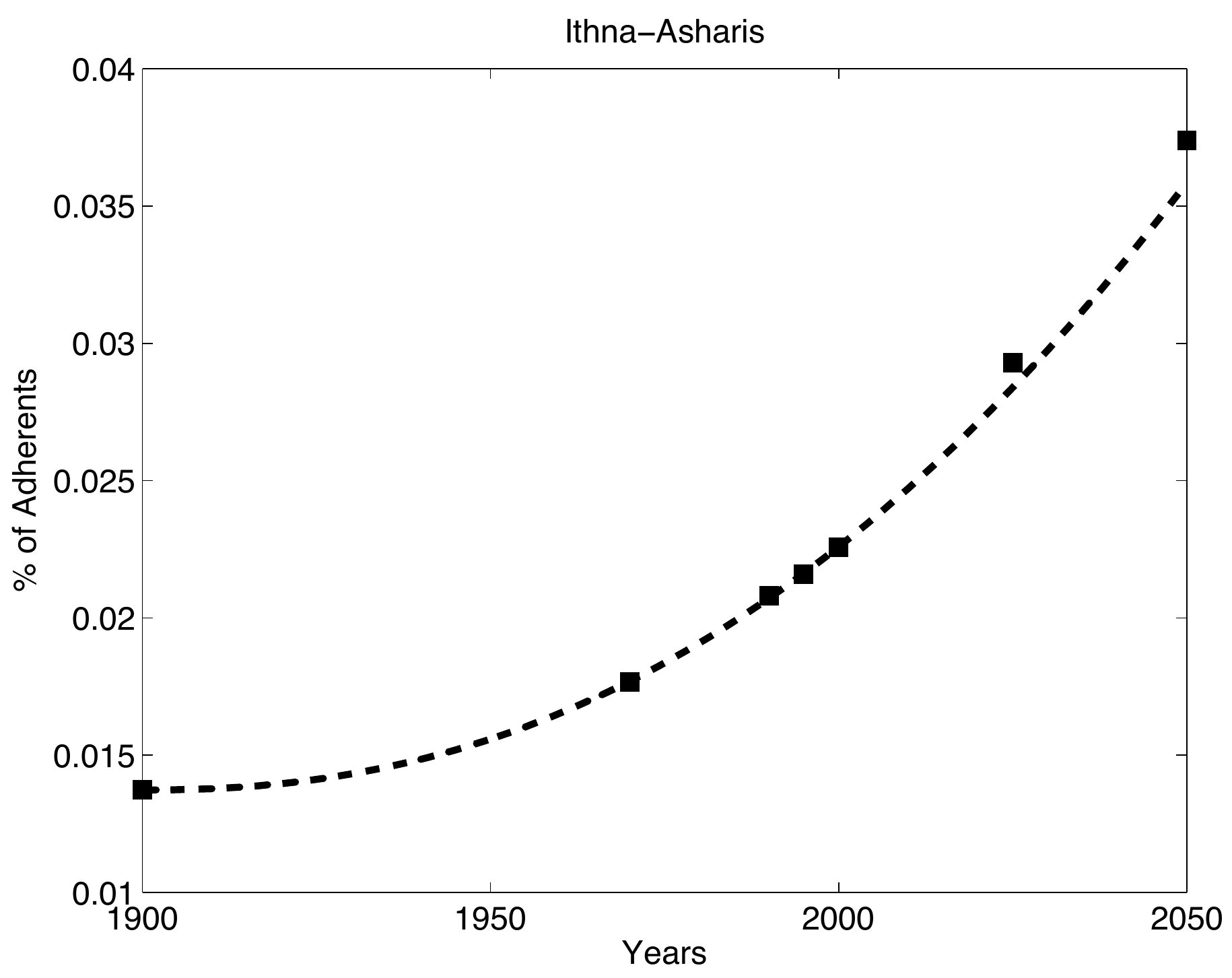}
\includegraphics[height=5cm,width=5cm]{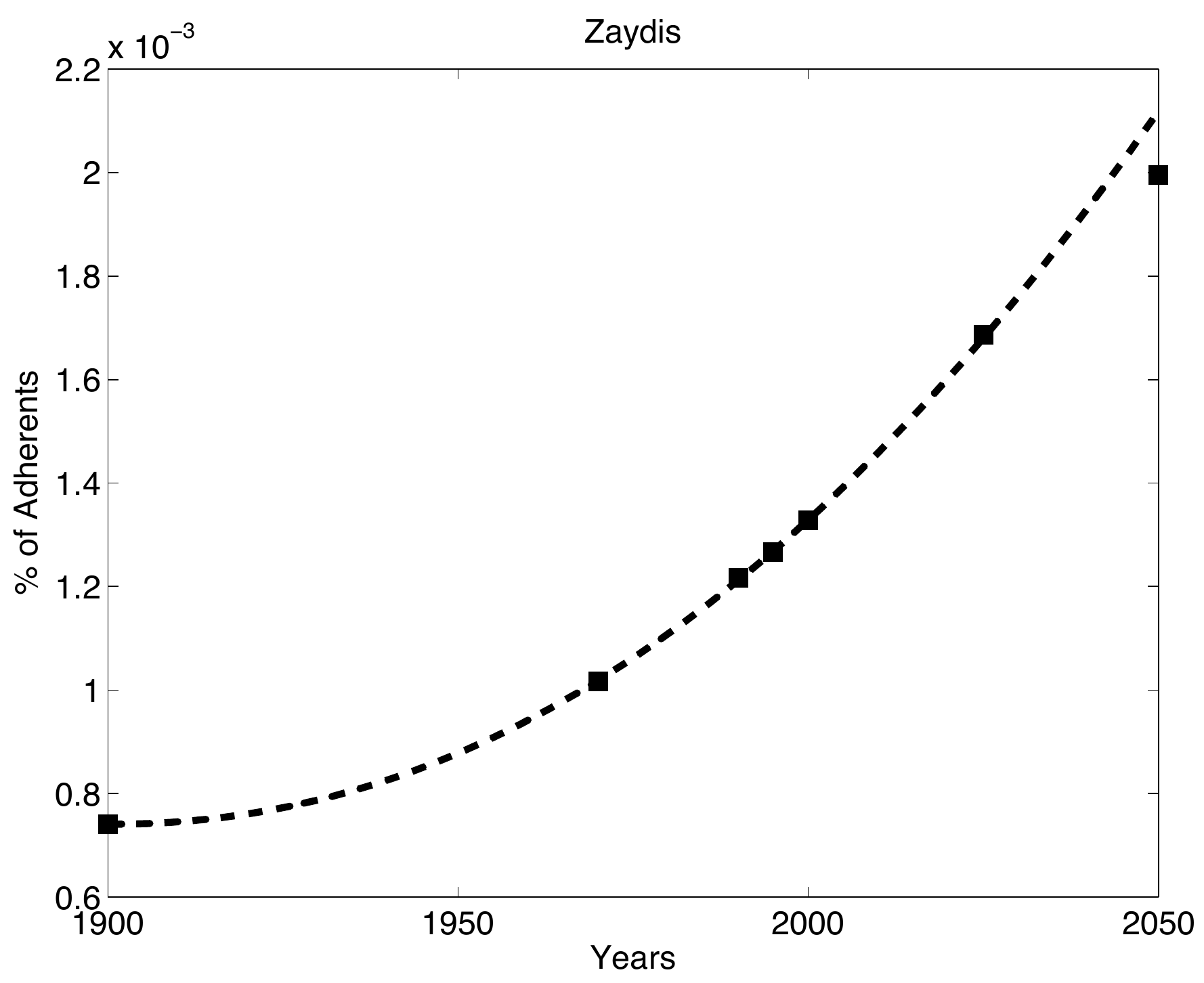}
\includegraphics[height=5cm,width=5cm]{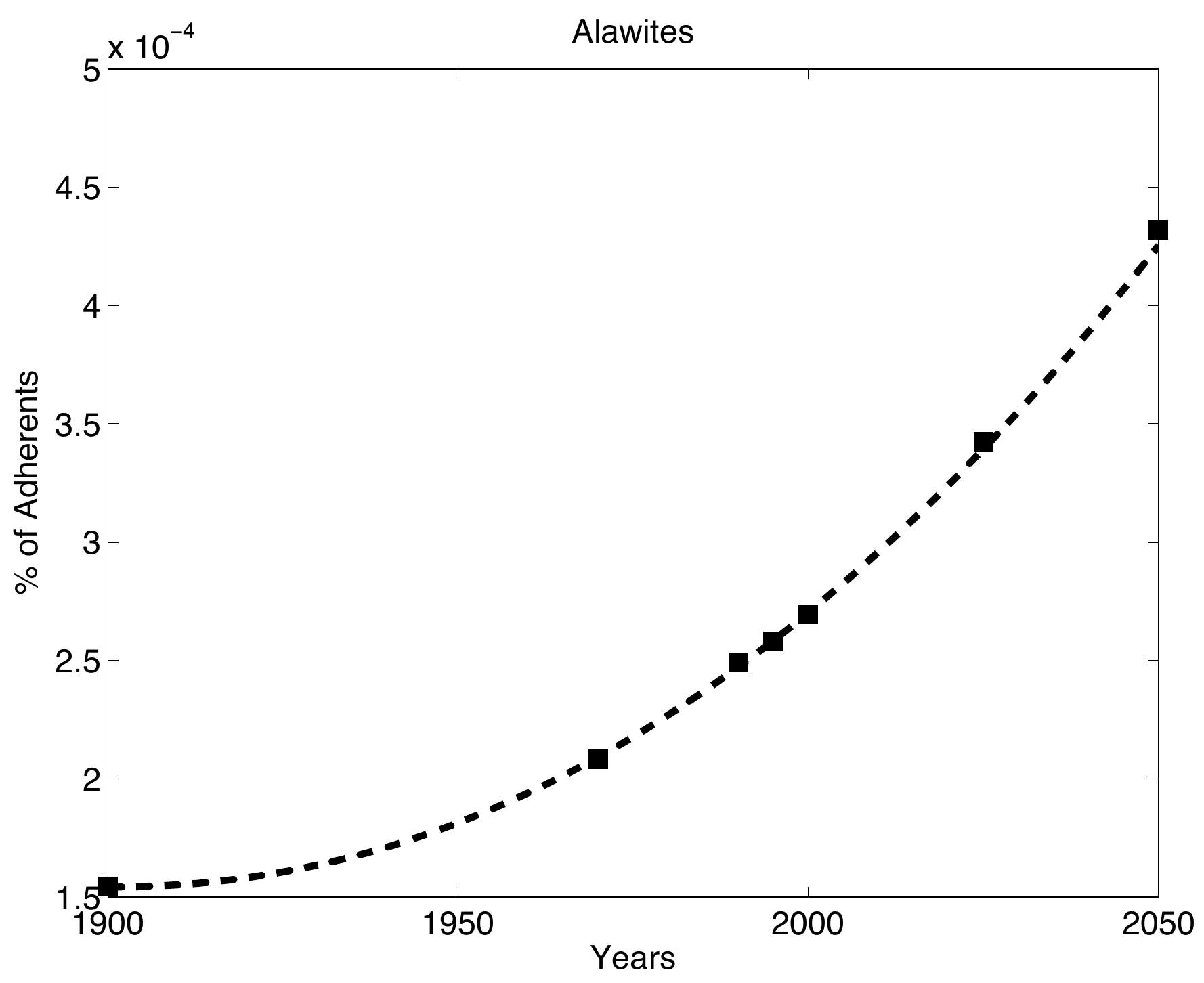}
\includegraphics[height=5cm,width=5cm]{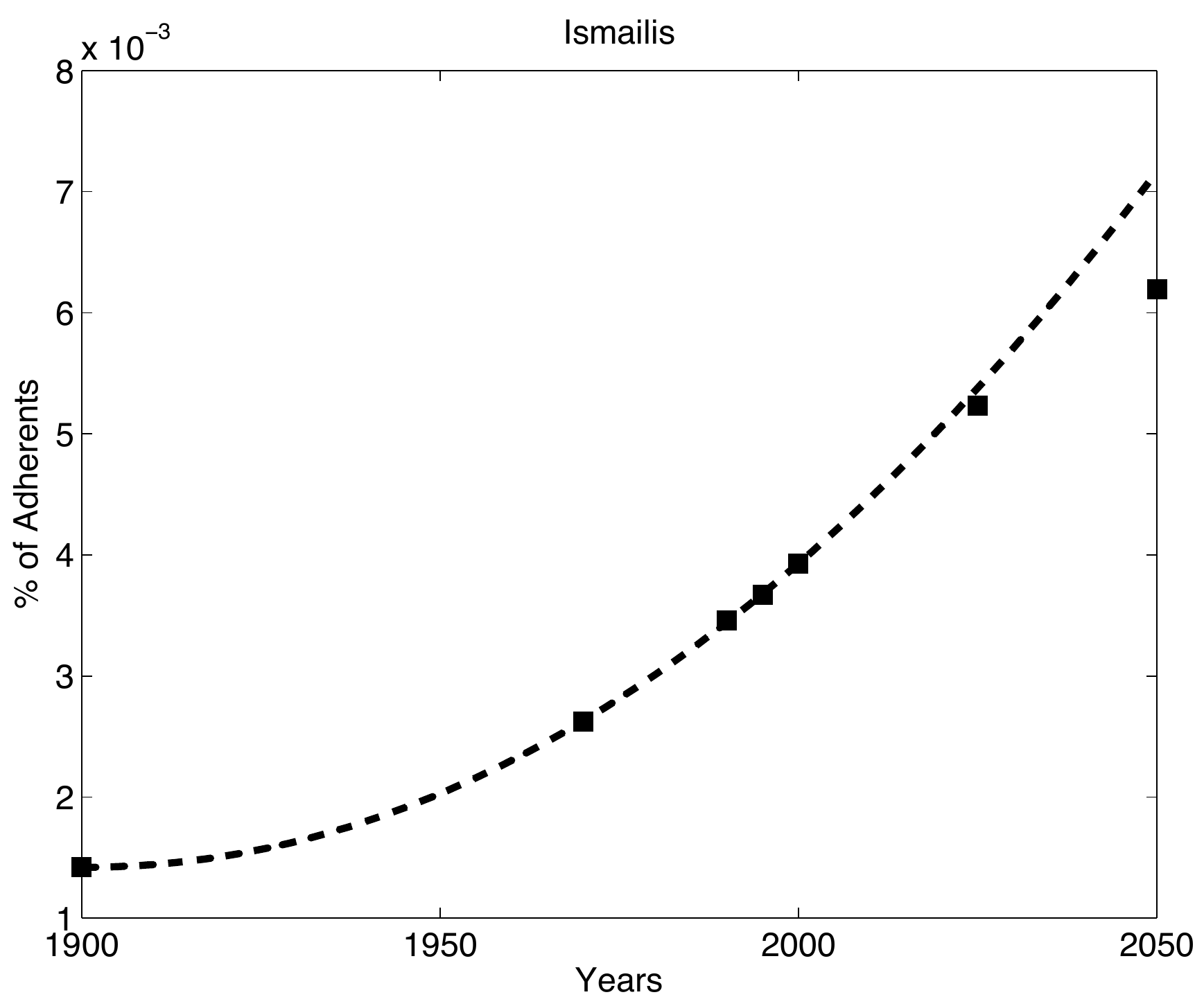}
\includegraphics[height=5cm,width=5cm]{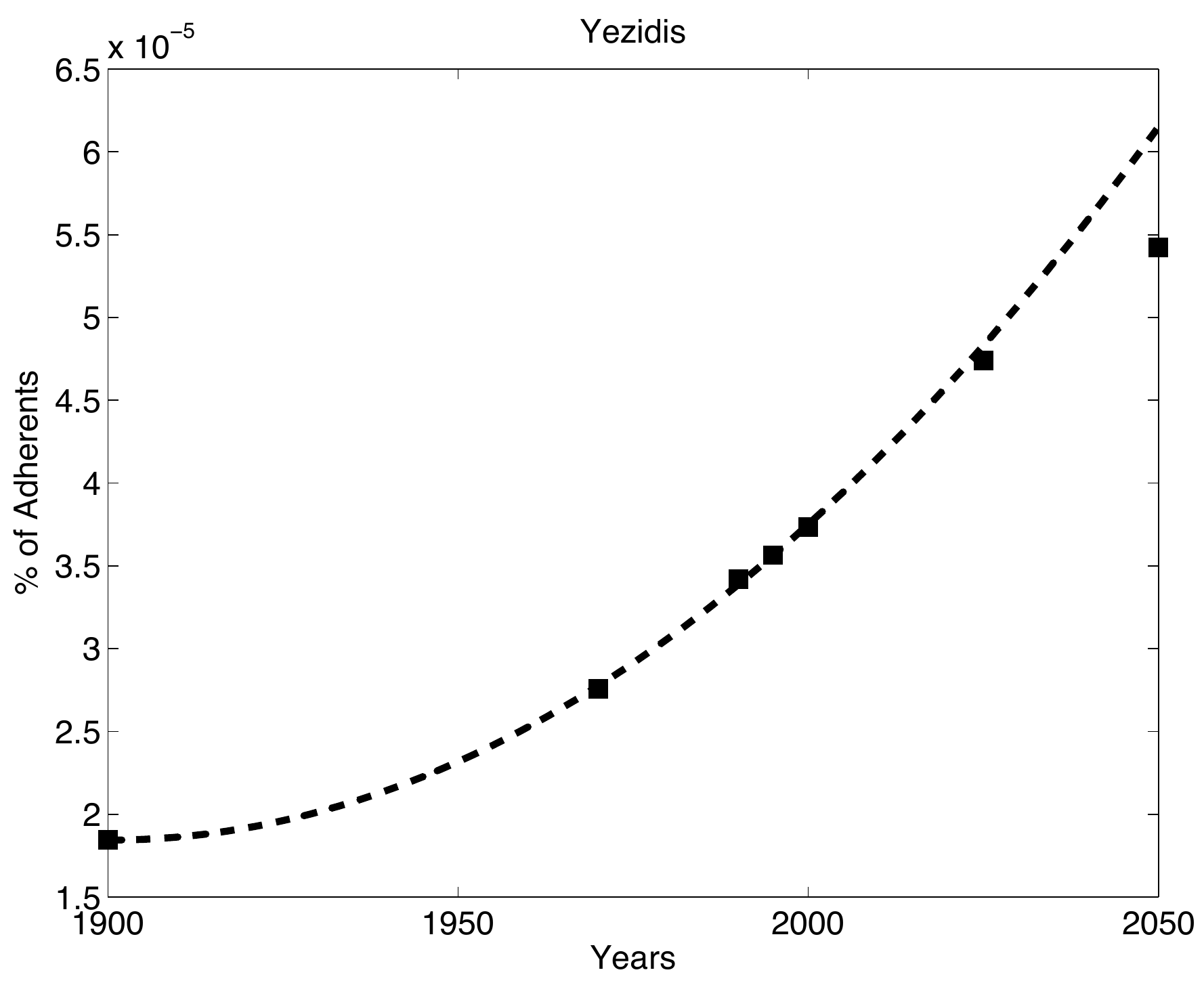}
\includegraphics[height=5cm,width=5cm]{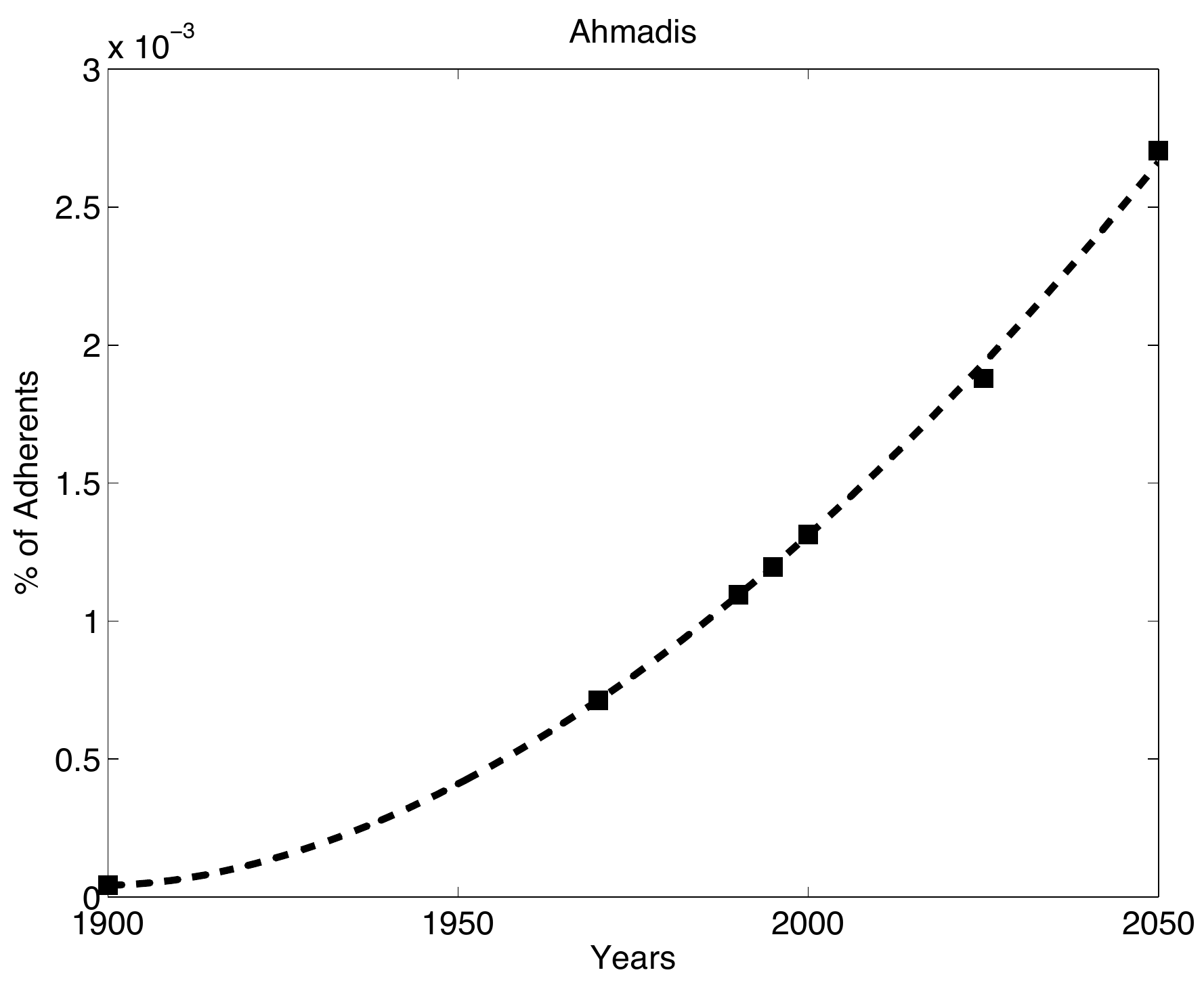}
\includegraphics[height=5cm,width=5cm]{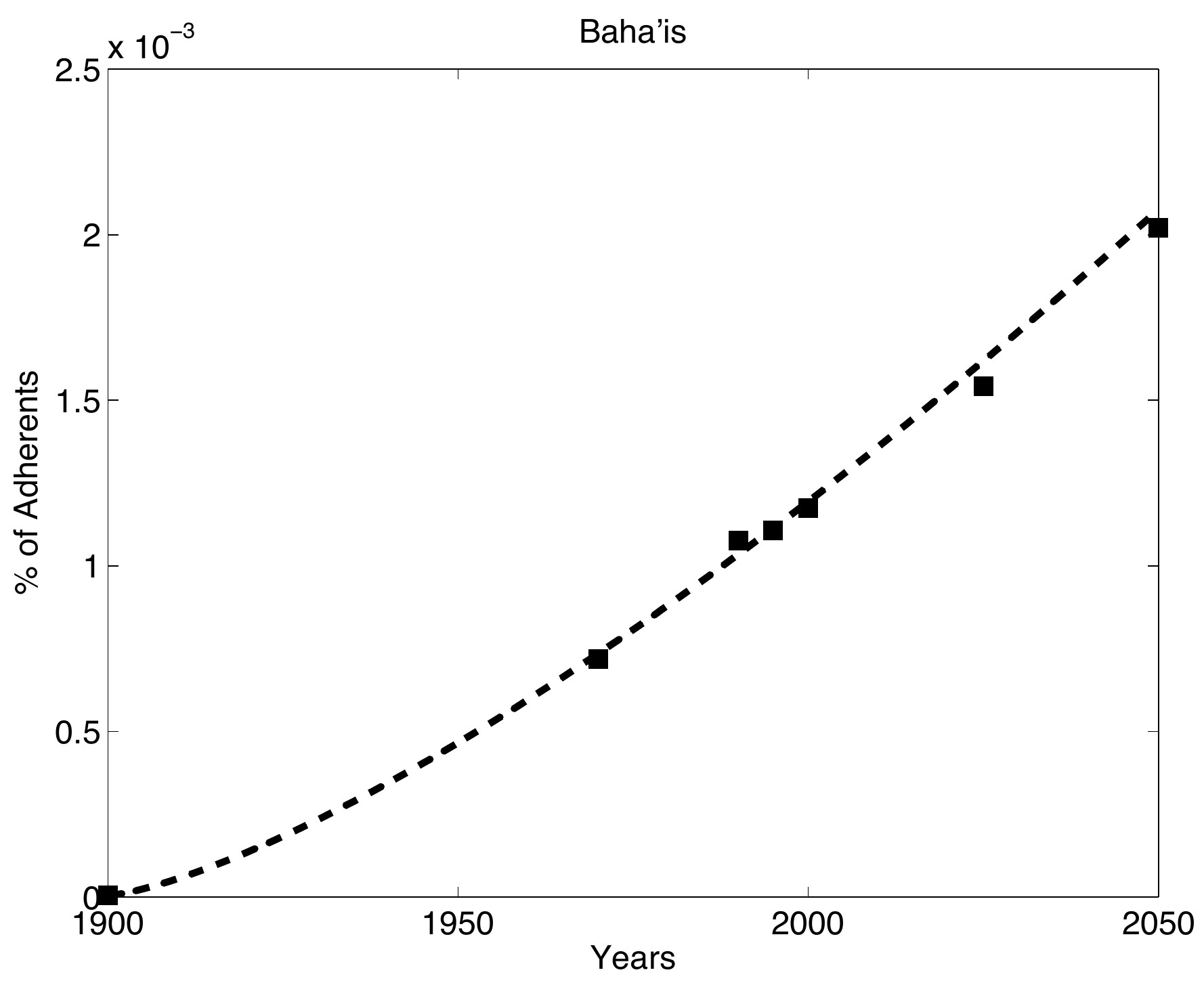}
\includegraphics[height=5cm,width=5cm]{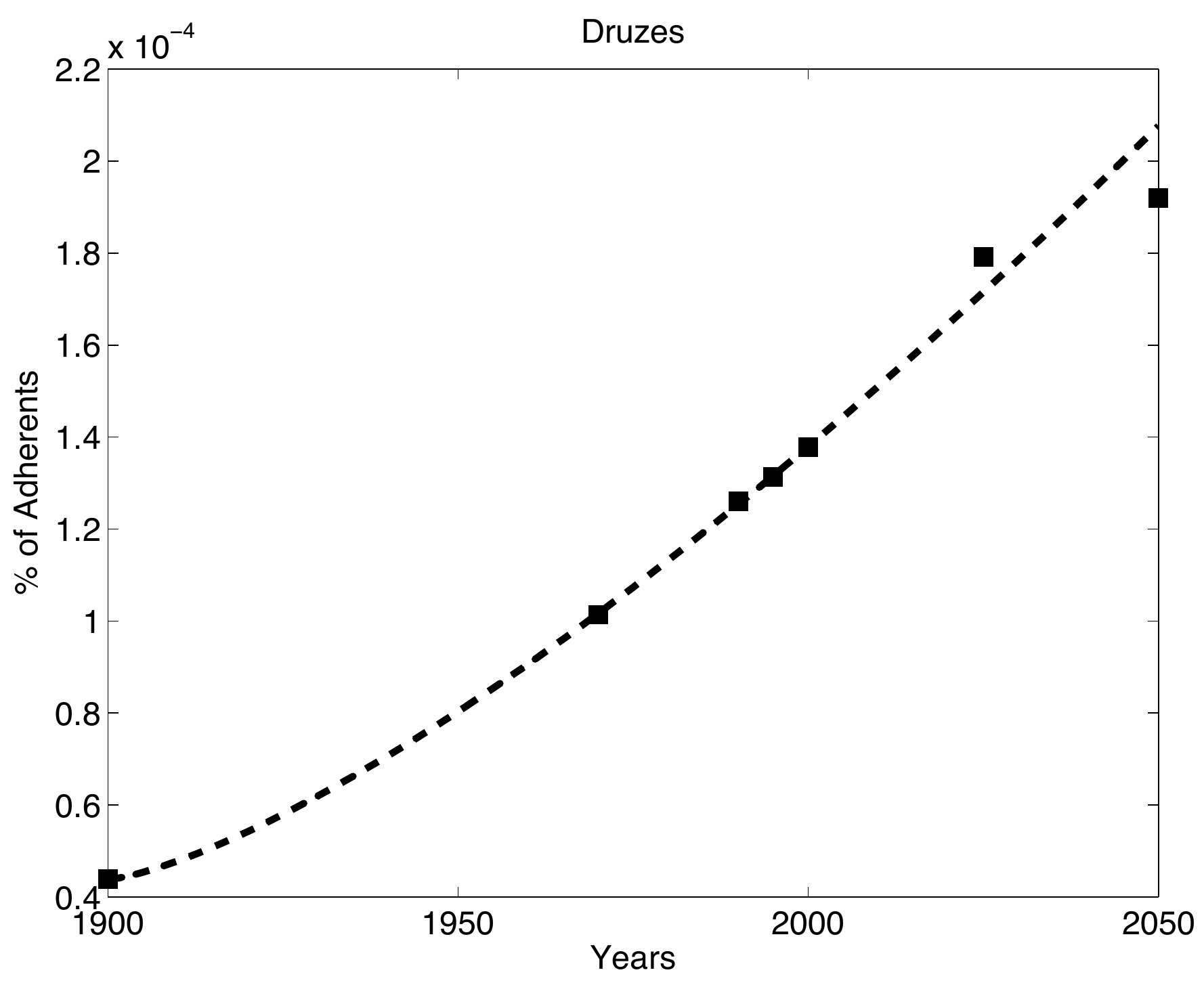}
\includegraphics[height=5cm,width=5cm]{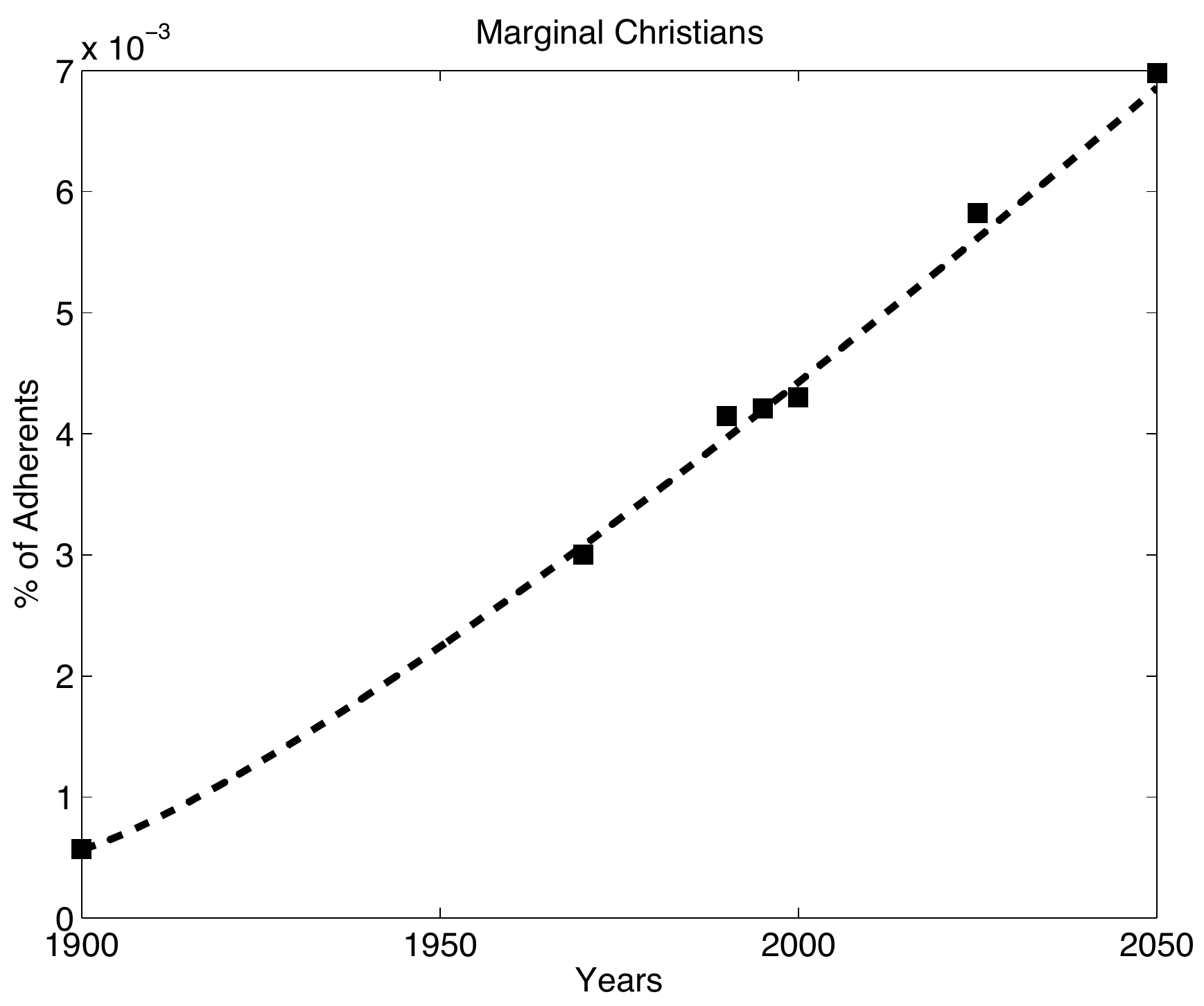}
\caption{\label{Fig3}   Six  illustrative cases of actually increasing (small size)  religions ... ; our empirical law confirms the WCT forecast}
\end{figure}

\begin{figure}
\centering
\includegraphics[height=5cm,width=5cm]{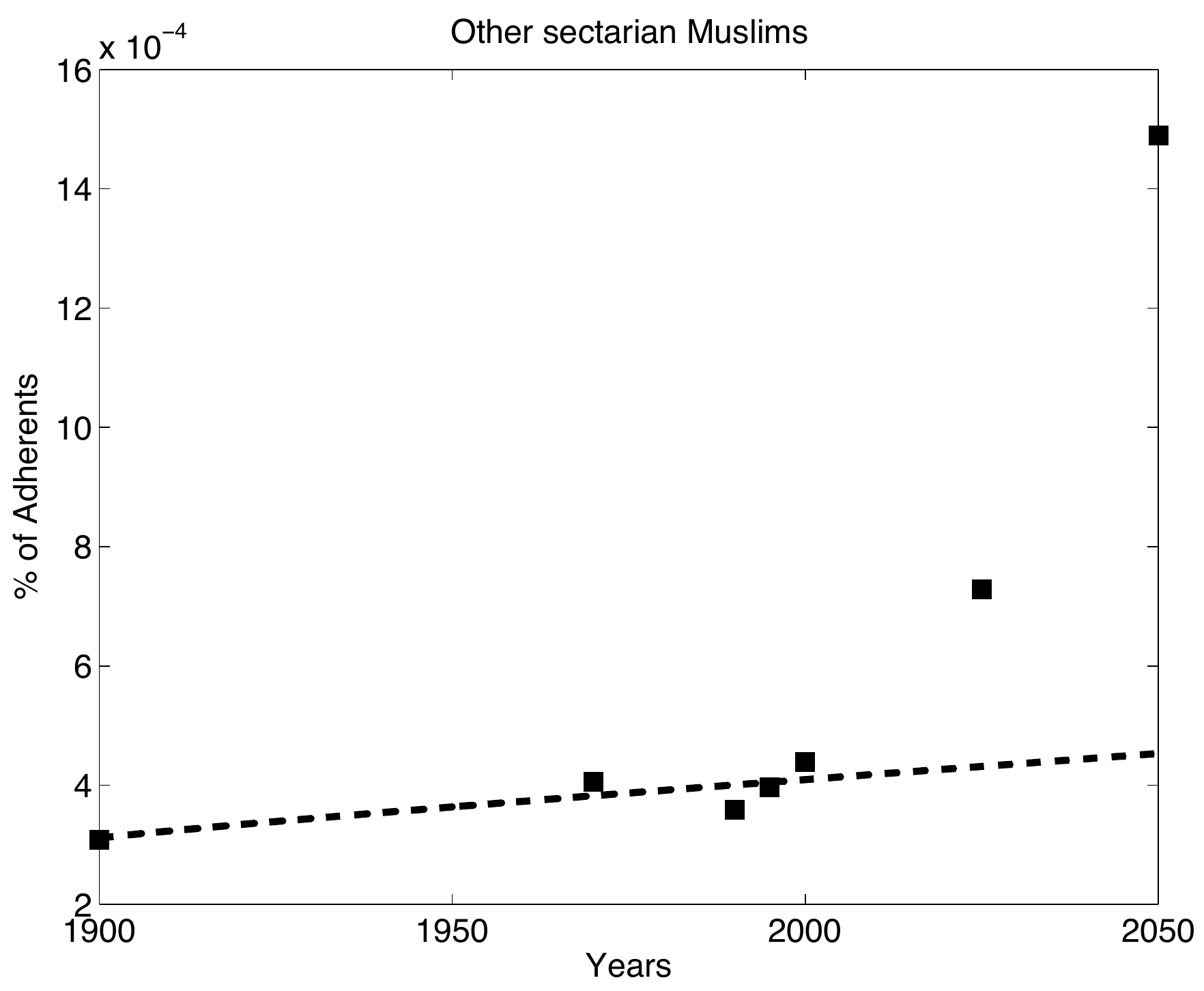}
\includegraphics[height=5cm,width=5cm]{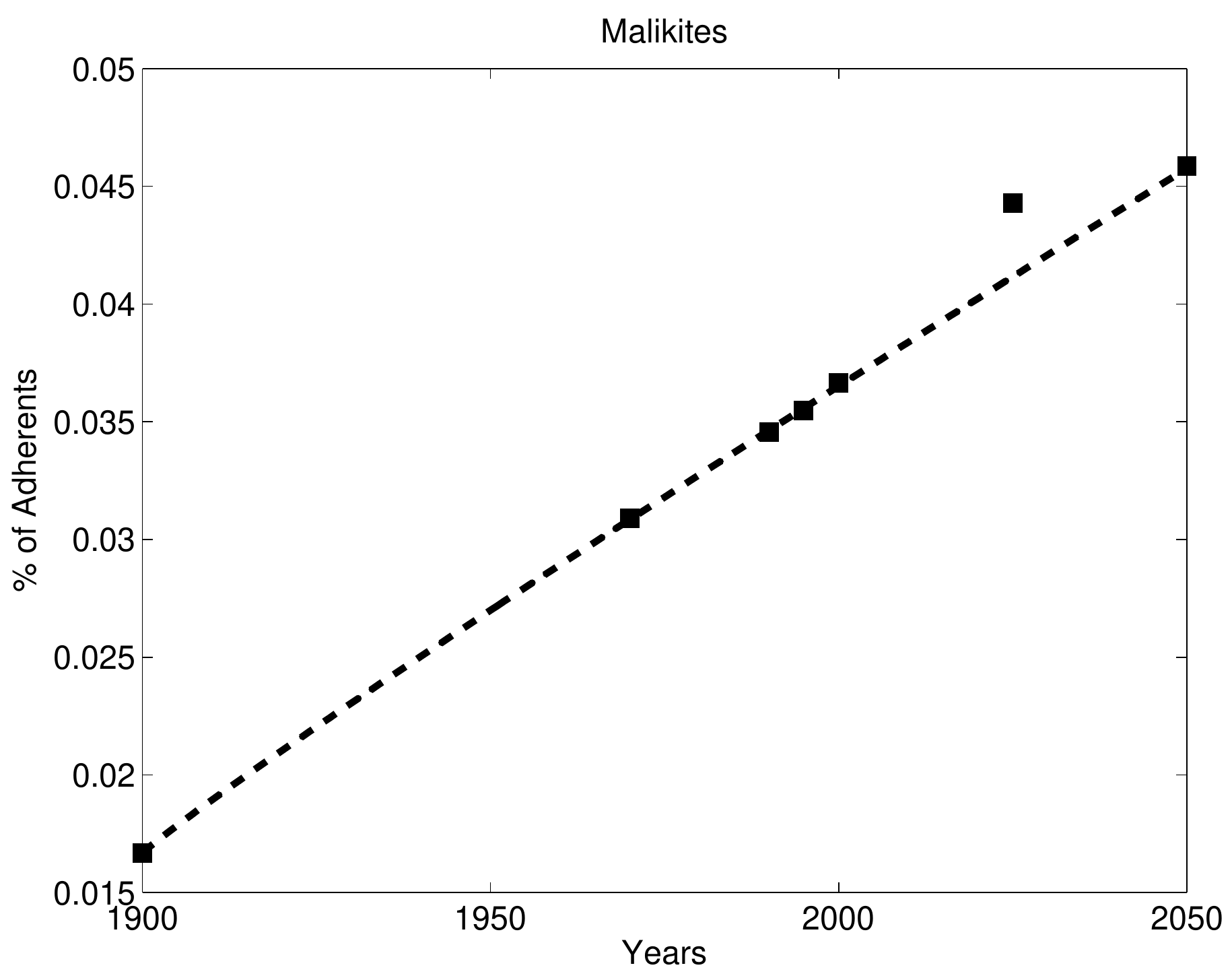}
\includegraphics[height=5cm,width=5cm]{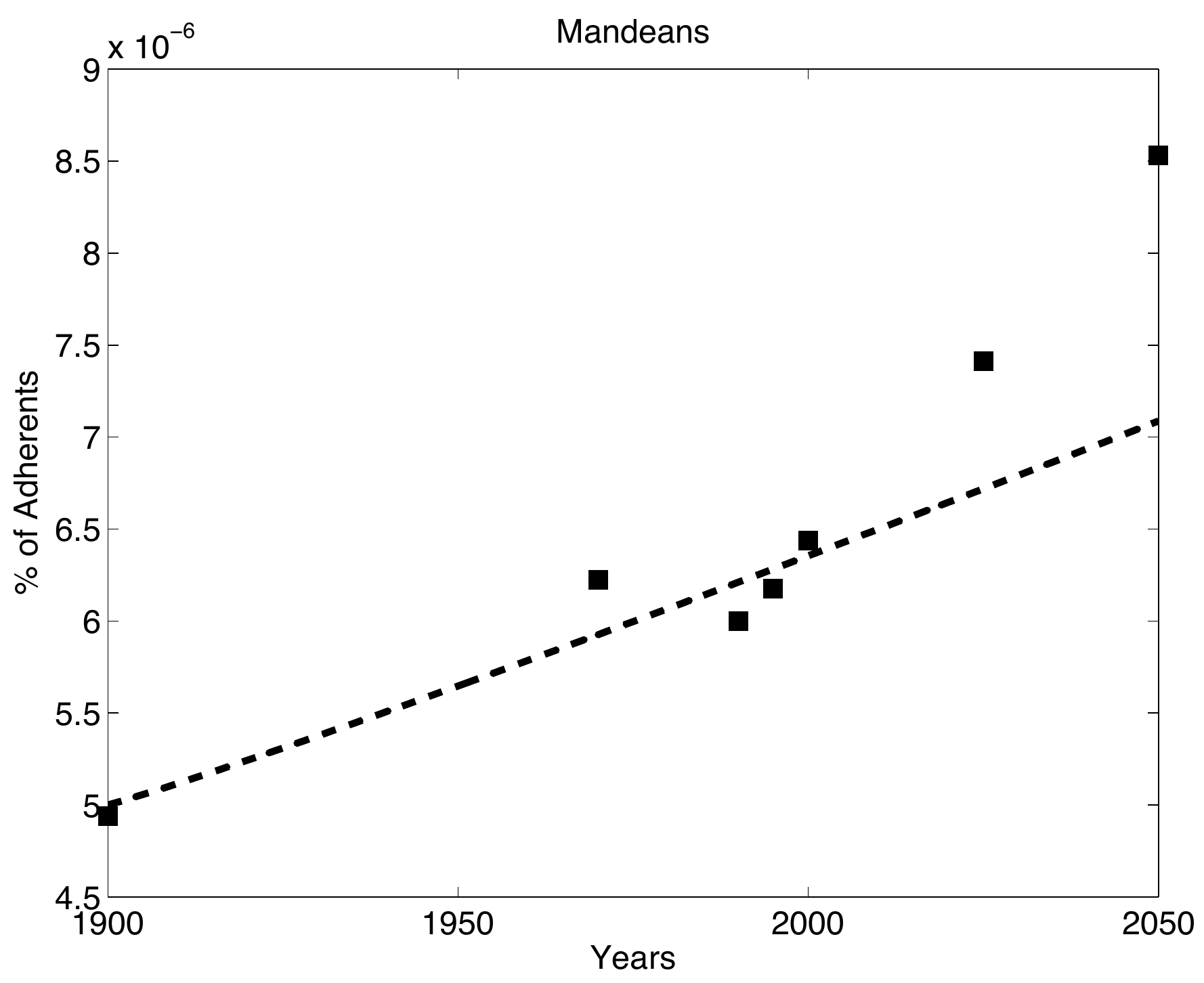}
\caption{\label{Fig6} Three small size religions, with increasing number of adherents; decreasing $h$ from left to right, with $h$ close  to 0; see Table I; our empirical law underestimates the WTE forecast}
\end{figure}

 \begin{figure}
\centering
\includegraphics[height=5cm,width=5cm]{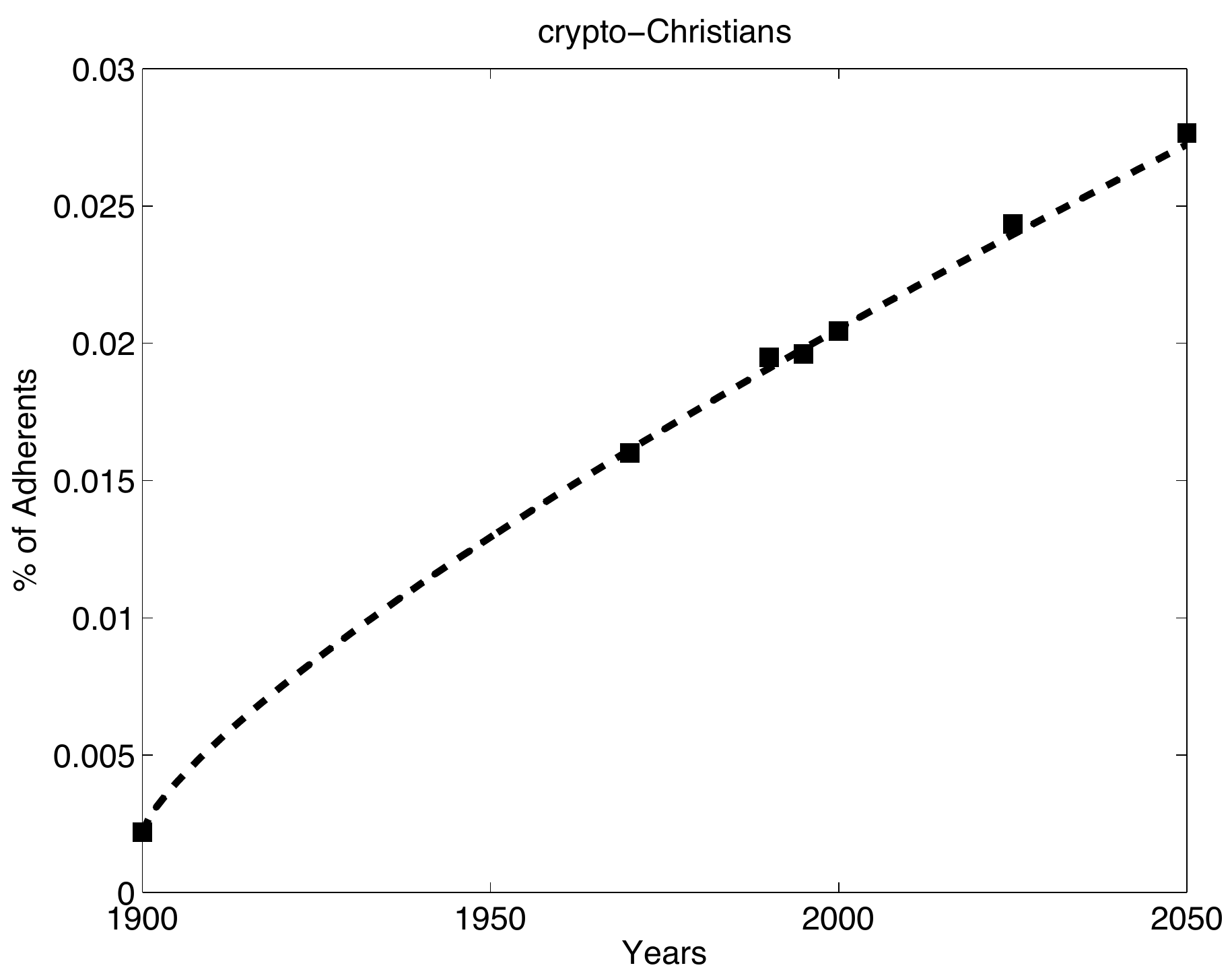}
\includegraphics[height=5cm,width=5cm]{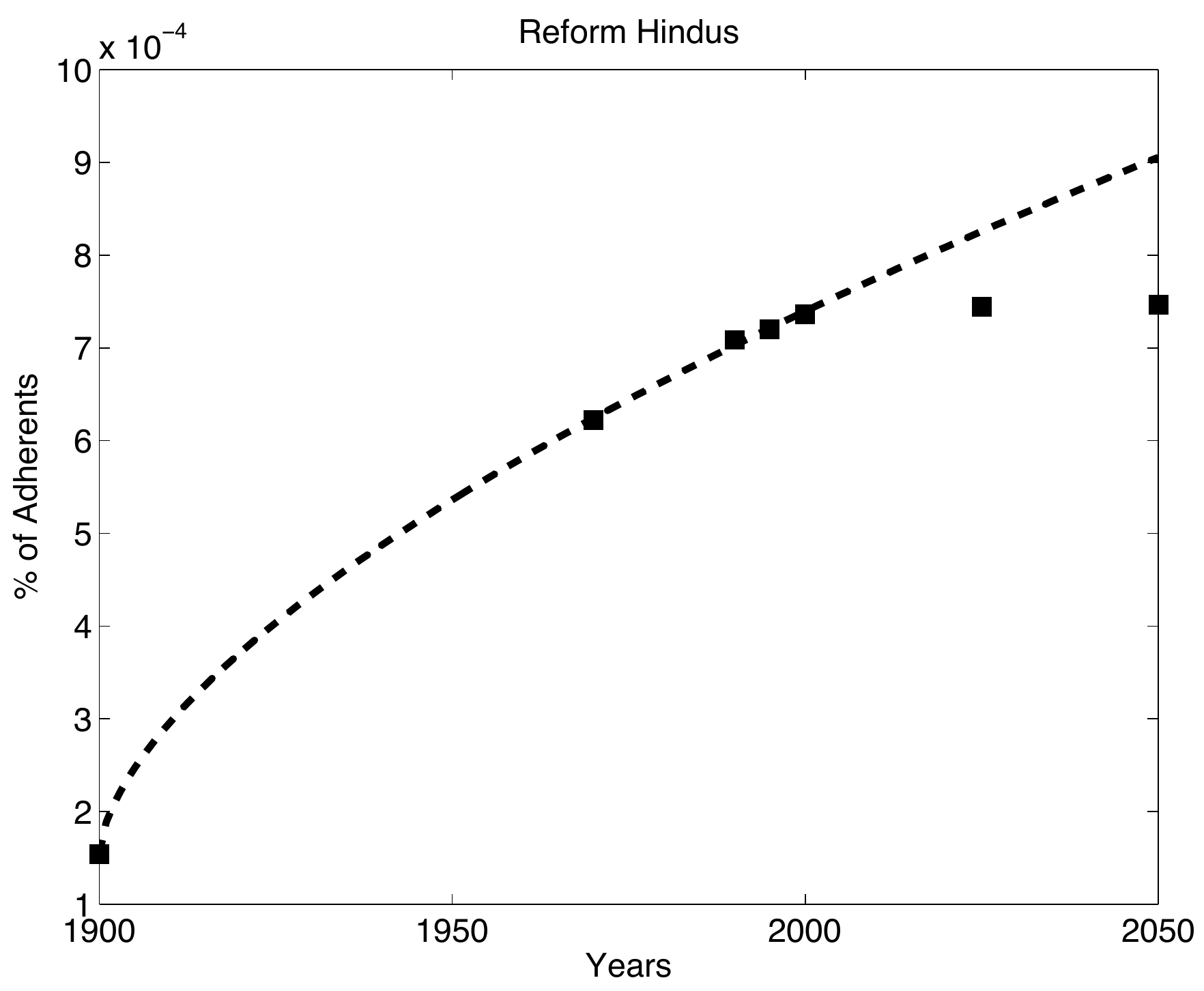}
\caption{\label{Fig7}    Two illustrative cases of actually increasing  (indicated) religions, with  saturating like forecast. Observe  the rather good fits, parameters in Table I,  $h$ positive and $\le$ 1, and even a rather  good confirmation of the WCT forecast}
\end{figure}

\begin{figure}
\centering
\includegraphics[height=5cm,width=5cm]{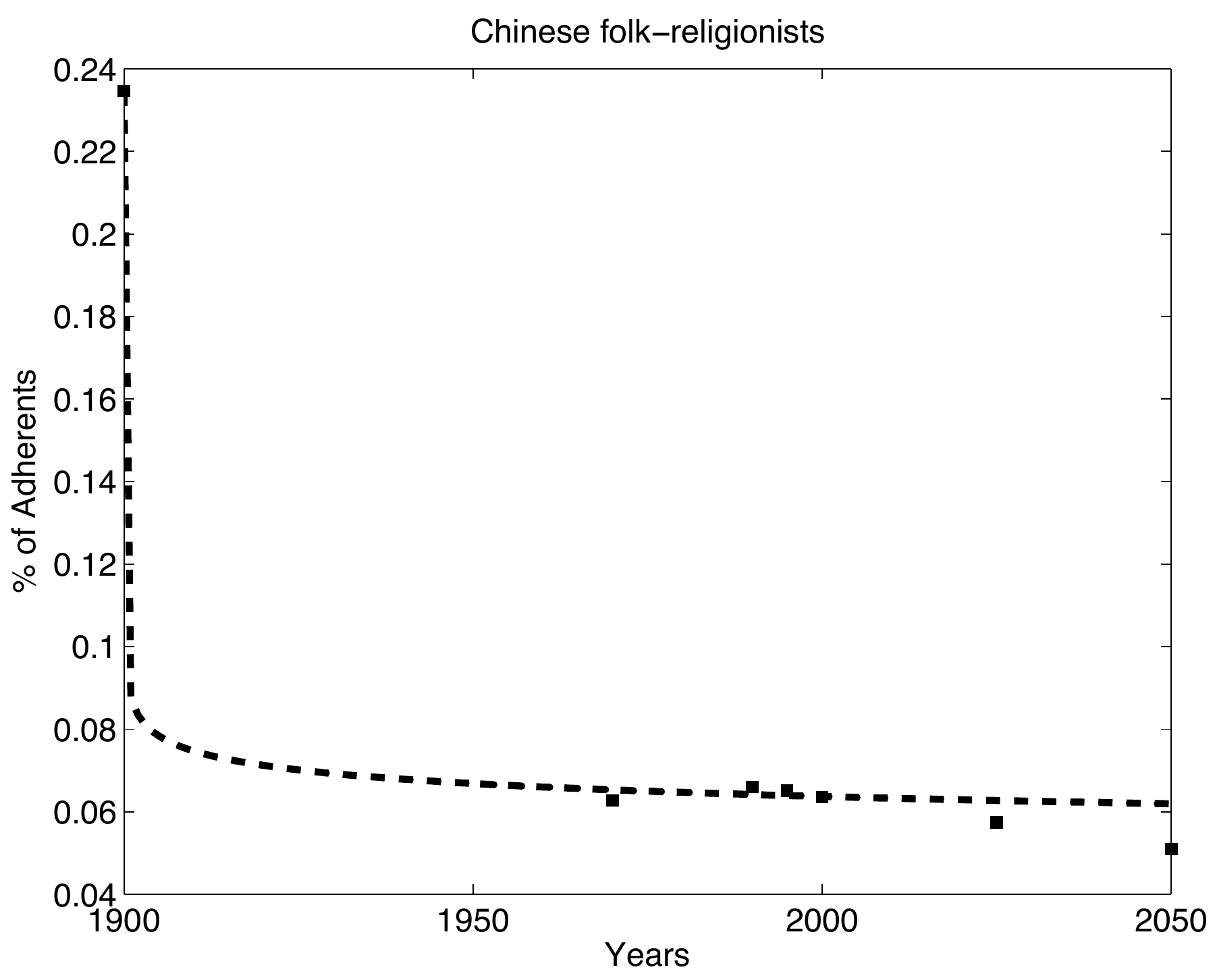}
\includegraphics[height=5cm,width=5cm]{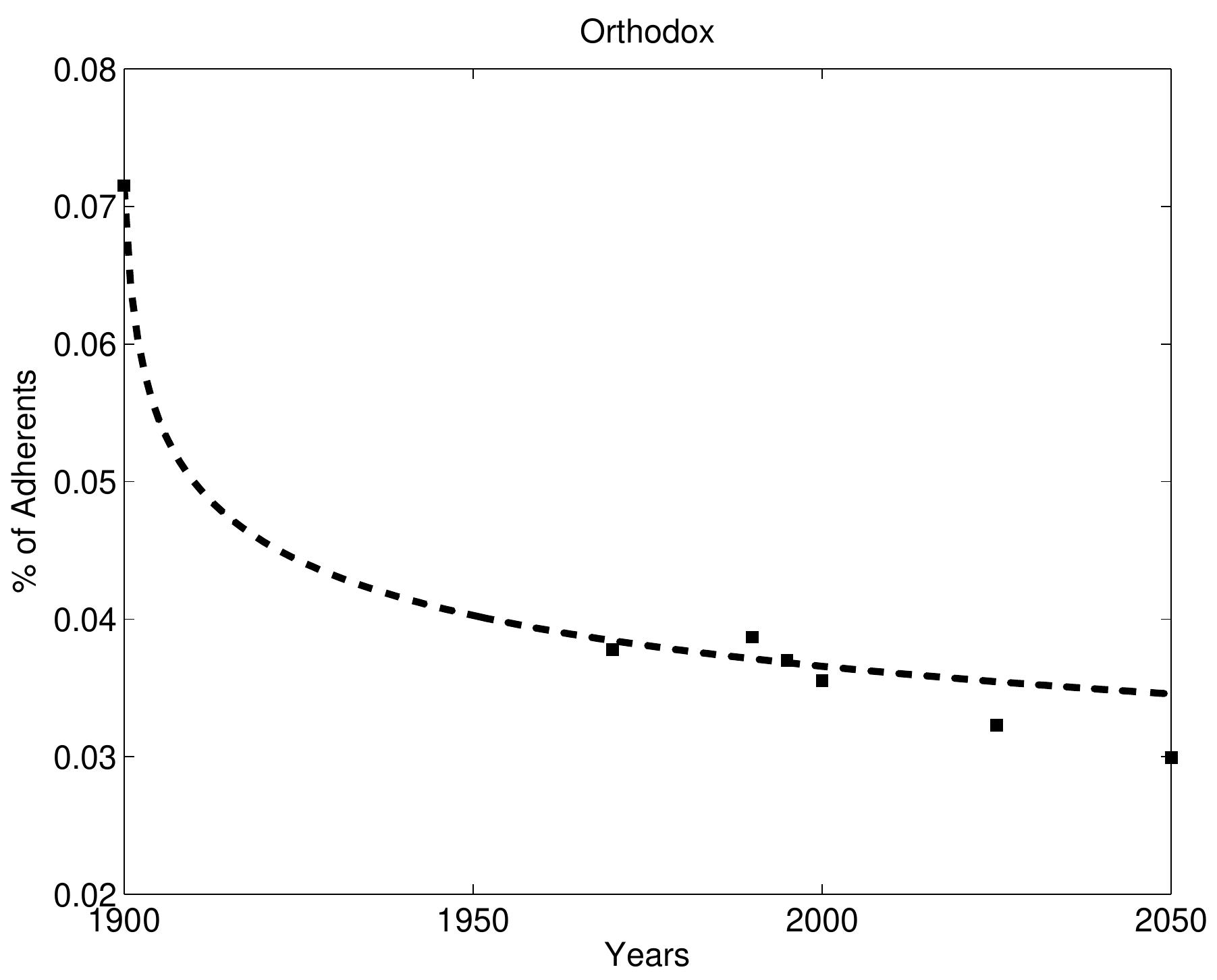}
\caption{\label{fig1.2.dec} Two sharply decaying religions, with very small $t_1$ and $h\ge1$}
\end{figure} 

\begin{figure}
\centering
\includegraphics[height=5cm,width=5cm]{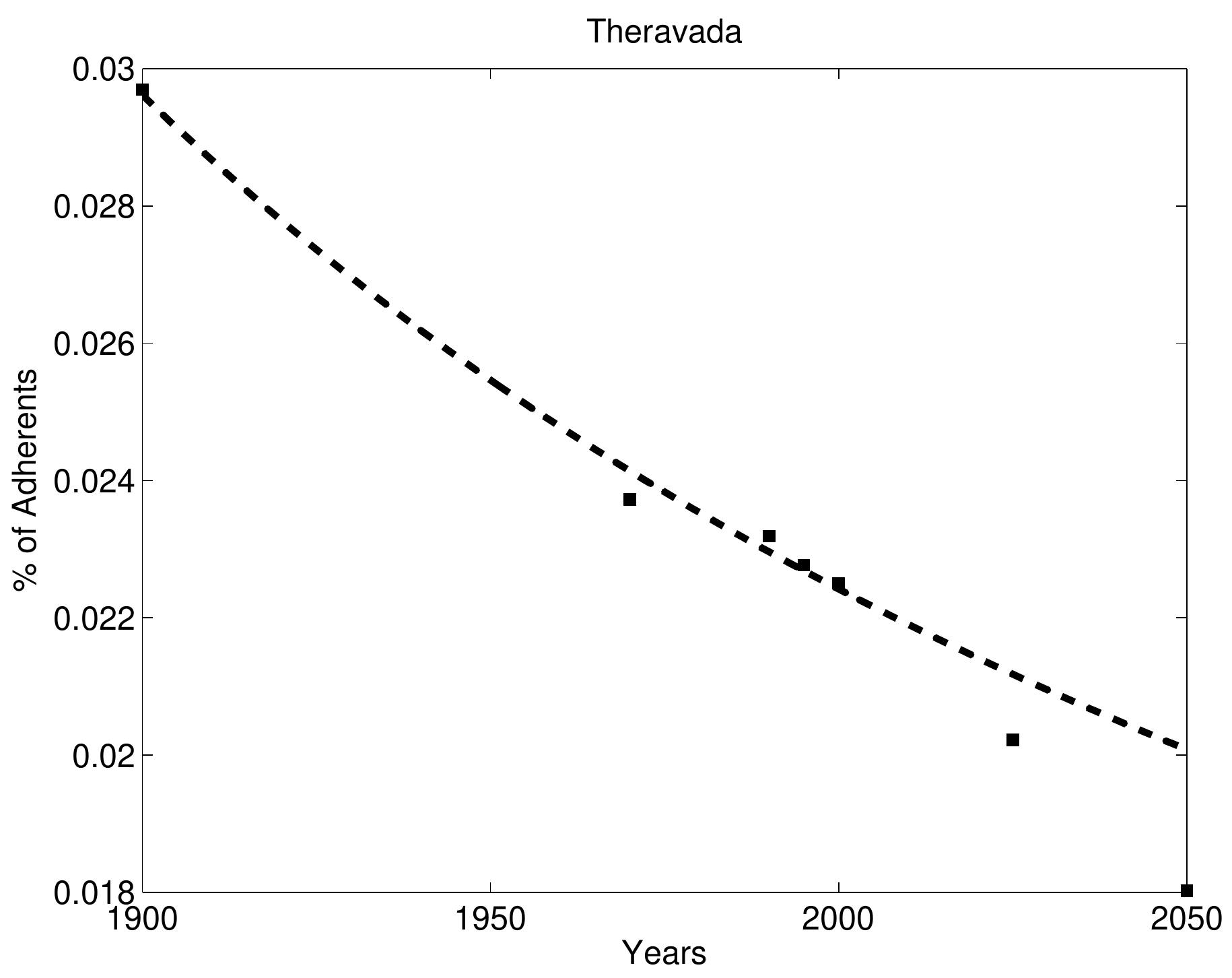}
\includegraphics[height=5cm,width=5cm]{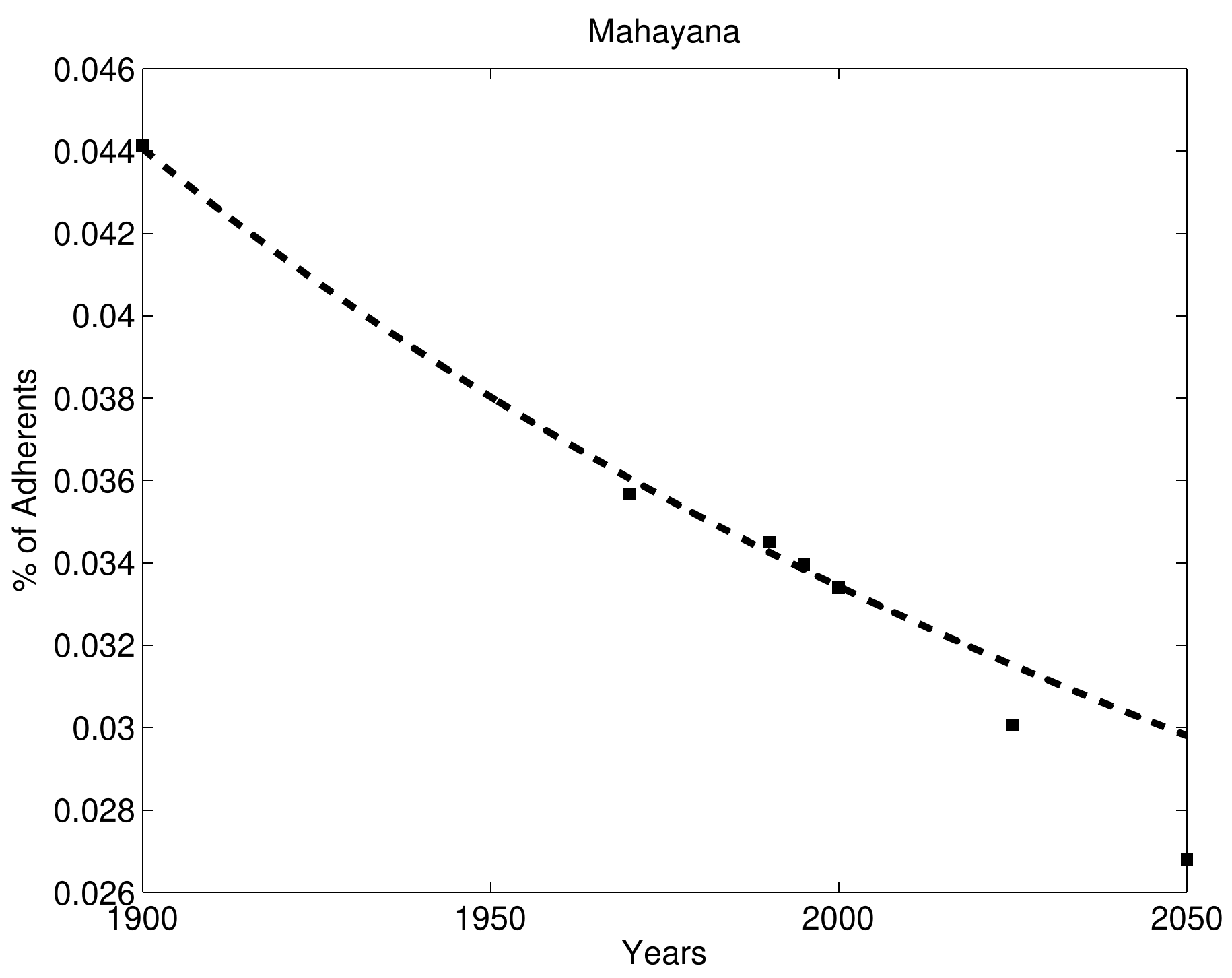}\\
\includegraphics[height=5cm,width=5cm]{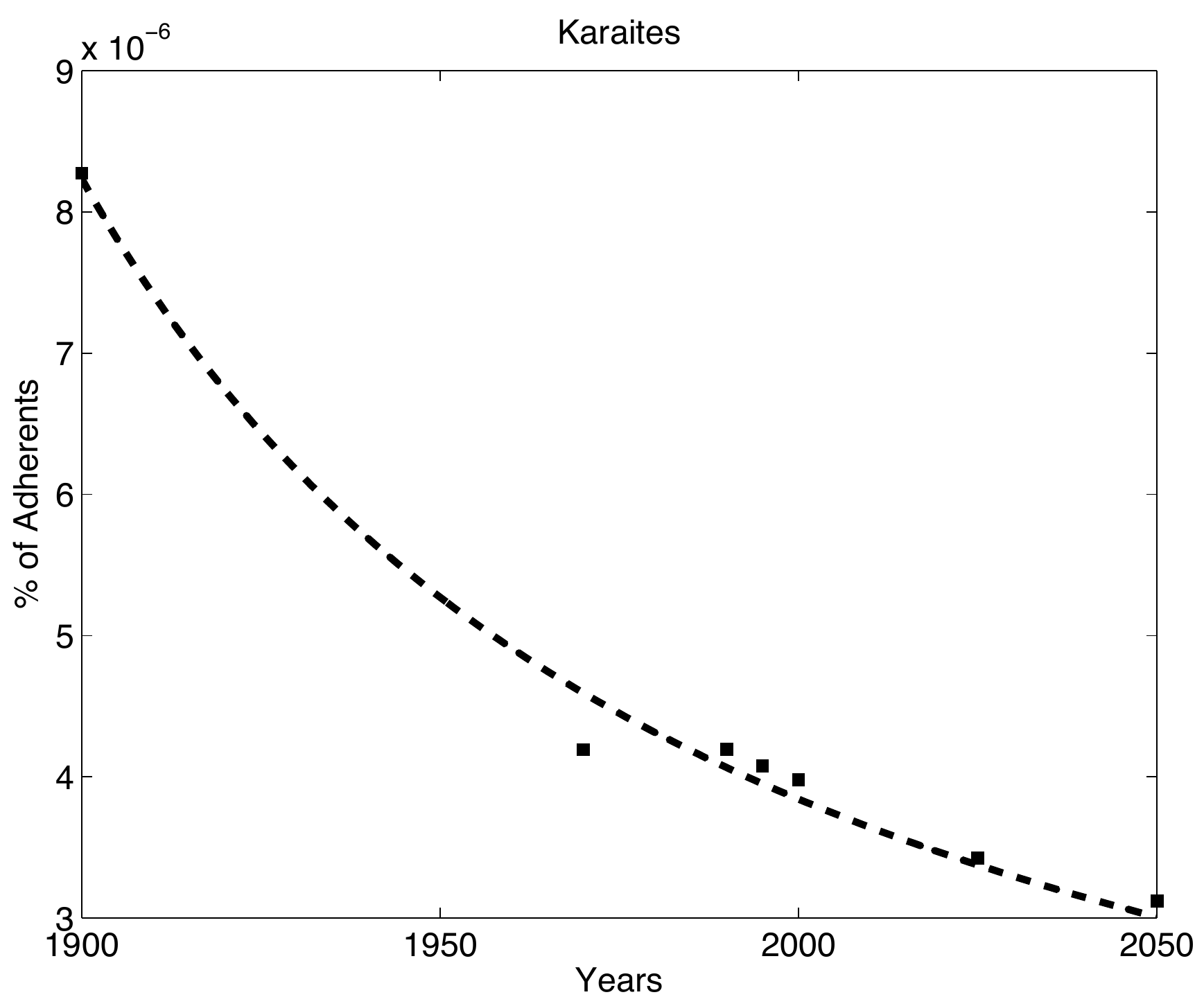}
\includegraphics[height=5cm,width=5cm]{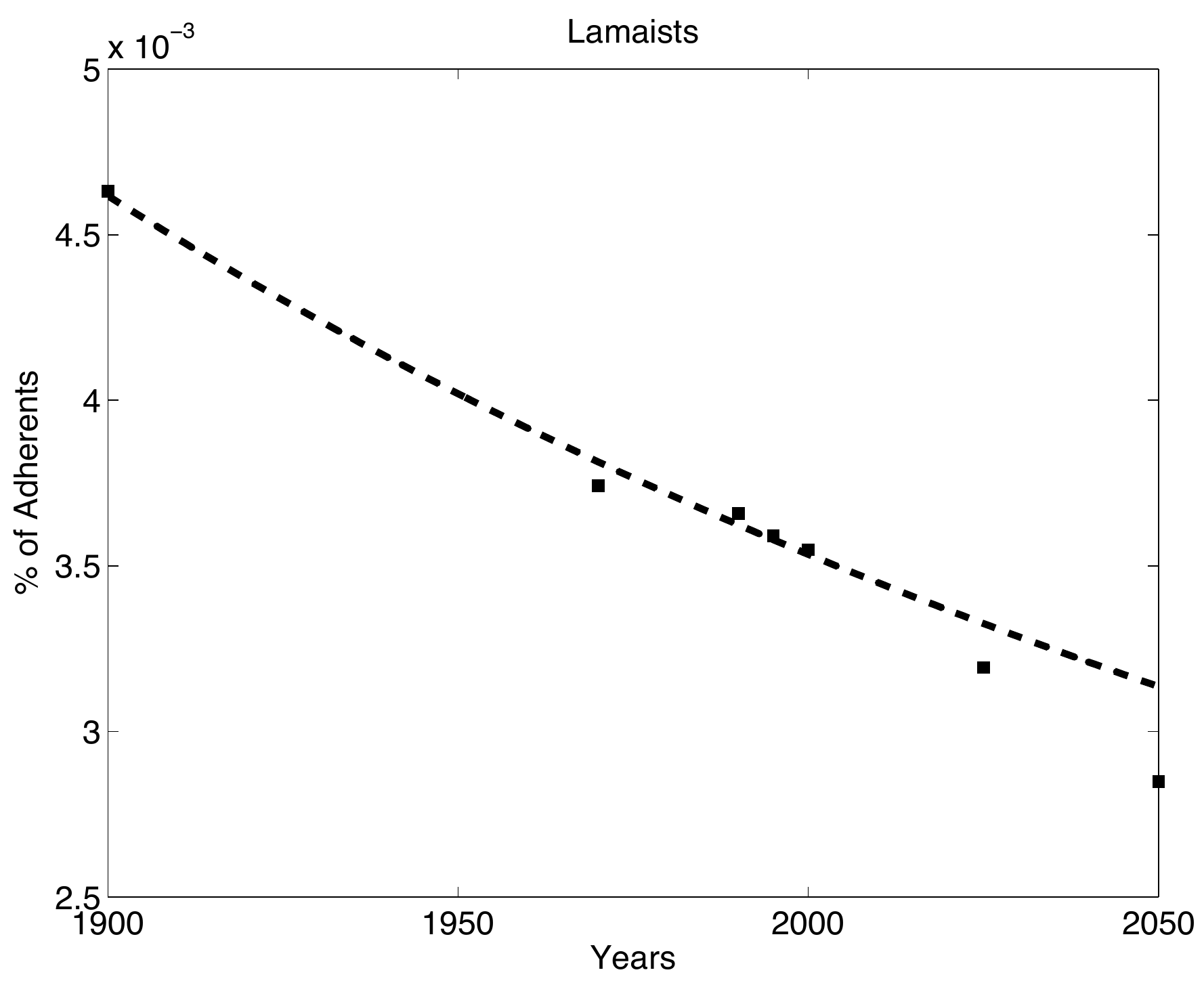}
\caption{\label{fig3.4.5.6.dec} Four   smoothly decaying religions with $h$ much larger than 1; our forecast being similar to that predicted  in WTE }
\end{figure} 

\subsection{Intermediary comments}
In order to read the figures, let us point out the way we have here chosen for their display.   It seems  somewhat obvious, from a mathematical or physics point of view that one should consider (i) strictly increasing or decreasing cases, (ii) cases of growth after a minimum, or (iii) of decay after a maximum. This hints also to consider the  curvature as a relevant indicator as for other (financial) time series  \cite{ausloosiVD}. Therefore we  have grossly ranked the figures and data according to whether  the religion number of adherents seems to be increasing (Table I), with $h\le0$, starting from the lowest value and increasing,  in a power law fit. Next we display the ''decreasing'' cases along an Avrami law, again ranking in order of increasing $h$, corresponding to fits with parameters given in (Table II).

Sometimes it is readily observed from the WCT tables that there are ''presently growing'' religions but for which a minimum is observed during the XX-th century, or a few are decaying after some maximum.  For such ''religions''   the number of adherents  can be in a first approximation fitted with
a second order polynomial  $y=A+Bx+Cx^2$ for which the parameters are 
given in  Table \ref{table3}). 

\begin{figure}
\centering
\includegraphics[height=5cm,width=5cm]{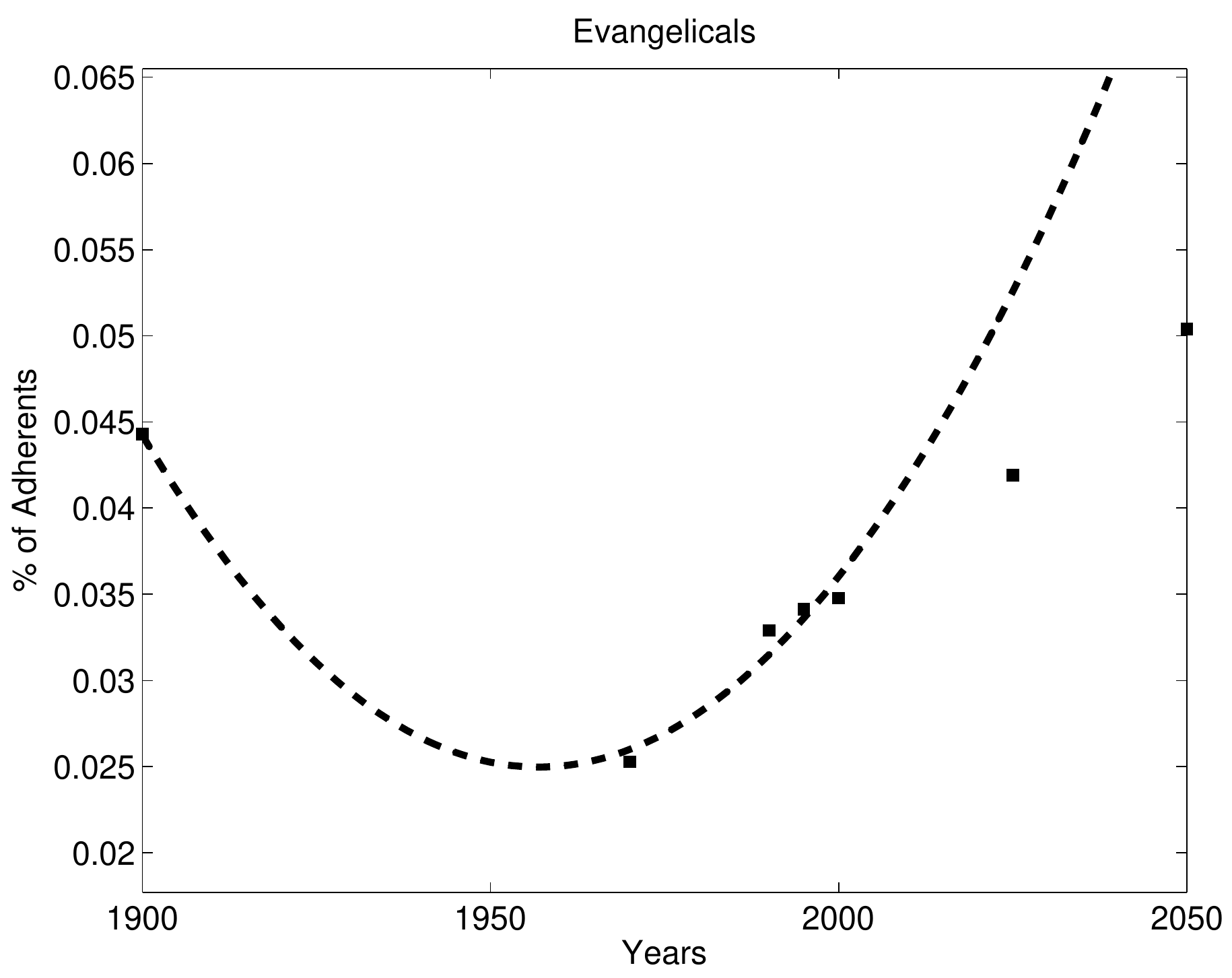}
\includegraphics[height=5cm,width=5cm]{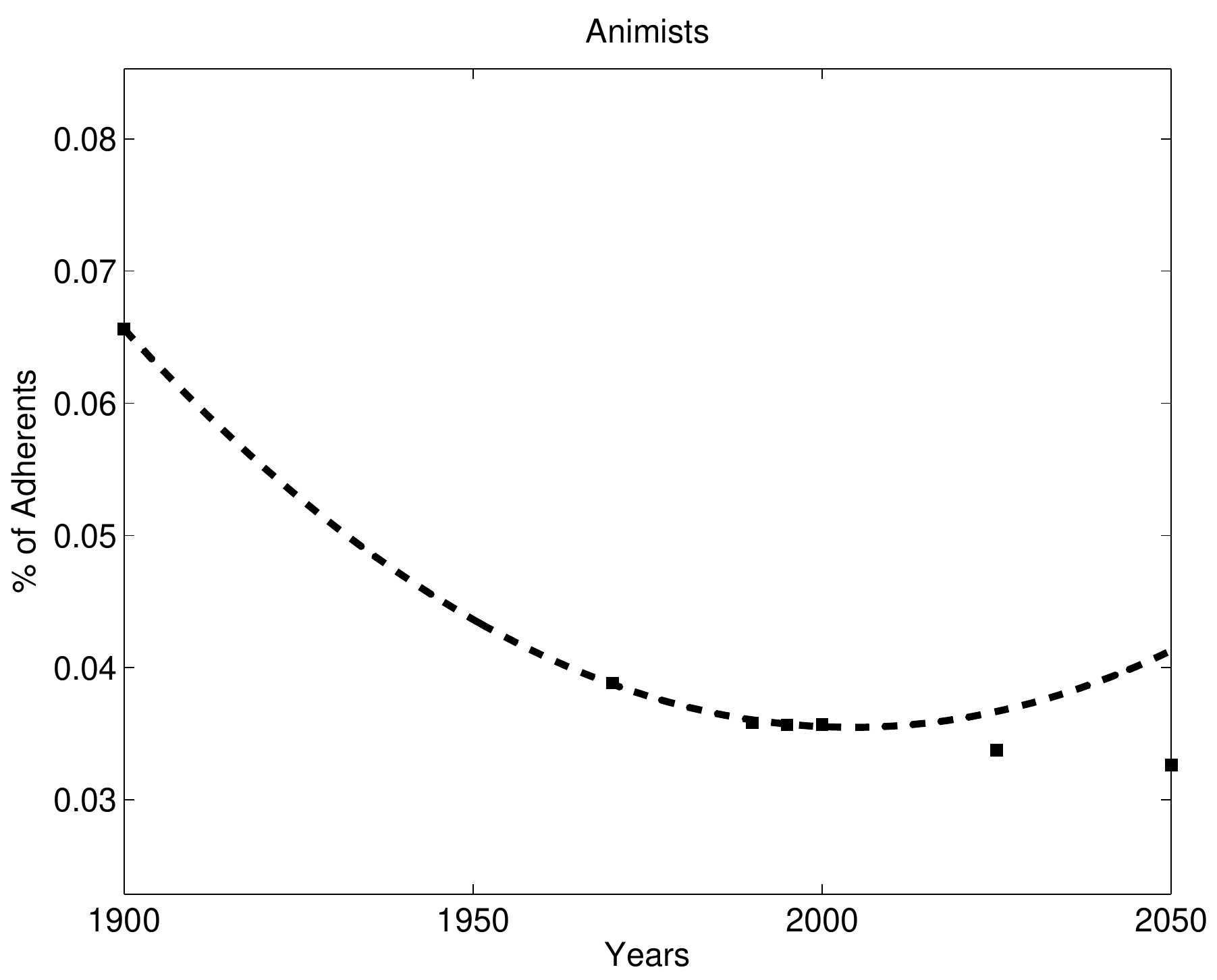}\\
\includegraphics[height=5cm,width=5cm]{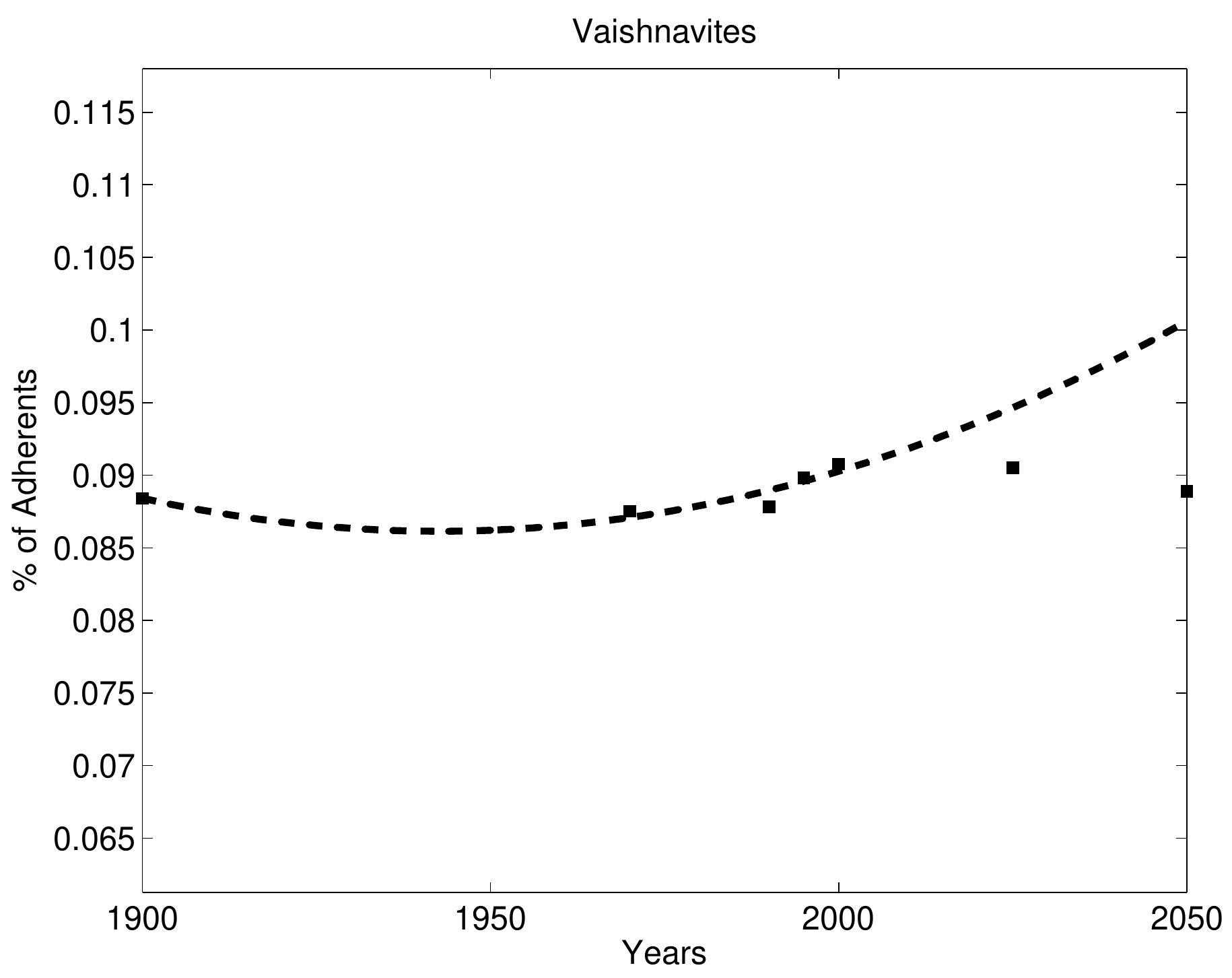}
\includegraphics[height=5cm,width=5cm]{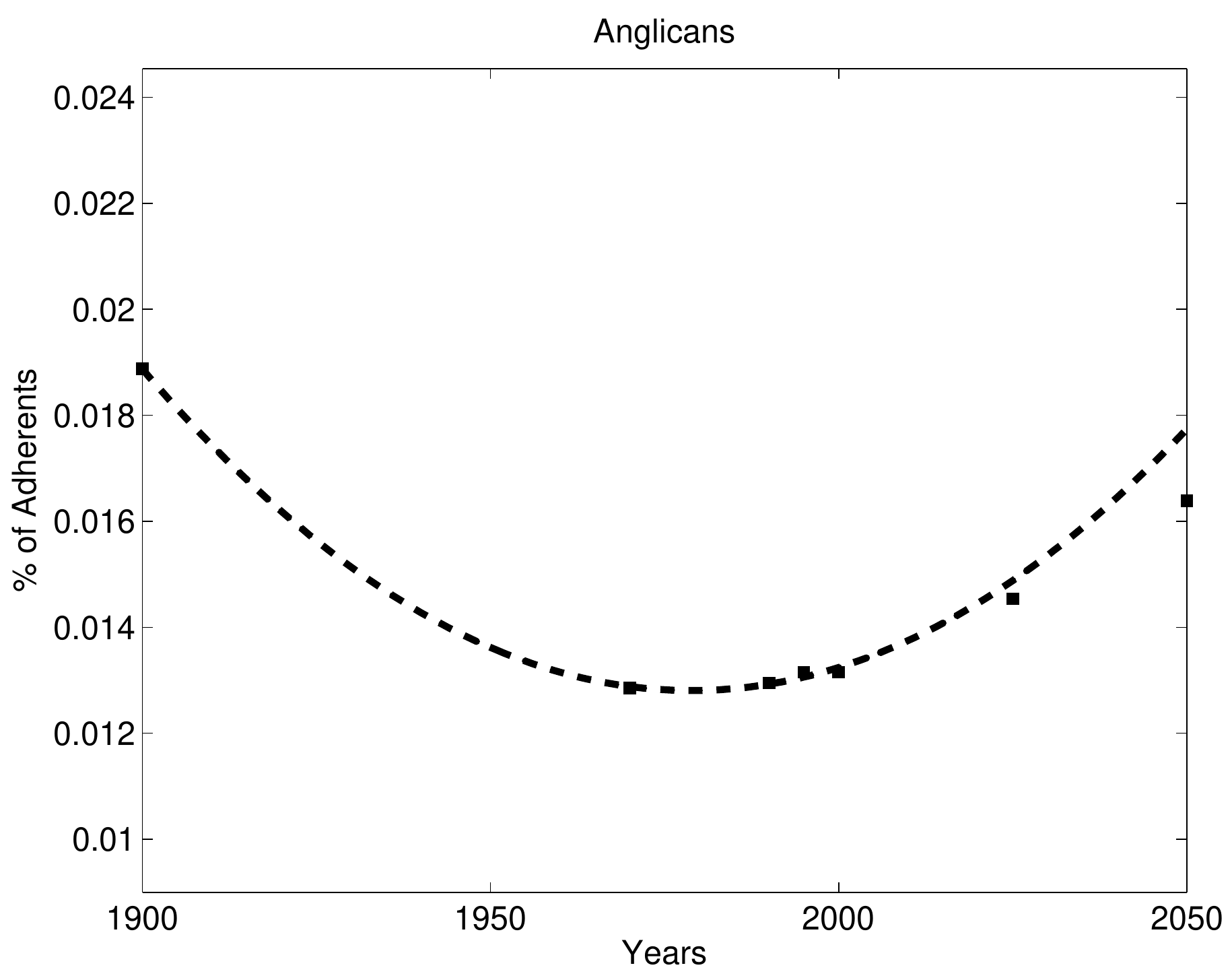}
\includegraphics[height=5cm,width=5cm]{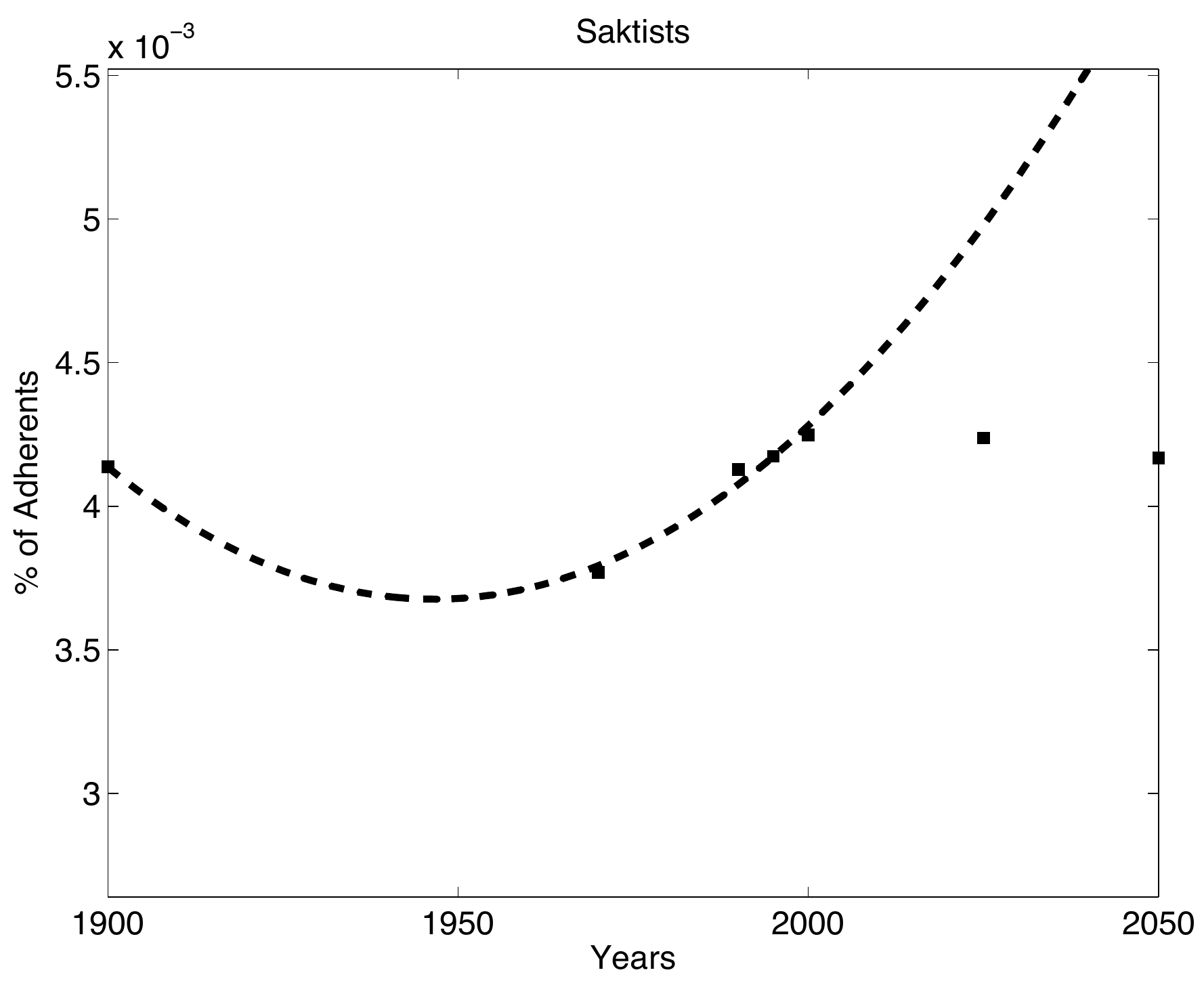}
\caption{\label{fig2.3.4.lq} Seven  large size religions indicating a turn over  with a minimum in XX-th century; theoretical forecasting with respect to WTE  is debatable though our fit slightly overshoots the WTE data}
\end{figure} 

\begin{figure}
\centering
\includegraphics[height=5cm,width=5cm]{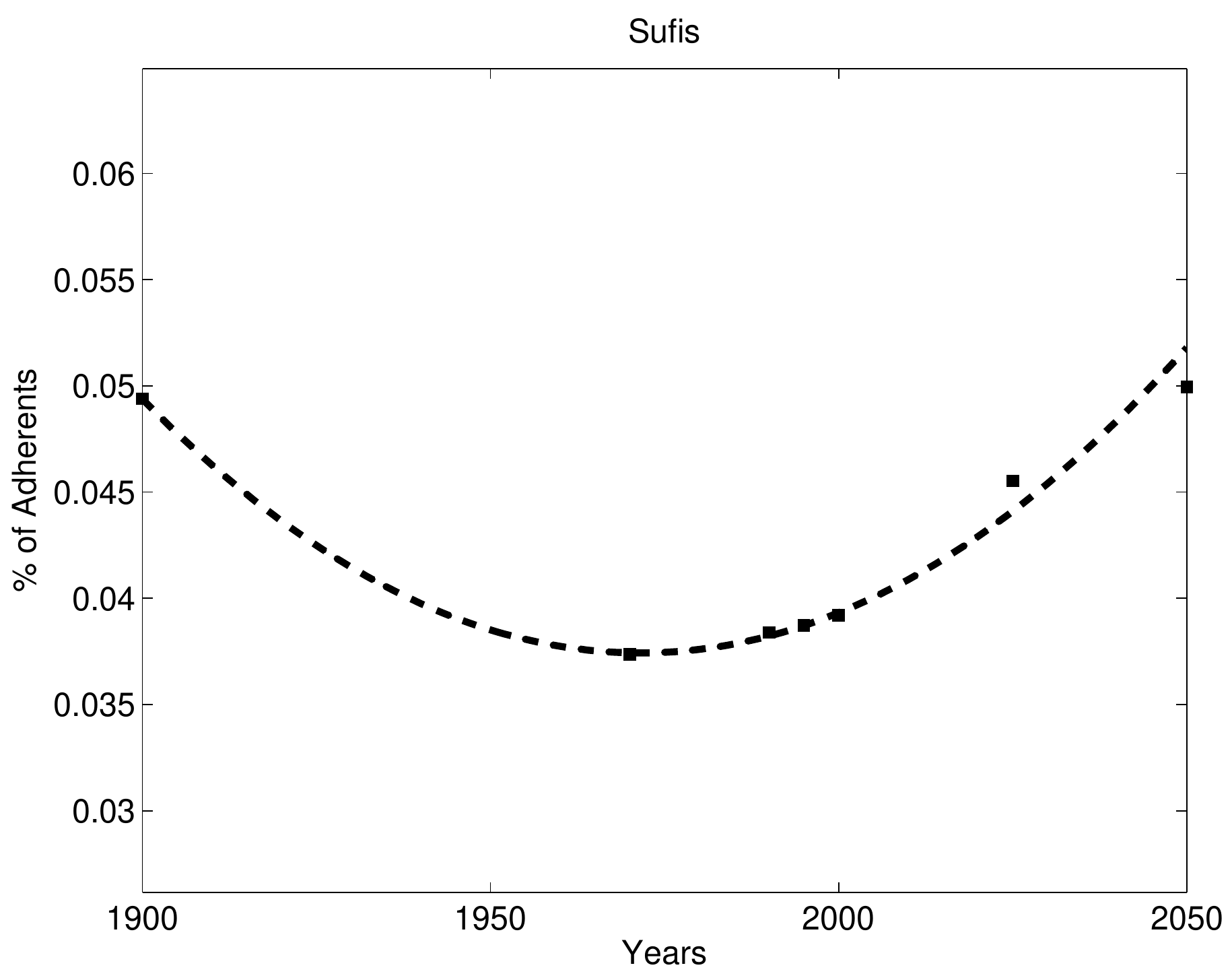}
\includegraphics[height=5cm,width=5cm]{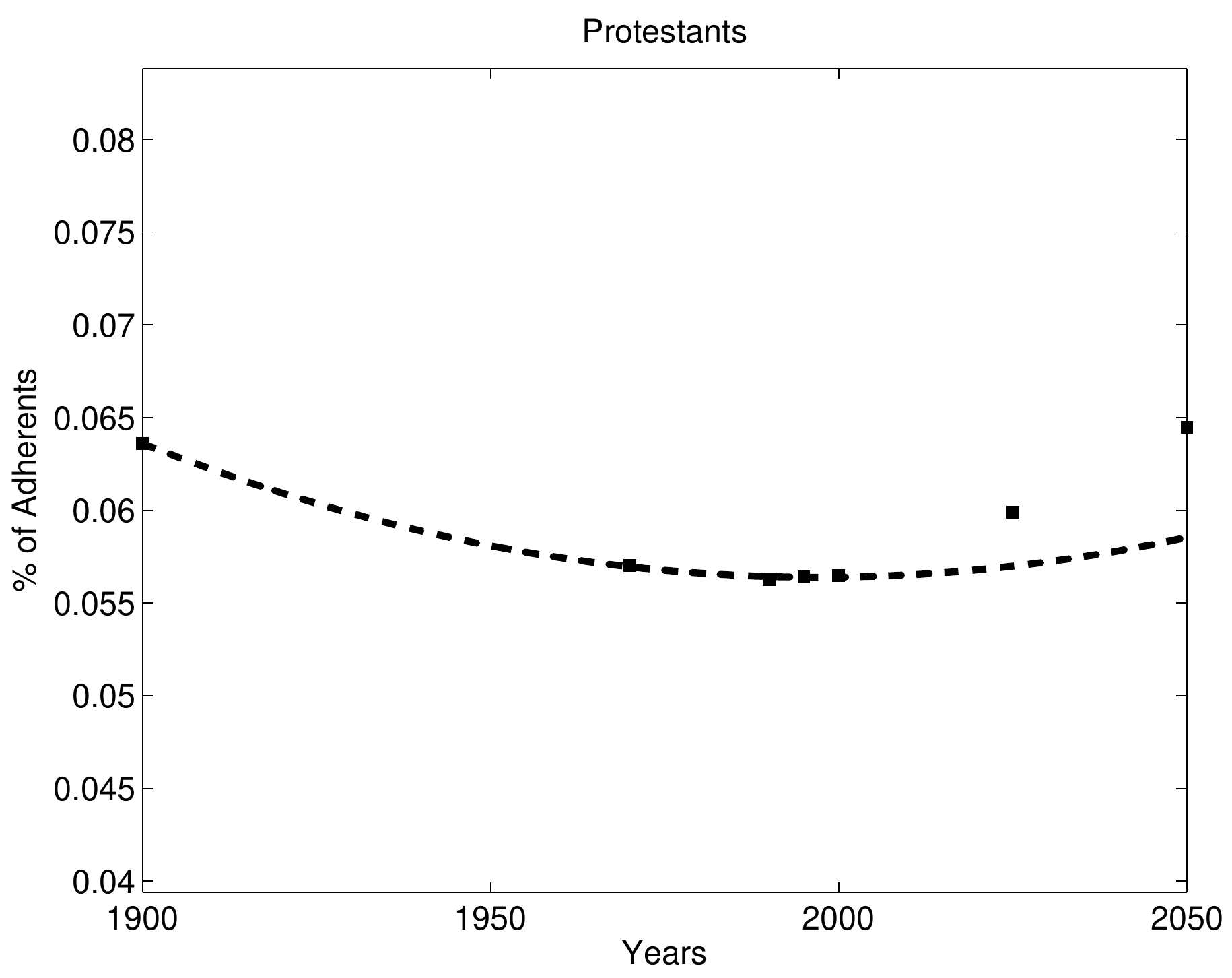}
\caption{\label{fig_lq_new}  Two large size religions indicating a turn over  with a minimum in XX-th century; our theoretical forecasting  slightly underestimates the WTE data}
\end{figure}

\subsection{Growing Religions}

In this subsection, we show cases of small or large size religions which are strictly increasing (Figs. 2-6). The illustrations are sufficiently readable and  understandable that we do not convey much more hereby than in the figure captions.  We distinguish the case of rather good fits (Fig. 4) , or not; we emphasize that we either overshoot  or underestimate the WTE forecast for 2025 and thereafter. We suggest to the reader to compare the figures with the $h$ values in the Table, and observe that the $y$-scales are evidently quite different from figure to figure depending on the rank of the religion.

We observe (Fig.5) that there are three cases where the growth appears to be linear and in all cases underestimate  the WTE  forecast.  We emphasize that growth does not mean lack of saturation, as it should be appreciated:  this is illustrated by two cases in Fig. 6.

\begin{figure}
\centering
\includegraphics[height=5cm,width=5cm]{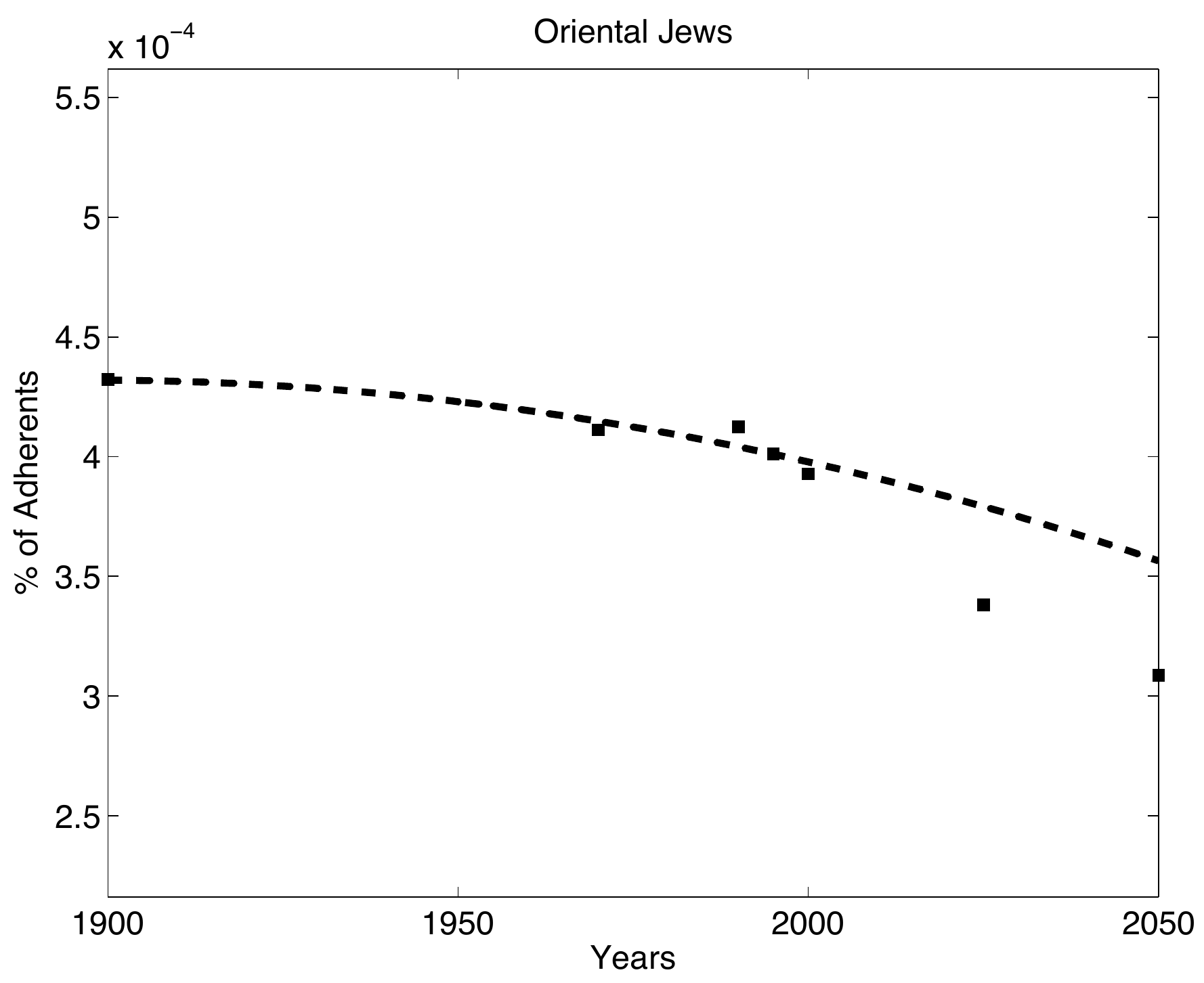}
\includegraphics[height=5cm,width=5cm]{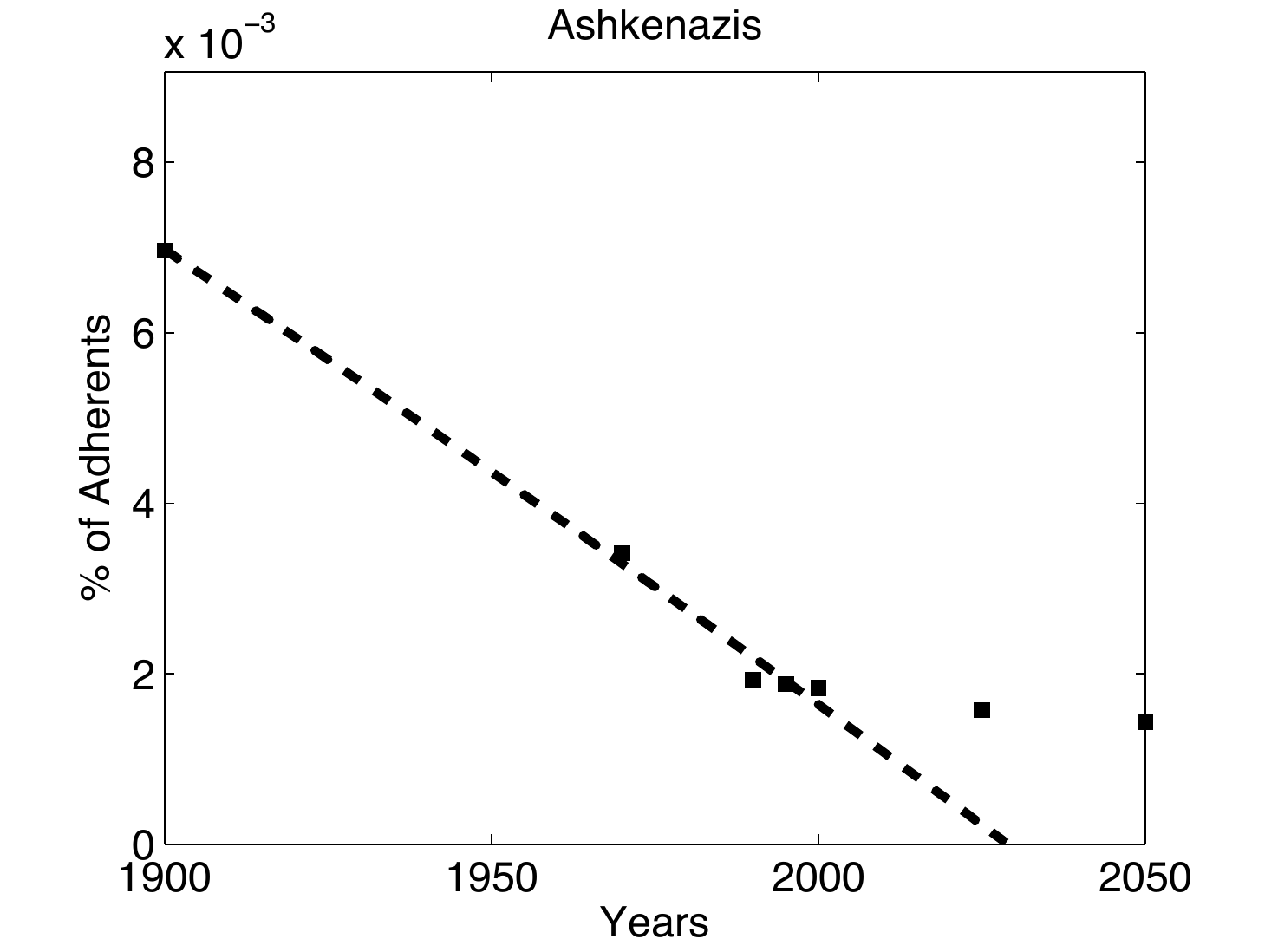}
\caption{\label{fig5.6.7.lq}  Two cases of religions having a markedly predicted collapse after having had a maximum in the XIX-th century}
\end{figure} 

\begin{figure}
\centering
\includegraphics[height=5cm,width=5cm]{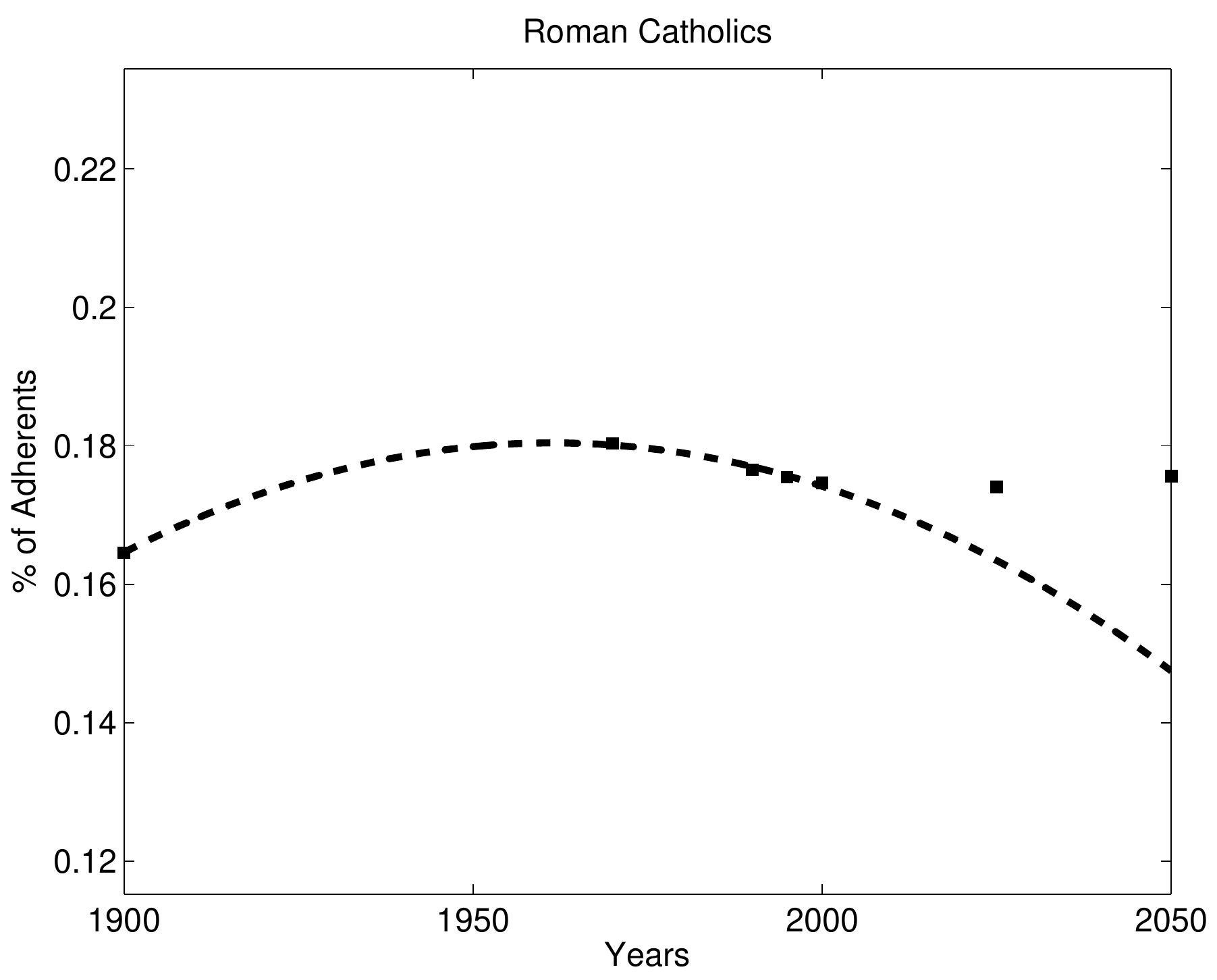}
\includegraphics[height=5cm,width=5cm]{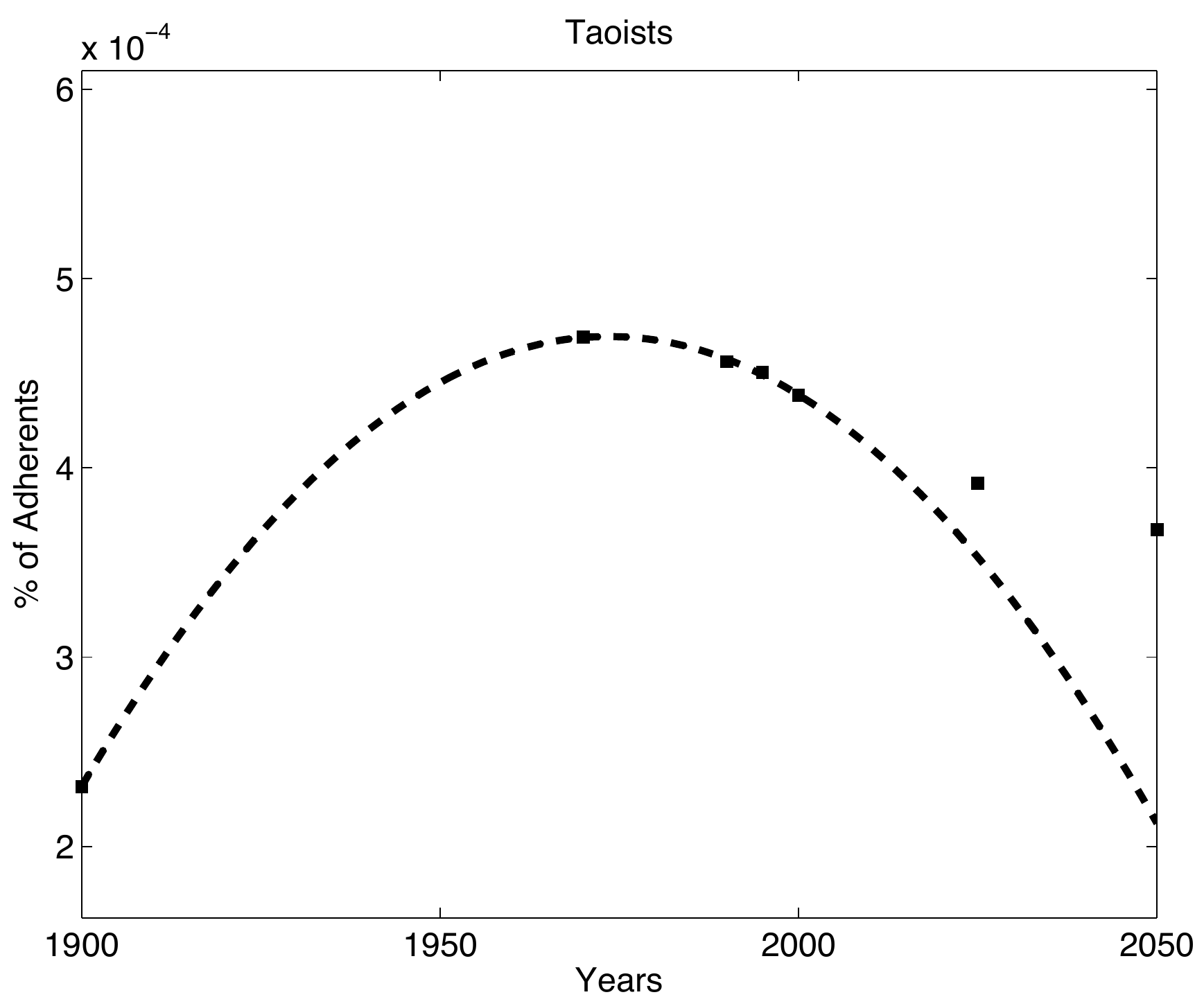}
\includegraphics[height=5cm,width=5cm]{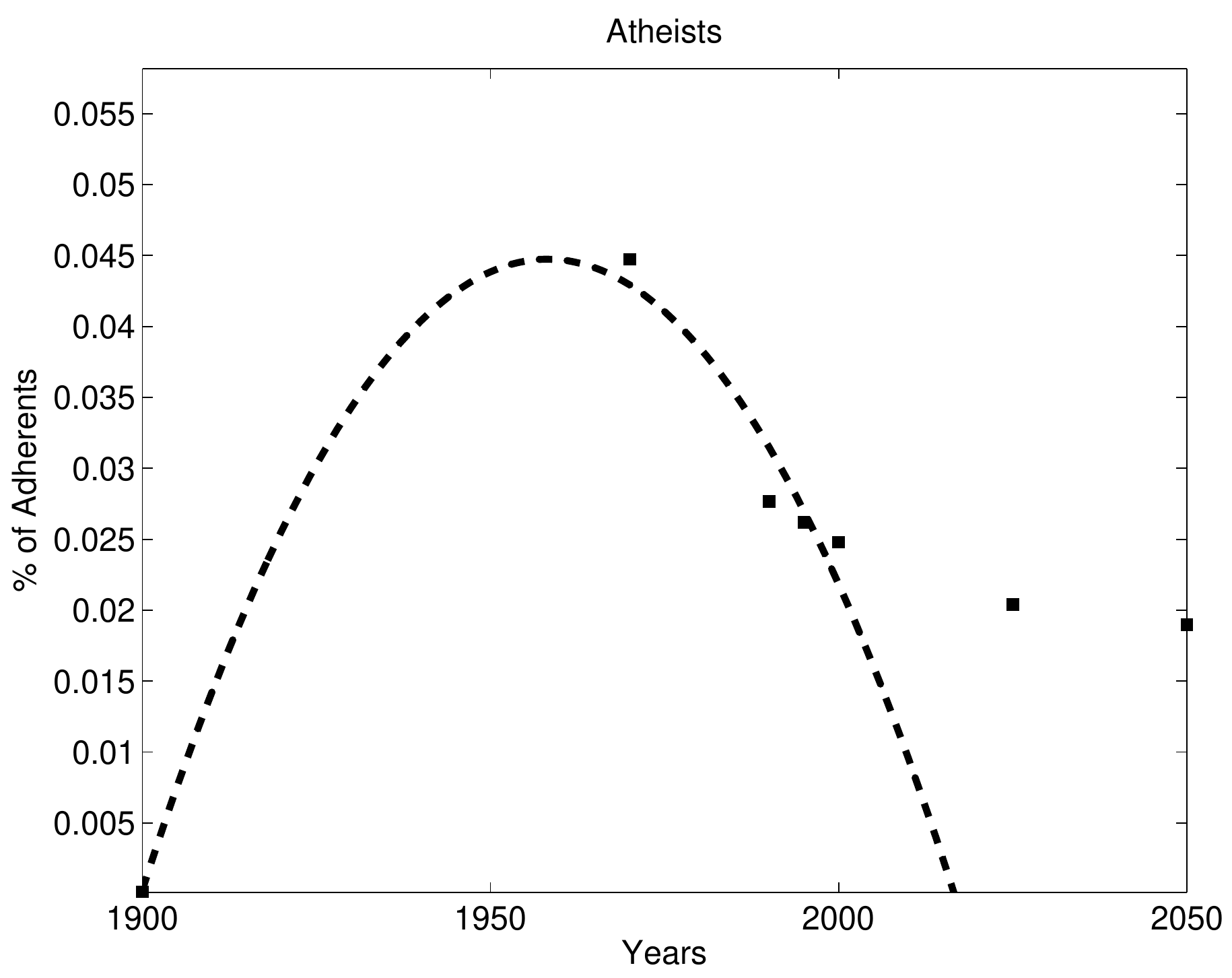}
\includegraphics[height=5cm,width=5cm]{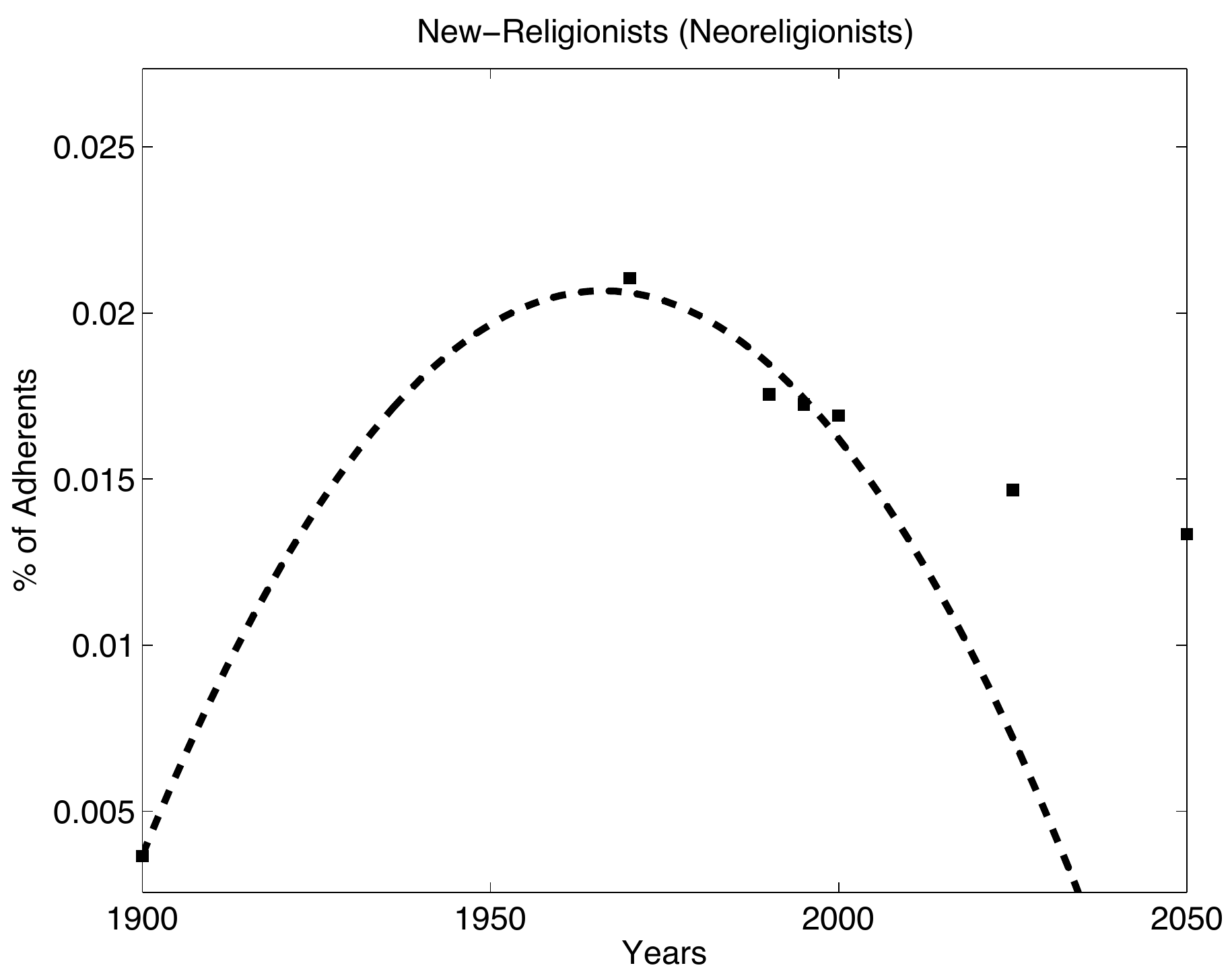}
\includegraphics[height=5cm,width=5cm]{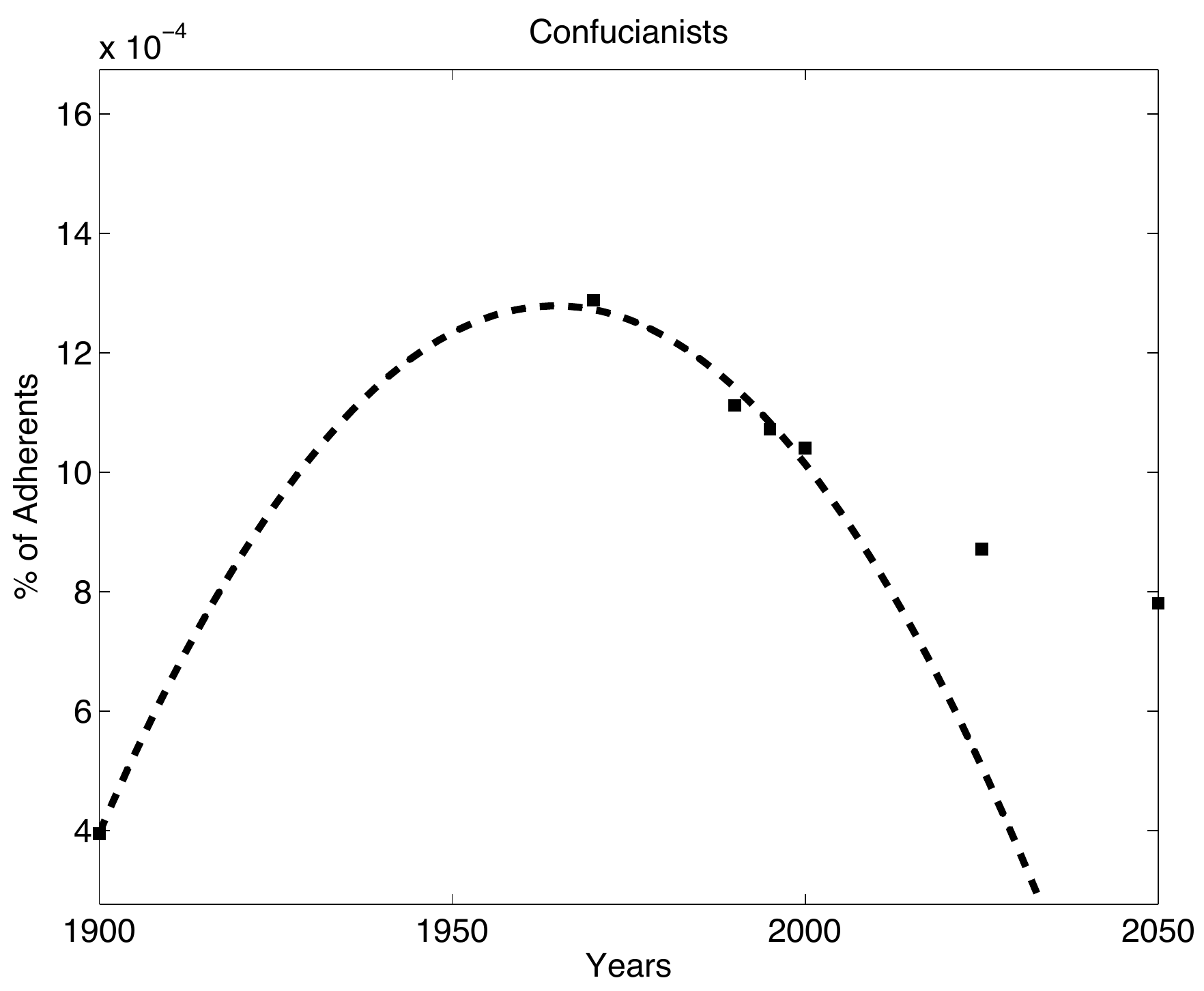}
\includegraphics[height=5cm,width=5cm]{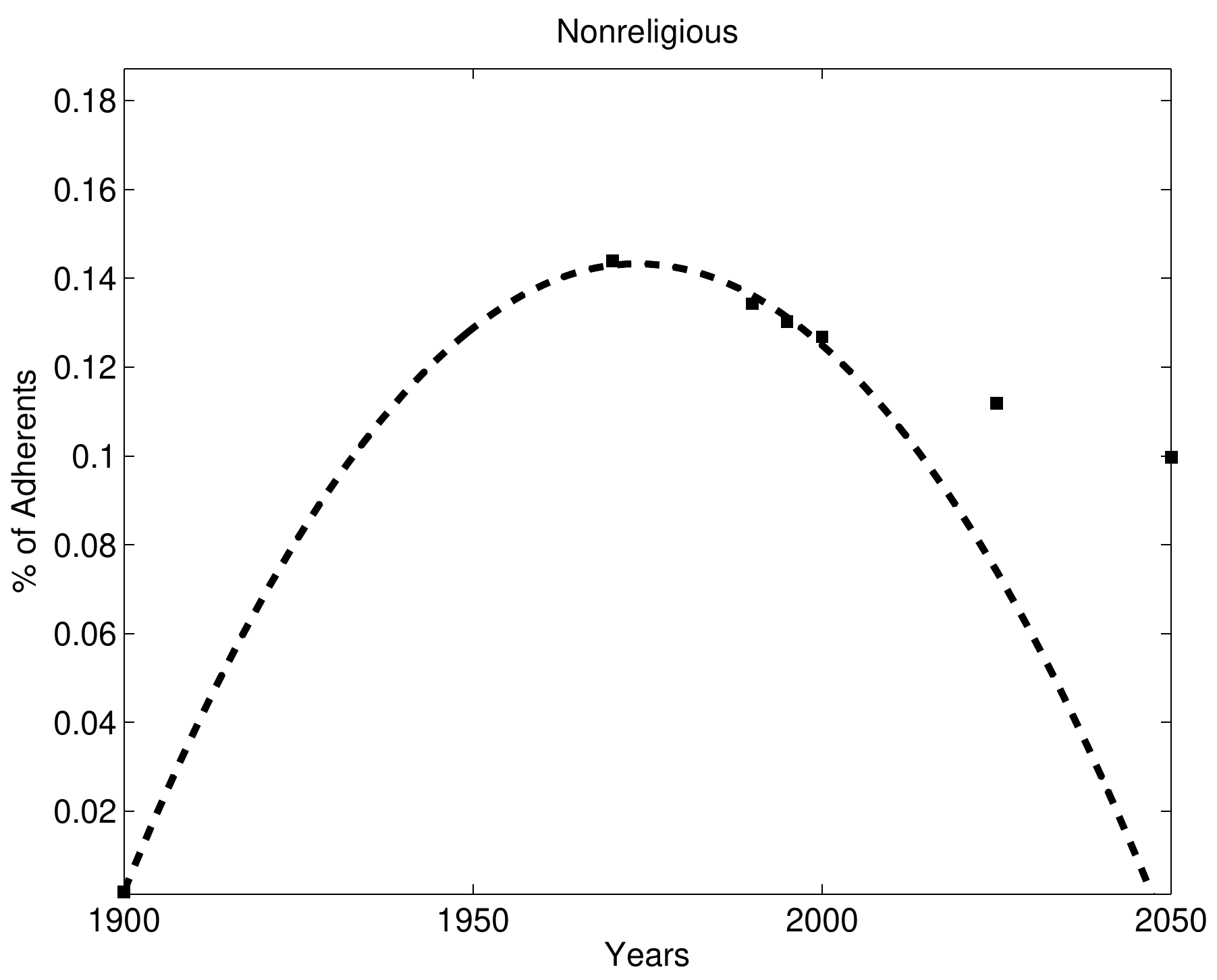}
\includegraphics[height=5cm,width=5cm]{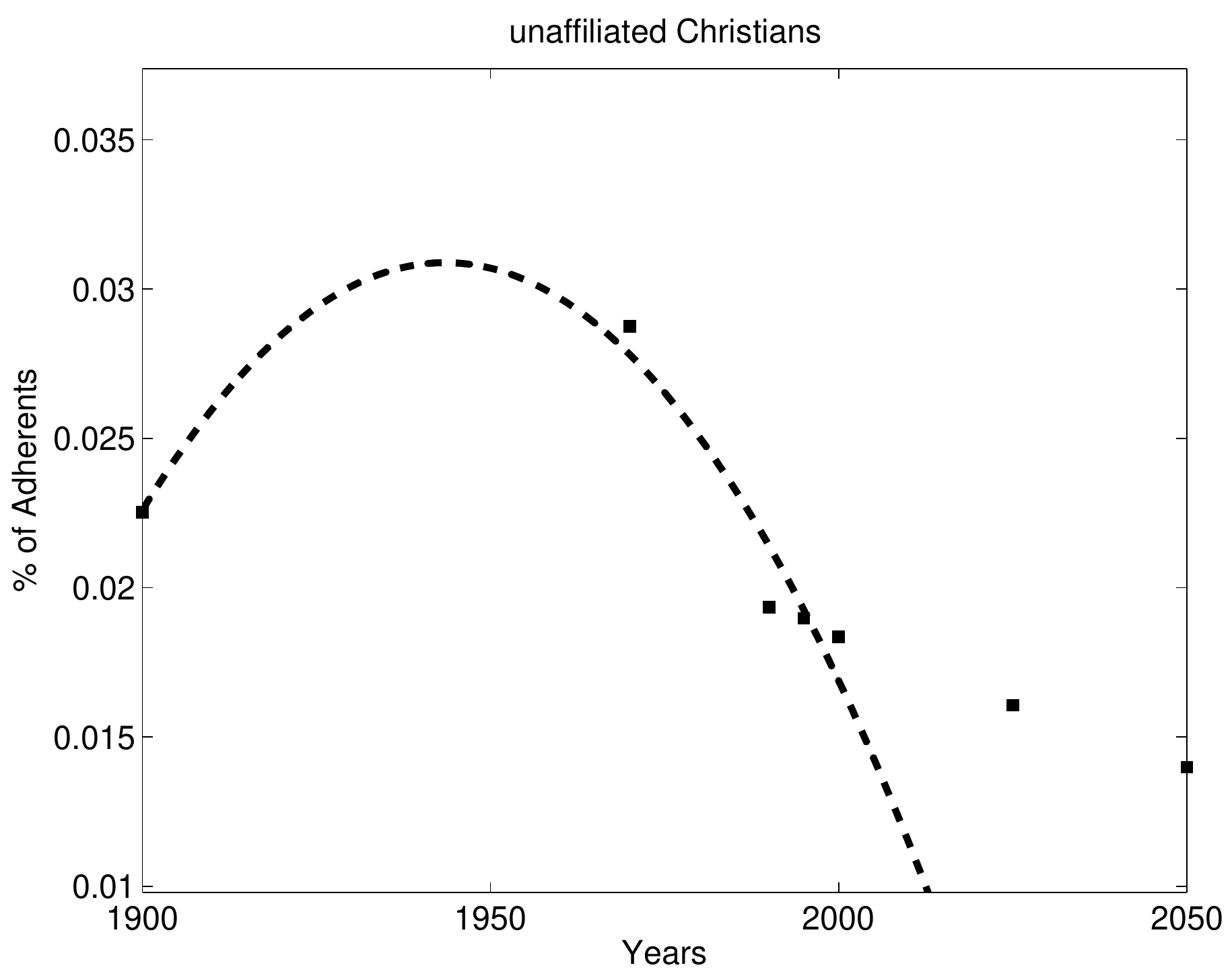}
\includegraphics[height=5cm,width=5cm]{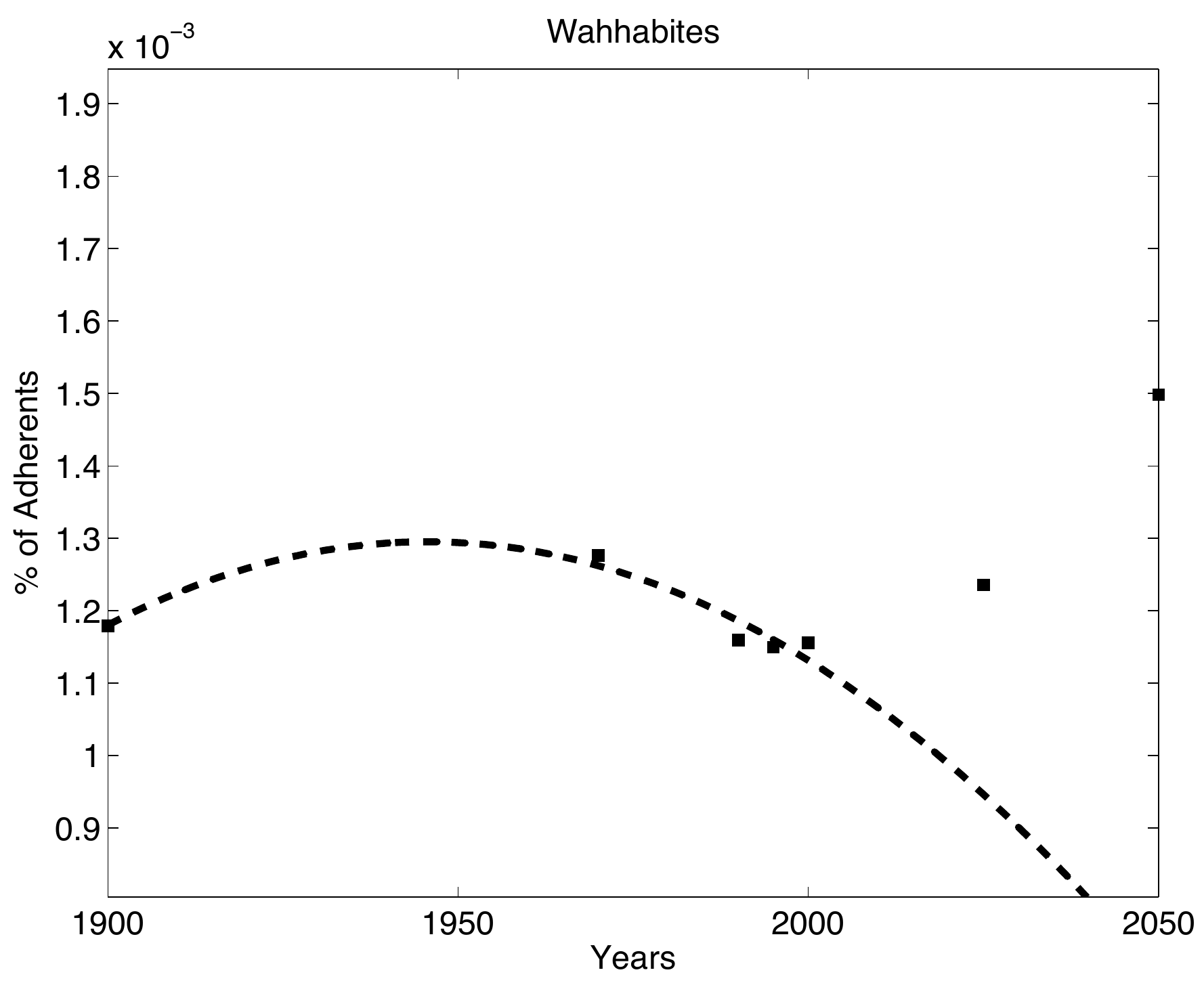}
\includegraphics[height=5cm,width=5cm]{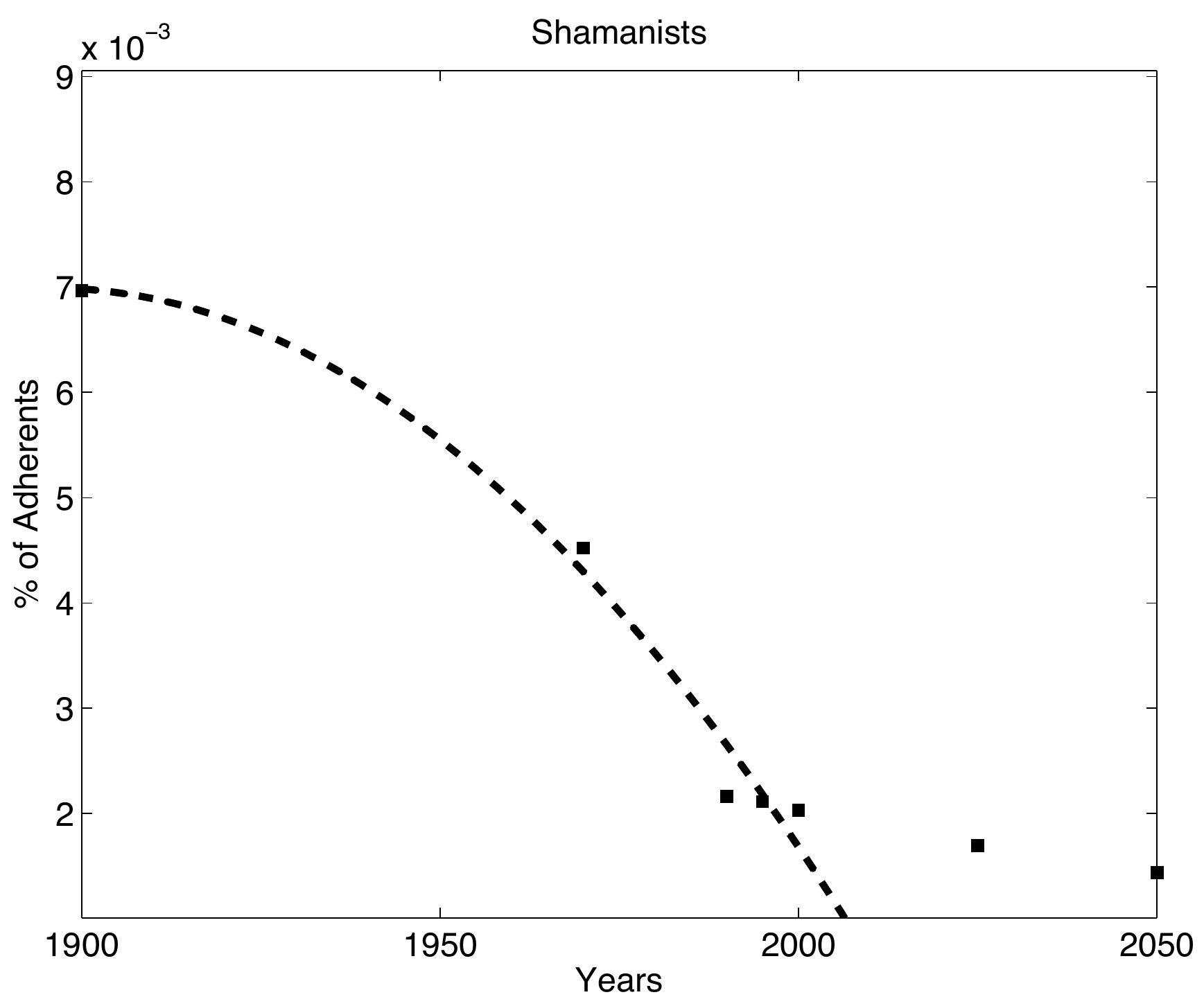}
\caption{\label{fig_lq_new} Nine religions having had a maximum during the XX-th century; the parabolic forecast undershoots the WTE expectation}
\end{figure} 

\begin{figure}
\centering
\includegraphics[height=5cm,width=5cm]{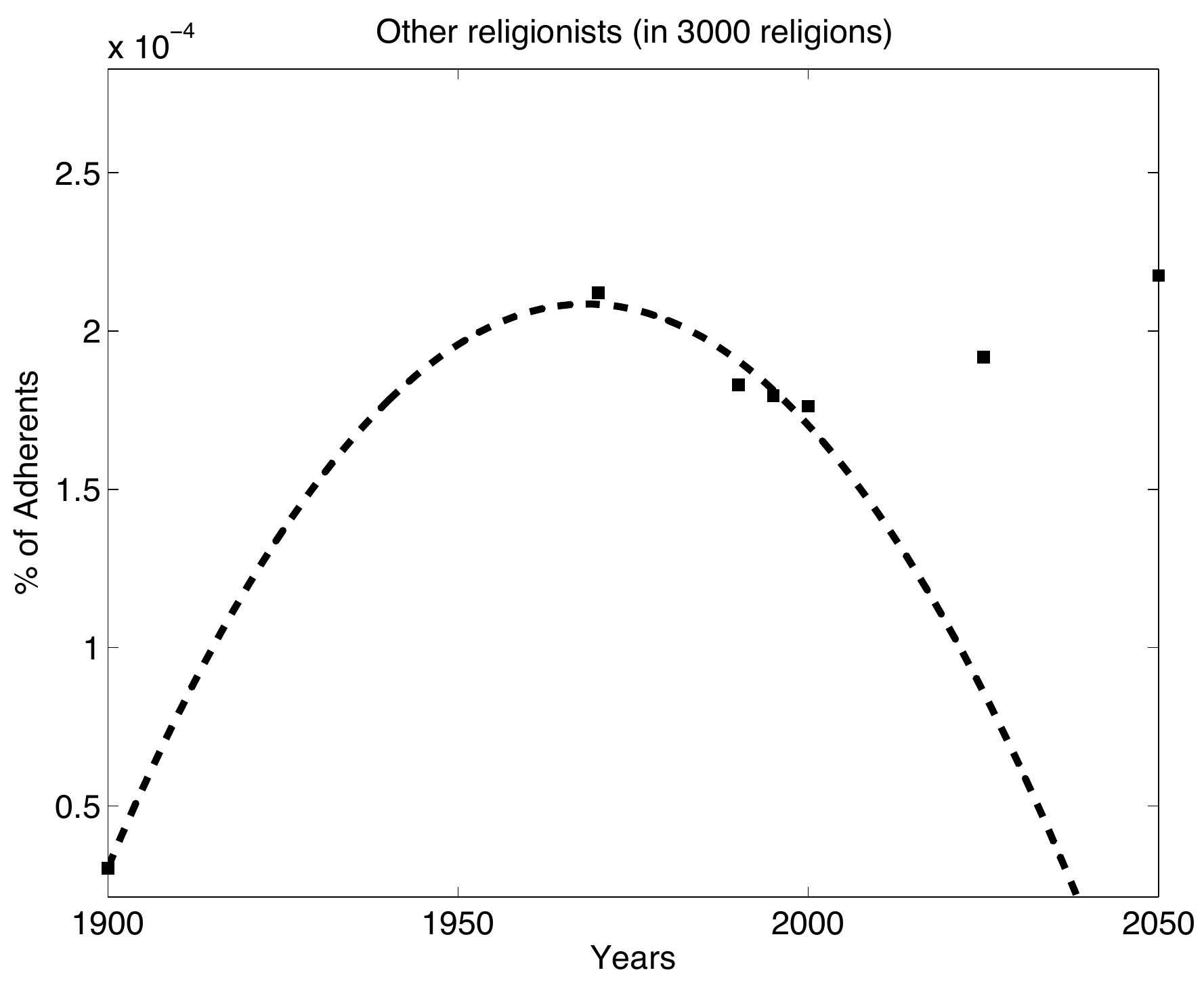}
\caption{\label{RCannd3000}  Case of so called 3000 other religions for which a decreasing behavior is observed; notice the marked underestimate of our forecast with respect to WTE, - predicting an increase in this XXI-th century}
\end{figure}

\begin{figure}
\centering
\includegraphics[height=5cm,width=5cm]{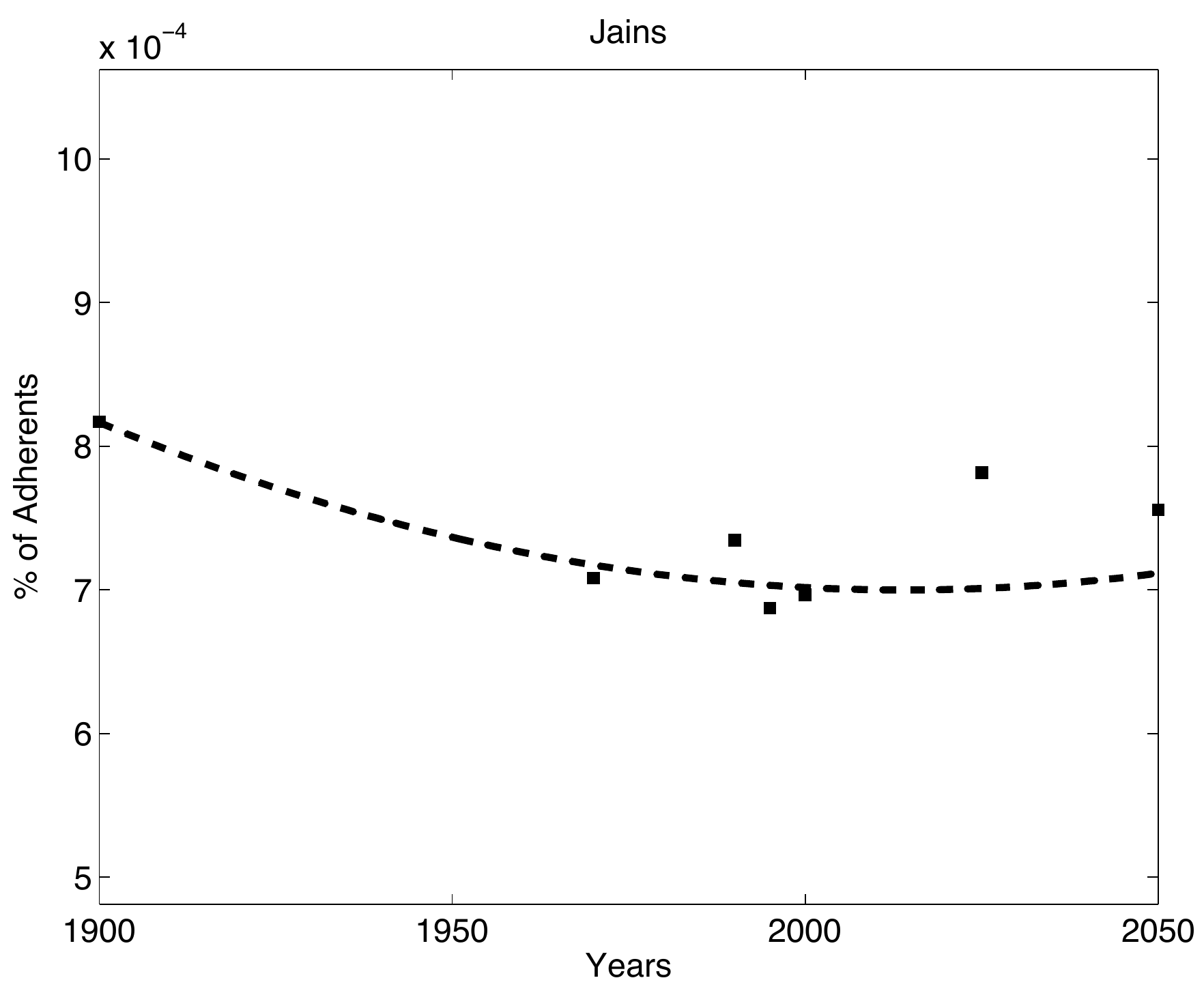}
\includegraphics[height=5cm,width=5cm]{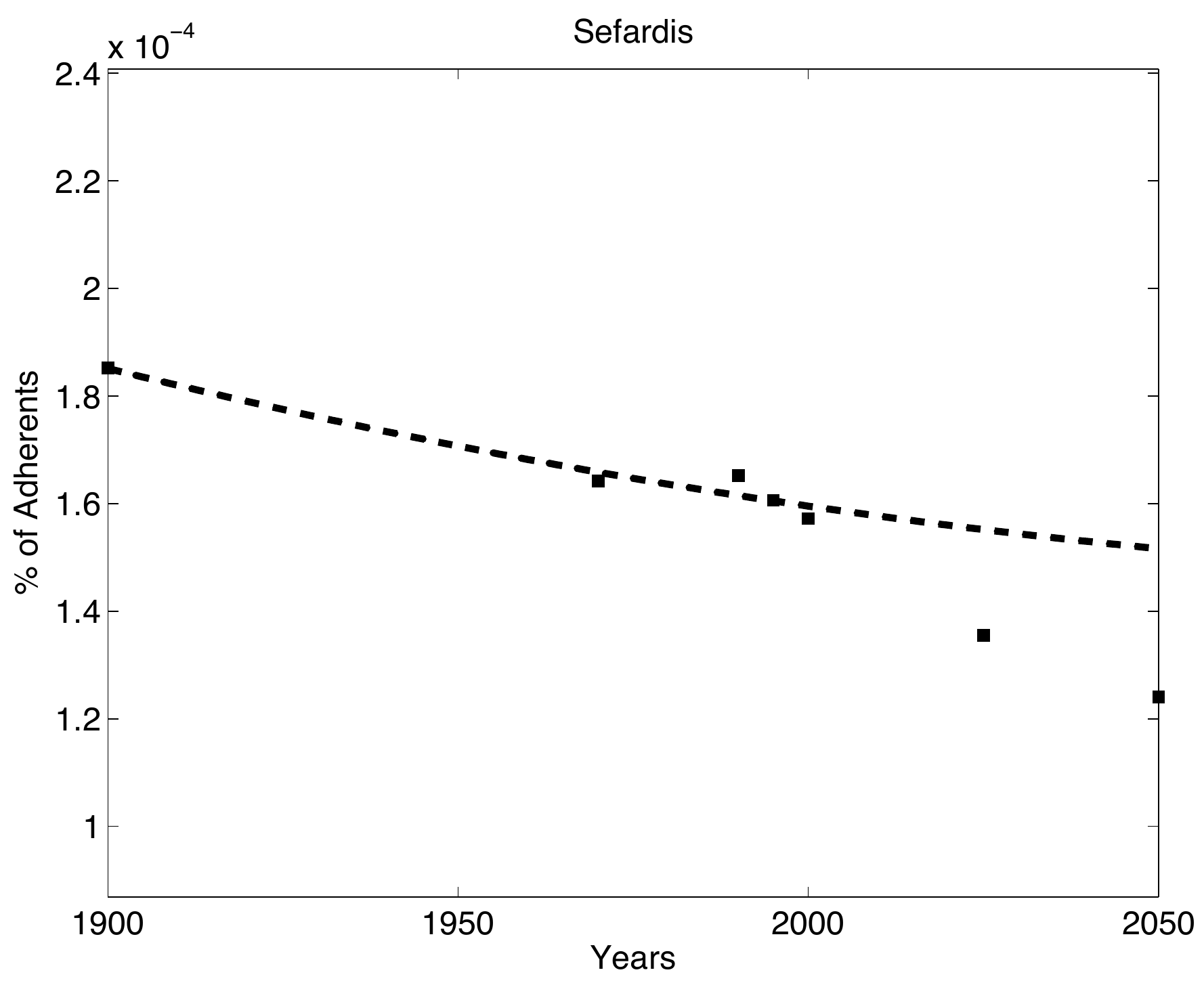}\\
\includegraphics[height=5cm,width=5cm]{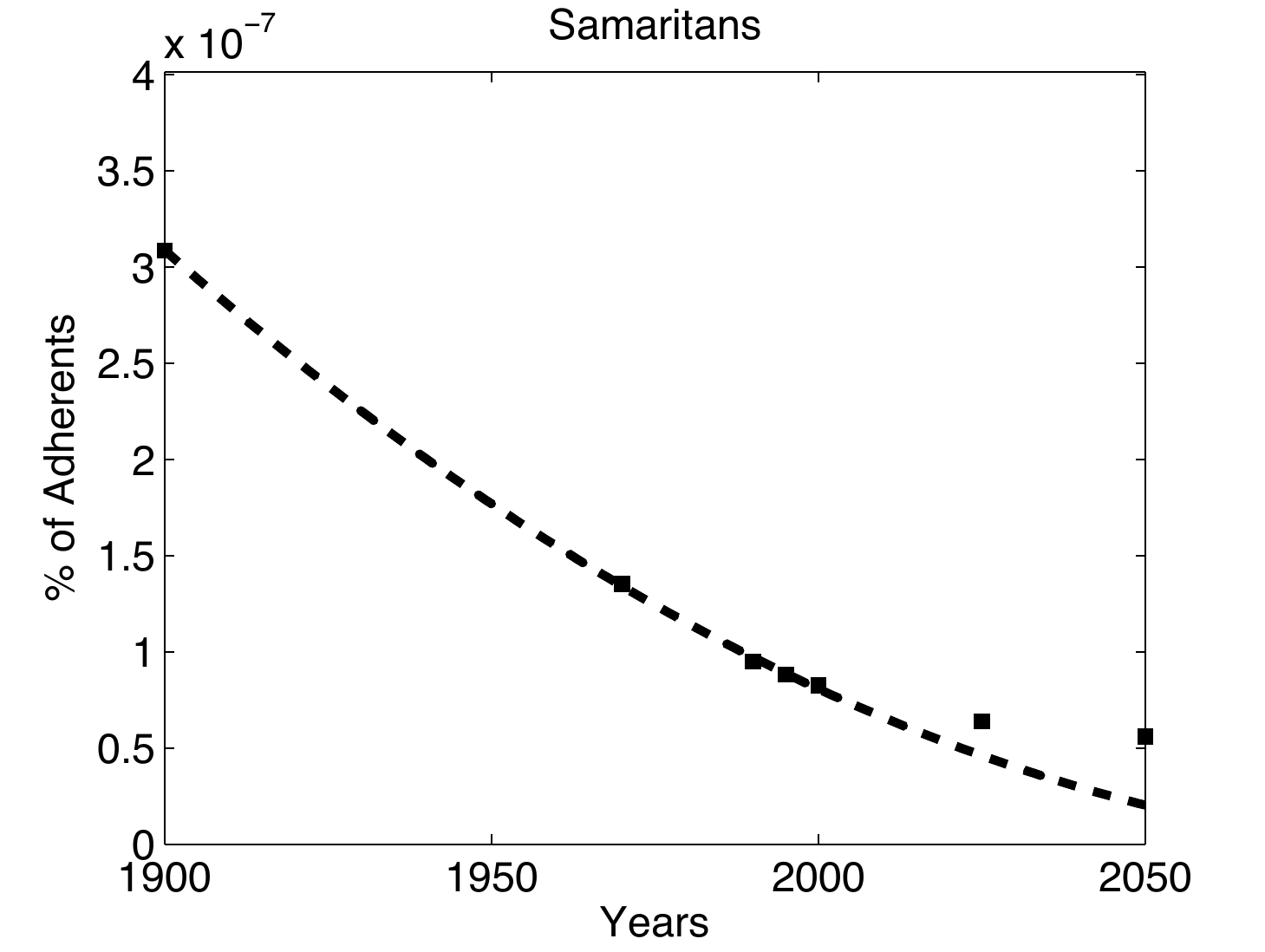}
\includegraphics[height=5cm,width=5cm]{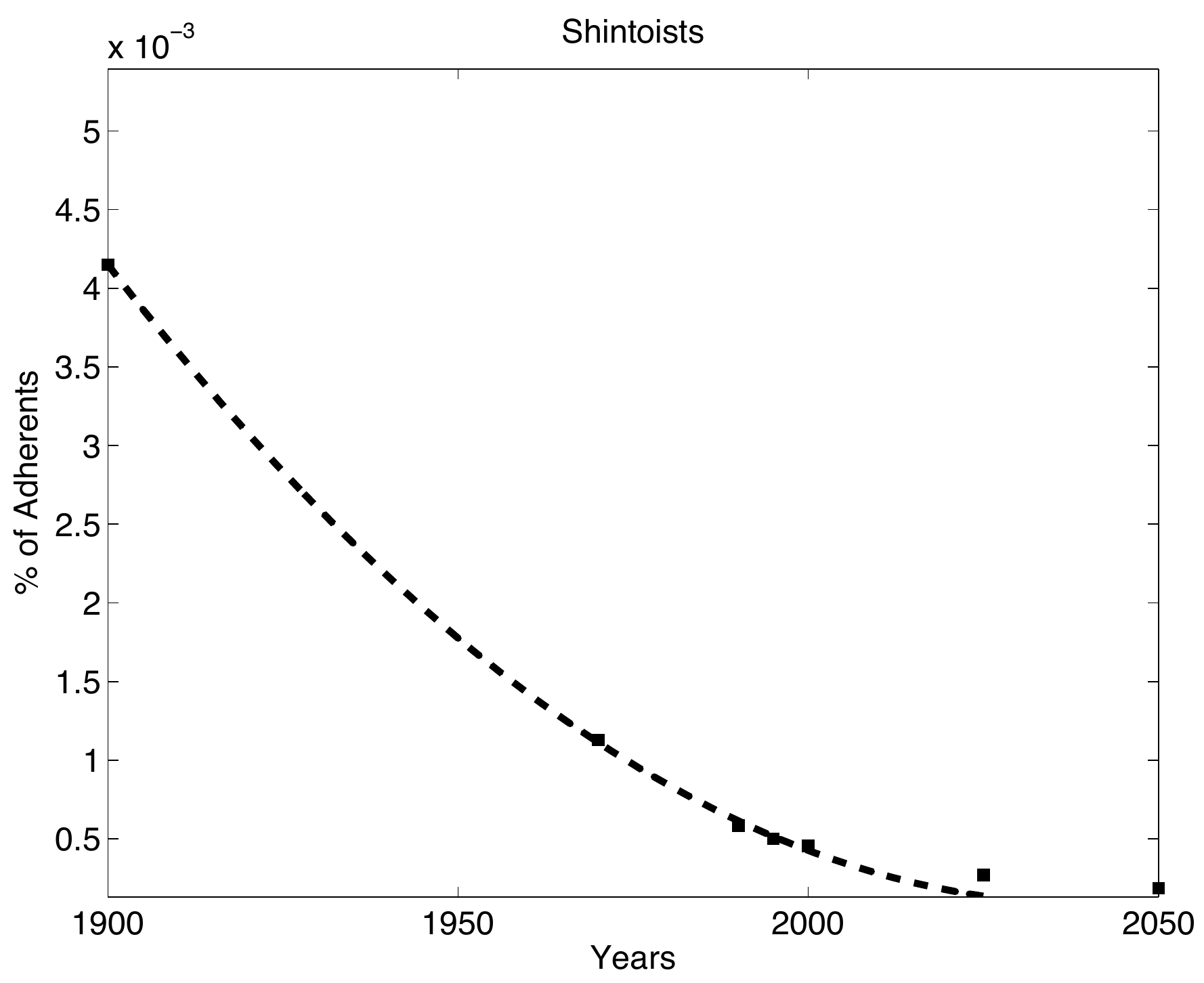}
\caption{\label{fig_lq_new} Four cases of equivalent size, but rather small, religions having a relatively complex behavior, apparently decaying during the XX-th century, but with debatable forecasting for the XIX-th century }
\end{figure} 

\subsection{Decaying religions}

Turning to apparently decaying religions (Figs. 7-8) , the same classification can be made: either there are  strongly decaying cases or smooth decays, with remarkable fits, even though there is no $a$ $priori$ knowledge of the time when the number of adherents is maximum. The empirical forecast is in almost all cases  in rather good agreement with the WTE one.

\subsection{Cases with presently observed extremum}

Sometimes  there are ''presently decaying'' (''growing'') religions for which a maximum (minimum) is observed during the present centuries.  For such ''religions''    the number of adherents  can be fitted with
a second order polynomial  
$y=A+Bx+Cx^2$ ; the relevant parameters for these cases are 
given in  Table \ref{table3}). Notice that these are rather large size religions. 
In the Fig. 9  cases   we overshoot the WTE forecast, - not in Fig. 10.
In Fig. 11, the maximum occurs during the XIX-th century, and apparently there is a strong prognosis for the collapse of these religions. In Fig.  12, we underestimate the WTE forecast, which is thus more optimistic than the XX-th century data evolution indicates.  In some (3) cases  (Fig. 13) the collapse seems rather obvious, though the tail of the evolution law has some (sometimes  large) error bar, not shown for data readability.   Observe that   the "other religions", containing 3000 or so smaller denominations present also a parabolic  convex shape. One can debate whether the parabola makes physical sense of course. It might happen that such religions will disappear, but it might be through a tail evolution, like presumably the cases of Fig. 14. More data should be of interest in such  cases.

\section{Discussion and Conclusions}

We consider that the religious practice is more likely more diverse than WCE and WTE surveys indicate. 
yet claiming the interest of the data we suggest to let religious adherence to be a degree of freedom of a population, and take it through statistical physics considerations for our enlightment. Therefore we have analyzed 58  cases of growing and decaying (so called)  religions  observing several groups through the analytical behaviors. We indicate that with an Avrami equation the fit can be quite often good, in particular for the growing cases.  Physically speaking that gives some support to the conjecture of  religions grow like crystals \cite{iop}. However we cannot expect that the Avrami equation holds true for ever; the system should saturate at some point, except if only a few religions are excessively predominating, and not ''allowing'' (in a thermodynamic sense) the probability of existence of others. 

The same is true for the parabolic fit, which either indicates a quite quickly forthcoming disappearance of a religion or allows for infinite growth.  We recognize that these are approximations. 

It seems that  we often  overestimate/underestimate the WTE theoretical trend in the decaying cases and in several growing cases, though we sometimes agree in the latter cases. Again we claim that the WTE  trends can be quite arbitrary, supposedly predicting a linear evolution from the last three data points in the surveys. It might be interesting to use other types of statistical analysis to conclude whether the forecasts so much differ from one another. ... As well perform detailed analyses taking into account error bars in the original data.
E.g., the case of Zoroastrians (in Fig. 3) indicates an anomalous point corresponding to 1975, while other cases seem to indicate major (but unknown)  error bars (like on Figs. 5)  on the data from the surveys. To resolve such questions is outside the scope of this report.

Turning to the  data displayed on different figures, a ''high'' growth is seen for  Hanfites, Shafiites and Malikites  which are all Sunnists. Maybe we should not need to add a comment based on ''political considerations'' here, but we may consider that the meaning of $h$ makes sense again. In fact this is emphasized when considering two of the highest growth rates, i.e. as found for Charsimatics and Independents, though a strict  {\it late growth stage theory} might be debated upon. 
One case where one case trust the data points is likely that of the black muslims (Fig. 3 ) since they are hardly existed before 1900, whence for which an Avrami equation would hold. It would be  very interesting to check soon the number of adherents in such a case. 

Finally observe that the ''non religion'' adherent data finds a remarkable position as the fourth growing "denomination''. Observe the maximum in the number of ''adherents'' in such a case near 1970, rendering the theory (or the data !) to be debated upon.

In conclusion, here above we have shown that we can attempt to make a statistical physics like analysis of the number of adherents in religions, going beyond our first paper \cite{religion1} on the subject. However the data seems sometimes barely reliable.
 
 Nevertheless one can, expecting better surveys, at a more limited scale, suggest further lines of research. One could suggest agent based models like for languages, including the role of external fields. One could try to have a Langevin equation connexion to Avrami equation;  of course  we need to define a hamiltonian $H$ and a current : that implies interactions thus competitions between entities; what we do not see here yet.   However the hamiltonian can be obtained following standard  ideas, like turning over the pdf into its log and defining some temperature. Religions seem to be an interesting field of study for statistical mechanics!

        \section{Appendix A. Languages vs. Religions}
        
        Through this      Appendix A we  wish to outline what we consider are a few aspects, i.e. ''differences'' , between languages and religions, from a physics point of view, perspective or input into modeling their sociological features; see  
        Table \ref{tab01AppA}, as a summary of to be considerations of interest.
        
        \begin{table}

\caption{Comparison : similarities and differences between languages and religions seen from a statistical physics point of view}
\smallskip
\begin{footnotesize}
\begin{center}
\begin{tabular}{c| c c c|  c c c c c c}

\\\hline 
\\&  & Languages &	 &	 &		Religions &	& \\
\\\hline
\\&  & more than 6000 &	 &	 &		more than 3000 & &	 \\
\\\hline
\\ agents & multilingual &	frequent &	  &	 &	polyreligious &	rare &	 \\
\\\hline
\\ variety &	 &	huge: &	dialects, slangs  & &	huge: &	denominations, sects &	\\
\\\hline
\\ time scales &	   &  &	  &	 &	 &	  & \\
\\ &	nucleation  &slow  &	  &	nucleation&	fast &	  & \\
\\  &	growth  &	slow&	  &	 &	fast & through	avalanches &	 \\

\\  &	decay &	fast&	  &	 &	slow&	  & \\
\\\hline
\\ semantics &	grammar&	vocabulary&	  &	& images&	 rituals  &	 \\
\\\hline
\\  	applied fields& &	 rare&	  &	 & many, strong&	 &	 \\
\\\hline

\end{tabular}
\label{tab01AppA} 
\end{center}
\end{footnotesize}
\end{table}
  
  We insist that  in physics one should study the response of the system to intrinsic or extrinsic fields.     We may describe  the population of agents through a free energy, Hamiltonian  formalism or Langevin equation indicates that all terms, ordered along  the increasing size of the cluster,  should be included
      
                \section{Appendix B. Indicators of Religion Status}
             
              The time dependence of the  number of adherents can be considered to be a very restrictive    way to ''measure''  the evolution of a religion.   One could also ''weight'' the level of adherence to a religion. For example, one could try as for languages to define a religion through its quantity of practitioners, rituals,  ....  Many other indicators are possible. One can measure diverse quantities related to the  religious efffect. As in physics one can search for the relation between causes and effects, the response to internal or/and external fields. 
              
              As there are  several definitions of a  language \cite{klinkenberg}, similarly one could also define what a religion ''is'' in different ways  \cite{Dennett}.

             First let us list a few definitions of religions form the conventional literature : 
    \begin{enumerate}
\item             
                Barns \& Noble (Cambridge) Encyclopedia (1990):
"...no single definition will suffice to encompass the varied sets of traditions, practices, and ideas which constitute different religions."
\item  The Concise Oxford Dictionary (1990):
"Human recognition of superhuman controlling power and especially of a personal God entitled to obedience.Ó
\item  Webster's New World Dictionary (Third College Edition):
"any specific system of belief and worship, often involving a code of ethics and a philosophy.Ó

\item  Merriam-Webster's Online Dictionary:
"a cause, principle, or system of beliefs held to with ardor and faith."
                         \end{enumerate}  
                         
                         In fact, we can admit that
                             \begin{enumerate}
  \item               Religion is any specific system of belief about deity, often involving rituals, a code of ethics, a philosophy of life, and a worldview. (!)
 \end{enumerate}  
 accepting that those not included in the above are ''non religious'', in which one can distinguish between atheists, agnostics, non-interested ones, etc. ,  while we can contain an adherent of whatever ''denomination'' into a definition like
    \begin{enumerate}
  \item  
An adept is an individual identified as having attained a specific level of knowledge, skill, or aptitude in doctrines relevant to a particular (author or) organization. 
         \end{enumerate}

                It is indeed clear that a religious adherent instead of being an analog of an up or down spin, is rather a vector for which each element can be a quantity of value like considered  in sociology, i.e. a ''quality''. Next one may imagine a Potts vector or ferrroelectric type of  (Hamiltonian) models for describing an ensemble of religious  agent evolution or state.  Quantitative and qualitative dynamical evolutions of agents and groups (''denominations'') can also find some basis in many competition and organization physics models.

                Moreover, one should consider religions from another ensemble of point of views also called sometimes indicators)
         \begin{enumerate}
\item  Number of groups, sects,
\item Number of churches, parishes,
\item Number of chapels, sites,
\item Number of  ÒpriestsÓ, (clergy)
\item Number of  ÒbelieversÓ, 
	sex, age, wealth, language,
\item Intensity of participations, in rituals, in practicing principles, 
\item Wealth and financing,
\item Type of hierarchy, ...
          \end{enumerate}

               No need to say that physicists are not the first ones to reflect on variability in religion distribution  or adherence level. We may find already such considerations in  books and papers by specialists of the history or sociology of religions \cite{Dennett}.

                \acknowledgments
The work by FP has been supported by European Commission Project
E2C2 FP6-2003-NEST-Path-012975  Extreme Events: Causes and Consequences.
Critical  comments by  A. Scharnhorst have to be mentioned.


\begin{thebibliography}{99}
\bibitem{footnote1} However one can recall the case of ''Matines Brugeoises'' , when flemish ''peasants'' killed the french nobles, recognized as such, because they could not pronounce correctly "schild and vriend''. Another case is that of the khmer rouges killing vietnamese educated intellectuals in Cambodia.
\bibitem{footnote2} It is sometimes hard to know or to be sure whether an adherent, a disciple, ... is truly a member of a religious denomination or church. However this caveat pertains to usual problems encountered in sociological investigations.
\bibitem{footnote3} ... admitting indifference, atheism, agnosticism, ... as a sort of religion, from our point of view.
\end{thebibliography}
\end{document}